%% file: LIF.tex
\documentclass[12pt]{article}
\usepackage{amsmath,amsfonts,amssymb}
\usepackage{graphicx}
\usepackage{epstopdf}
\usepackage{placeins}

\newcommand{\be}{\begin{equation}}
\newcommand{\ee}{\end{equation}}
\newcommand{\bea}{\begin{eqnarray}}
\newcommand{\eea}{\end{eqnarray}}

\begin{document}
\newcommand{\todo}[1]{{\em \small {#1}}\marginpar{$\Longleftarrow$}}   
\newcommand{\labell}[1]{\label{#1}\qquad_{#1}} 

\rightline{DCPT-11/41}   
\vskip 1cm 

\begin{center} {\Large \bf Flows involving Lifshitz solutions}
\end{center} 
\vskip 1cm   
  
\renewcommand{\thefootnote}{\fnsymbol{footnote}} 
\centerline{\bf Harry Braviner\footnote{harry.braviner@durham.ac.uk}, Ruth Gregory\footnote{r.a.w.gregory@durham.ac.uk}  and
  Simon F. Ross\footnote{s.f.ross@durham.ac.uk} }
\vskip .5cm 
\centerline{\it Centre for Particle Theory, Department of Mathematical Sciences}
\centerline{\it Durham University, South Road, Durham DH1 3LE, U.K.}

\setcounter{footnote}{0}   
\renewcommand{\thefootnote}{\arabic{footnote}}

\begin{abstract}
We construct gravity solutions describing renormalization group flows 
relating relativistic and non-relativistic conformal theories. We work both 
in a simple phenomenological theory with a massive vector field, and in 
an $\mathcal N=4$, $d=6$ gauged supergravity theory, which can be 
consistently embedded in string theory. These flows offer some further 
insight into holography for Lifshitz geometries: in particular, they enable 
us to give a description of the field theory dual to the Lifshitz solutions in 
the latter theory. We also note that some of the AdS and Lifshitz solutions 
in the $\mathcal N=4$, $d=6$ gauged supergravity theory are 
dynamically unstable.
\end{abstract}

\newcommand{\zed}{$\mathbb{Z}_2$}

\section{Introduction}

The use of gravitational duals to study strongly-coupled field theories 
\cite{Maldacena:1997re,Aharony:1999ti} has provided a unique 
calculational tool which has shed light on a number of important 
questions concerning such field theories. The domain to which this 
holographic approach has been applied has recently been 
substantially enlarged to include applications to field theories of 
interest to condensed matter physics (see 
\cite{Hartnoll:2009sz,McGreevy:2009xe} for useful reviews). 
In particular, models have been developed which exhibit
an anisotropic scaling symmetry, $t \to \lambda^z t, 
x^i \to \lambda x^i$, with $z$ being referred to as the dynamical 
exponent. Systems with this scaling 
symmetry can arise as critical points in condensed matter systems. 
If the field theory has this scaling symmetry, translation and spatial 
rotations as symmetries, but no boost symmetry, it is commonly 
referred to as a Lifshitz field theory. 

A holographic duality for these field theories was proposed in 
\cite{Kachru:2008yh}. The proposal is that the dual of the field 
theory vacuum is a bulk metric
\begin{equation} \label{lif} 
ds^2 =  L^2 \left [ -e^{2zr} dt^2 + e^{2r} d\vec{x}^2 +dr^2\right ] ,
\end{equation}
where $L^2$ represents the overall curvature scale, and the spacetime 
has $d+1$ dimensions, so there are $d-1$ spatial dimensions $\vec{x}$. 
This metric is referred to as a Lifshitz geometry; the anisotropic scaling 
symmetry is realised as an isometry throughout the bulk geometry, 
analogous to the 
conformal symmetry in the original AdS/CFT correspondence 
\cite{Maldacena:1997re}. Such a metric can be realised as a solution 
in a variety of bulk gravitational theories with different matter content. 
In \cite{Kachru:2008yh}, the bulk theory involved two $p$-form fields 
with a Chern-Simons coupling. A simpler theory with a massive vector 
(which is on-shell equivalent to the previous theory) was introduced 
in \cite{Taylor:2008tg}. More recently, \eqref{lif} was realised as a 
solution in string theory in a number of different truncations 
\cite{Balasubramanian:2010uk,Gregory:2010gx,Donos:2010tu,
Donos:2010ax,Cassani:2011sv} (see \cite{Hartnoll:2009ns} for 
earlier attempts). 

Here, we will focus both on the phenomenological massive vector 
theory of \cite{Taylor:2008tg}, which provides the simplest context 
for studying this geometry, as well as the embedding of four-dimensional 
Lifshitz geometries in the six-dimensional $F(4)$ gauged supergravity 
in \cite{Gregory:2010gx}, which provides a realisation in string theory 
which allows for all values of $z$, up to issues of flux quantisation. 

In both the simple massive vector model of \cite{Taylor:2008tg} and in 
the $F(4)$ gauged supergravity \cite{Romans:1985tw}, there are multiple 
anti-de Sitter (AdS) and Lifshitz solutions. If we apply the usual holographic 
dictionary, the anti-de Sitter solutions would be interpreted as dual to 
the vacuum state in different conformal field theories, and the Lifshitz 
metrics as dual to the vacuum state in different non-relativistic Lifshitz 
theories. It is then naturally interesting to investigate the relations 
between these different field theories. 

In this paper, we study this question by constructing domain-wall 
like solutions which interpolate between the different AdS and/or Lifshitz 
geometries. These domain wall solutions can be  interpreted as dual to 
renormalization group flows between the corresponding field theories.  
In the AdS/CFT context, such geometries were first considered in \cite{Girardello:1998pd}
(supersymmetry-preserving flows were considered in \cite{Freedman:1999gp}). A number of interesting results 
were obtained, including a holographic $c$-theorem. Since the 
theories we are considering have multiple AdS and Lifshitz solutions, 
there are a variety of possible flows: solutions may interpolate between 
AdS or Lifshitz in the UV and AdS or Lifshitz in the IR. The simplest case 
is a spacetime approaching different AdS solutions at large and 
small distances, corresponding to flows between ordinary relativistic 
conformal field theories as in \cite{Girardello:1998pd,Freedman:1999gp}. These can be 
considered as a warm-up exercise for the more complicated cases 
involving Lifshitz solutions. They also provide a nice simple example 
exhibiting a type of IR singularity which was recently pointed out in 
\cite{Copsey:2010ya}. We will comment briefly on the appearance 
of these singularities in the context of the gauged supergravity 
model, but we leave its further investigation for future study. 

The flows can provide insight into holographic renormalization for the 
Lifshitz field theories in a number of ways. Flows which approach an 
AdS solution in the UV and approach a Lifshitz solution in the IR 
provide a relation between the more familiar AdS/CFT correspondence 
and holography for Lifshitz spacetimes. This can be used to understand 
elements of the holographic interpretation of asymptotically Lifshitz 
spacetimes (for example, black holes) by considering instead an 
asymptotically AdS spacetime which approaches Lifshitz at some 
intermediate distance scale, as in \cite{Bertoldi:2010ca}. We also 
note that a number of the contexts in which Lifshitz solutions have 
appeared in the recent literature involve such asymptotically AdS 
solutions, such as \cite{Gubser:2009cg,Hartnoll:2010gu}, or the Lifshitz-like solutions in \cite{Goldstein:2010aw,Hartnoll:2010gu}. 
Our solutions provide relatively simple examples of such interpolations. 
Note some of that the embeddings in string theory which 
give $z=2$ Lifshitz solutions 
\cite{Balasubramanian:2010uk,Donos:2010tu,Donos:2010ax} 
do not also have AdS solutions of the same dimension as the Lifshitz solution, 
so we cannot study such flows in that context (although they may have 
flows between AdS$_{d+1}$ and d-dimensional Lifshitz geometries). The more elaborate model studied in \cite{Cassani:2011sv} does however have such AdS solutions in addition to the $z=2$ Lifshitz solution, and it would be interesting to study this example as well.

If we can identify the conformal field theory dual to the UV AdS solution, this construction 
can also allow us to define the field theory dual to the IR Lifshitz solution as the 
corresponding relevant deformation of the former conformal field theory. To identify the dual 
field theory in this way we need to work in a top-down model; we do not know the field 
theory dual of the AdS solution in the simple massive vector model. In the gauged 
supergravity of  \cite{Romans:1985tw}, by contrast, the dual description of the AdS$_4$ 
solutions was explored in \cite{Nunez:2001pt}, making use of the twisted field theory 
ideas of \cite{Maldacena:2000mw}. We show that this twisted field theory construction 
can easily be extended to provide the first description of a field theory dual to a Lifshitz 
geometry. A detailed study of this description is left for future work.

Flows which are Lifshitz in the UV provide simple examples of asymptotically Lifshitz 
spacetimes, corresponding to deformations of the field theory by a relevant operator. 
These deformations are one of the simplest extensions of the dictionary beyond the 
consideration of its vacuum state.  Examples of such asymptotically Lifshitz solutions 
were previously constructed in the original paper \cite{Kachru:2008yh}. We give a 
systematic discussion of such solutions in the context of the theories we consider.

We study the interpolating solutions both analytically and numerically, matching a 
linearised perturbation expansion about the AdS and Lifshitz solutions to numerical 
solutions. In the course of the perturbative analysis, we have noted that both the AdS and 
Lifshitz solutions in the six-dimensional supergravity of \cite{Romans:1985tw} can have 
excitations that violate the Breitenlohner-Freedman bound \cite{Breitenlohner:1982bm}. 
That is, some of these solutions are unstable. We leave exploration of this instability for 
future work. 

The remainder of the paper is organised to first consider the solutions in the simple 
massive vector theory of \cite{Taylor:2008tg}, and then consider the solutions in the six-
dimensional supergravity of  \cite{Romans:1985tw} following \cite{Gregory:2010gx}. In 
the next section, we review relevant aspects of the massive vector theory, and identify its 
AdS and Lifshitz solutions. In section \ref{mvmflows}, we construct domain walls dual to 
renormalization group flows interpolating between these solutions. We first consider the 
linearised analysis about each of the AdS and Lifshitz solutions, and then construct the 
full interpolating solutions numerically. In section \ref{romans}, we discuss the six-
dimensional supergravity theory and the four-dimensional AdS and Lifshitz solutions 
obtained from compactification on a compact hyperbolic space with flux, and obtain a consistent truncation to a four-dimensional theory. We then discuss 
interpolating between these solutions in section \ref{romansflows}. In this section we also 
discuss the identification of field theories dual to the Lifshitz solutions. We conclude with 
some remarks and discussion of open problems in section \ref{concl}. 

\section{Massive vector theory}

The simplest context in which to consider the Lifshitz metric is the massive vector theory 
introduced in  \cite{Taylor:2008tg}. This is a phenomenological model, so it is not {\it a 
priori} clear that there is a well-defined quantum theory of gravity that reduces to this 
theory in a low-energy limit.\footnote{Qualitatively similar models with additional scalar 
fields can be embedded in string theory, as in \cite{Donos:2010tu}, but the presence of 
the additional scalar fields can make significant differences to the physics; for example, 
the black hole solutions of \cite{Amado:2011nd} are quite different from the ones in the 
massive vector model obtained in \cite{Danielsson:2009gi}.} That is, it is not clear that the 
Lifshitz geometry here is genuinely dual to a well-defined quantum field theory. 
Nonetheless, this is a simple theory, so it provides a useful warmup before we turn to the 
more complicated theory considered in \cite{Gregory:2010gx}, and it will turn out that the 
structure of the interpolating solutions in these two theories is actually strikingly similar.

The bulk spacetime action for the massive vector theory is 
\begin{equation} \label{Dact}
S =-\frac{1}{16 \pi G} \int d^{d+1}x \sqrt{-g} (R - 2\Lambda -
\frac{1}{4} F_{\mu\nu} F^{\mu\nu} - \frac{1}{2} m^2 A_\mu A^\mu). 
\end{equation}
To have the solution \eqref{lif}, we need the cosmological constant and mass to be 
related to the dynamical exponent $z$ by
\begin{equation} \label{Lm}
\Lambda = -\frac{1}{2L^2}(z^2+(d-2) z+(d-1)^2), \quad m^2 L^2=(d-1) z,
\end{equation}
which implies
\begin{equation} \label{param}
\frac{\Lambda}{m^2} = -\frac{1}{2(d-1)z}(z^2+(d-2)z+(d-1)^2).
\end{equation}
The theory then has a solution with metric \eqref{lif} and 
\begin{equation} \label{vecl}
A = \alpha e^{zr} dt = L \sqrt{\frac{2(z-1)}{z}} e^{zr} dt.
\end{equation}

Since \eqref{param} is quadratic in $z$, if we regard the Lagrangian parameters $
\Lambda$, $m$ as fixed, there will be generically either two or no real solutions for $z$. 
The quadratic has two real roots for $\Lambda < -\frac{\left(3d-4\right)}{2\left
(d-1\right)} m^2$. If we call the smaller root $z_1$, the larger root will be $z_2 = (d-1)^2/z_1$.  
The form of the vector field \eqref{vecl} restricts us to considering only solutions with 
$z>1$, and we find that $z_2 >1$for all $d$, whereas $z_1 > 1$ for $\Lambda/m^2 > -d/2$. 

\begin{figure}
\begin{center}
\input{MVM_vacua.tex}
\caption{The vacua of the massive vector model in $d+1$ dimensions, labelled according 
to whether or not they possess an irrelevant perturbation within our ansatz. This plot was 
made using $d=3$, but is qualitatively the same at higher $d$.  The arrows denote the 
holographic RG flows. }
\label{fig:MVM_flows}
\end{center}
\end{figure}
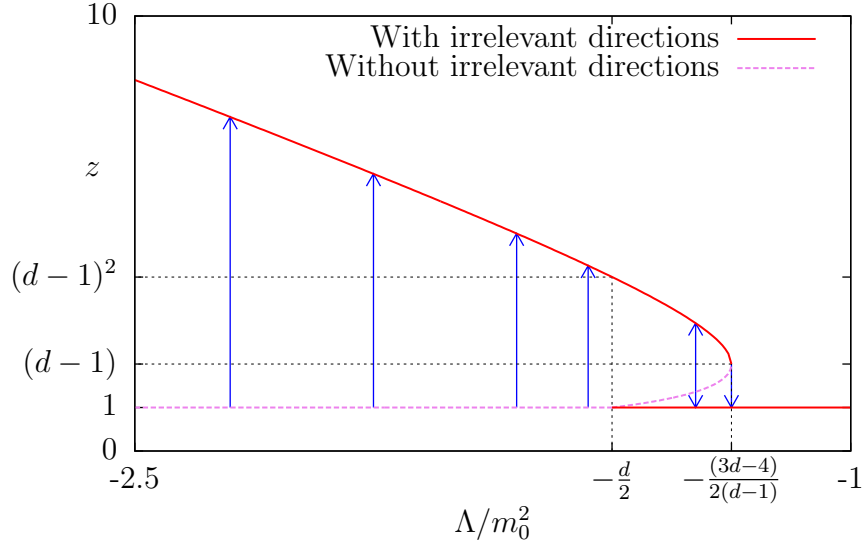

Thus, this theory will have a single Lifshitz solution for $\Lambda/m^2 \le - d/2$, two 
Lifshitz solutions for  $-d/2 < \Lambda/m^2 < -\frac{\left(3d-4\right)}{2\left(d-1\right)}$, a 
single degenerate solution with $z = d-1$ for $\Lambda/m^2 = -\frac{\left(3d-4\right)}{2\left
(d-1\right)}$, and none for $\Lambda/m^2 > \frac{\left(3d-4\right)}{2\left(d-1\right)}$. It also 
has an AdS$_{d+1}$ solution with no vector field for all $\Lambda <0$. These different 
solutions are depicted in figure \ref{fig:MVM_flows}. The flows depicted in this figure will 
be explained in the next section.

\section{Flows in the massive vector theory}
\label{mvmflows}

We now consider the construction of domain wall geometries which interpolate between 
these different solutions. Such a solution was previously found in \cite{Kachru:2008yh}, 
who numerically found a spacetime that is asymptotically AdS at small $r$ and 
asymptotically Lifshitz with $z=2$ at large $r$. Since the theory considered in \cite
{Kachru:2008yh} is on-shell equivalent to the massive vector model, their solution will be 
a special case of the solutions we find here.\footnote{Note that the flow considered in  
\cite{Kachru:2008yh} is a somewhat special case - we see from figure \ref{fig:MVM_flows} 
that this is at the value of $\Lambda/m^2$ for which there is only one Lifshitz solution.} 
Our aim in this section is to extend this to give a comprehensive survey of the flows 
relating all the different solutions for arbitrary $z$ and $d$. 

Since all of the solutions preserve translation invariance and spatial rotations, we 
assume that these symmetries are preserved in the domain wall solutions, and hence 
consider an ansatz 
\begin{equation} \label{mvmans}
 ds^2 = L^2\left(-e^{2F\left(r\right)} dt^2 + e^{2D(r)}\frac{dr^2}{r^2} + r^2 \sum_{i=1}^{d-1} 
dx_i^2\right),
\end{equation}
\begin{equation}
A = \alpha\left(r\right) e^{F(r)} dt,
\end{equation}
where $L$ is given by \eqref{Lm}. The equations of motion are:
\bea
\frac{(d-1)}{L^2e^{2D}} \left [\, r D' - d\, \right ] = 
 \frac{(\alpha'+\alpha F')^2\,r^2}{L^4\,e^{2D}}
+ \frac{m^2\alpha^2}{4L^2} + \Lambda \;\;\;\;\;&& \\
-\frac{(d-1)}{L^2e^{2D}}  \left [\, r F' - (d-2)\, \right ] =  
\frac{(\alpha'+\alpha F')^2\,r^2}{L^4\,e^{2D}}
-  \frac{m^2\alpha^2}{4L^2} + \Lambda \;\;\;\;\;&&\\
m^2 L^2 e^{2D} \alpha = r^2 (\alpha'+\alpha F')'
+ r(\alpha'+\alpha F') (d - r D') \;\;\;\;\; &&
\eea
plus a Bianchi identity not shown here. The Lifshitz solutions correspond 
to $F = z \ln r$, $D=0$, $\alpha =L\sqrt{ \frac{2(z-1)}{z}}$, and the AdS solution 
is the $z\to 1$ limit of the Lifshitz solution.

\subsection{Linearized equations of motion}

We start by considering the linearised equations of motion around each of these 
solutions. This will enable us to identify the conformal dimensions of the corresponding 
operators in the field theory. Around the AdS solution, the linearisation is very simple, as 
many of the equations decouple. If we write $F = \ln r + \delta F$, the linearised equations 
of motion are
\begin{equation}
r\delta F' = -r D'  = d D, \quad r(r\alpha')' = (L^2 m^2 - (d-1)) \alpha -d r \alpha', 
\end{equation}
where prime denotes derivatives with respect to $r$, and the solutions are
\begin{equation}
\delta F = F_0 + F_1 r^{-d}, \quad D = -F_1 r^{-d}, \quad \alpha = \alpha_1 r^{-\Delta_1} + 
\alpha_2 r^{-\Delta_2},
\end{equation}
where 
\begin{equation}
\Delta_{1,2} = \frac{d}{2} \mp \frac{\sqrt{4 m^2 L^2 + (d-2)^2}}{2}.
\end{equation}
The constant $F_0$ corresponds to the timelike component of the boundary metric in the 
field theory. The mode $F_1$ corresponds to the expectation value of the field theory 
energy density (more precisely, the tracelessness of the boundary stress tensor implies a 
relation between the boundary energy density and pressure, so this mode represents a 
non-zero expectation value for both energy density and pressure). If we impose 
boundary conditions on the vector field that fix the slow fall-off mode $\alpha_1$, the $
\alpha_1, \alpha_2$ modes correspond respectively to the source and expectation value 
for the operator dual to the massive vector. The dimension of this operator is
\begin{equation}
\Delta = \frac{d}{2} + \frac{\sqrt{4 m^2 L^2 + (d-2)^2}}{2}. 
\end{equation}

For $m^2 L^2 < (d-1)$, the operator dual to the massive vector is relevant. This 
corresponds to $\Lambda/m^2 < - d/2$.  Thus for $\Lambda/m^2 < - d/2$, deforming the 
conformal field theory dual to the AdS solution by this relevant operator will generate a 
flow from this theory in the UV. In the bulk, the interpolating solution corresponding to this 
RG flow will be an asymptotically AdS spacetime with a perturbation with non-zero $
\alpha_1$ at large $r$. Since this corresponds to turning on the vector field which 
sources the Lifshitz solutions, the natural expectation is that this flow will approach 
Lifshitz in the IR. For this range of parameters, there is a unique Lifshitz solution. This 
flow is indicated by the vertical arrows to the left in figure \ref{fig:MVM_flows}, and will be 
constructed numerically below. Note that such flows from AdS in the UV to Lifshitz in the 
IR have not previously been constructed for this theory. 

For the Lifshitz solutions, the analysis of the linearised equations of motion was 
performed for $d=3$ in \cite{Bertoldi:2009vn,Ross:2009ar}. Here we extend this analysis 
to general $d$. The solutions we are interested in here correspond to the scalar parts of 
the constant perturbations in the previous analysis. We write $F = z \ln r + \delta F$, $
\alpha = \alpha_0(1 + \delta \alpha)$, with $\alpha_0$ the background value given by 
\eqref{vecl}. There is then a simple two-parameter set of solutions given by 
\begin{equation}
\delta F = F_0 + \frac{d-1-z}{d-1+z} \frac{F_1}{r^{z+d-1}}, \quad D = \frac{F_1}{r^{z+d-1}}, 
\quad \delta \alpha = - \frac{d-2+z}{z-1} \frac{F_1}{r^{z+d-1}}. 
\end{equation}
As before, $F_0$ corresponds to the source for and $F_1$ corresponds to the 
expectation value of the field theory energy density, as discussed in detail for $d=3$ in 
\cite{Ross:2009ar}.\footnote{The geometrical description of the boundary data $F_0$ was recently discussed in \cite{ALL}.} This is a 
marginal operator, since in a Lifshitz solution, the dimension of a marginal operator is $z 
+ (d-1)$, because of the different scaling of the time direction.

We have two other solutions, given by
\begin{align}
\delta F =& \alpha_1 \frac{(z+d-2)(z+d-1+\beta)}{ r^{\frac{1}{2}(z+d-1 -\beta)}} + \alpha_2 
\frac{(z+d-2)(z+d-1-\beta)}{ r^{\frac{1}{2}(z+d-1 +\beta)}}\\
D =& \alpha_1 \frac{(z-1)(z-3d+3+\beta)}{  r^{\frac{1}{2}(z+d-1 -\beta)}} + \alpha_2 \frac
{(z-1)(z-3d+3-\beta)}{ r^{\frac{1}{2}(z+d-1 +\beta)}}\\
\delta \alpha = & \alpha_1 \frac{(z+d-2)(3z-d+1-\beta)}{ r^{\frac{1}{2}(z+d-1 -\beta)}} + 
\alpha_2 \frac{(z+d-2)(3z-d+1+\beta)}{ r^{\frac{1}{2}(z+d-1 +\beta)}}
\end{align}
where
\begin{equation}
\beta(z,d)^2 = (z+d-1)^2 + 8(z-1)(z-d+1).
\end{equation}
We would expect to interpret these as the source for and the expectation value of the 
operator dual to the massive vector field. We will assume this interpretation is valid, but 
note that in $d=3$ there were some unanswered questions about the calculation of the 
expectation value for $z\geq 2$ \cite{Ross:2009ar}. This then corresponds to an operator 
of dimension
\begin{equation}
\Delta = \frac{1}{2}(z + d-1 + \beta(z,d)).
\end{equation}
Thus, this operator is relevant if $\beta(z,d) < z+d-1$, that is for $z< (d-1)$. It is always a 
relevant perturbation on the branch of Lifshitz solutions with the smaller value of $z$, and 
always an irrelevant perturbation on the branch with the larger value. 

The perturbation of the smaller $z$ Lifshitz solution at large $r$ by the mode $\alpha_1$ 
which corresponds to the source for this operator then sources a flow to the IR. This 
perturbation represents the leading deformation of the massive vector field, so 
depending on the sign, we expect this to terminate either at the AdS solution or the larger 
$z$ Lifshitz solution, which has a larger value for the vector. We also see that the larger 
$z$ solution has an irrelevant direction, which we expect to correspond to the flow from 
AdS or smaller $z$. These expectations are confirmed in the next subsection. 

\subsection{Numerical flows}

We now turn to numerics to confirm the existence of the interpolating geometries 
predicted from our analysis of the linearised equations of motion in the previous section. 
As in previous work, starting with \cite{Freedman:1999gp}, we construct these solutions 
by starting from the IR (small $r$ region of the geometry). This is convenient because we 
want to consider geometries dual to renormalization group flows, so we are not 
interested in exciting the modes corresponding to the expectation value of the dual 
operators. Since those modes grow towards small $r$ (the IR), we can most easily 
construct the flows numerically by starting from a candidate IR geometry at small $r$ and 
following the effect of a small deformation by an irrelevant operator as we integrate out to 
larger $r$. Since the linearised analysis told us that within the ansatz we are considering 
these solutions have at most a single irrelevant direction, all we can choose is the sign of 
the perturbation.


The AdS solutions had an irrelevant perturbation for $\Lambda/m^2 > -d/2$. Based on 
the analysis presented in figure \ref{fig:MVM_flows}, we expect there to be flows from the 
small $z$ Lifshitz solution at large $r$ which approach the AdS solution at small $r$ 
along this deformation. The flow found in \cite{Kachru:2008yh} in the case $z=2$ $d=3$ 
is a special case of this class. Since the AdS vacuum has $\alpha = 0$ and the irrelevant 
perturbation  involves only the vector field, the $\alpha \mapsto -\alpha$ symmetry  
means that the sign of the perturbation does not matter here. 

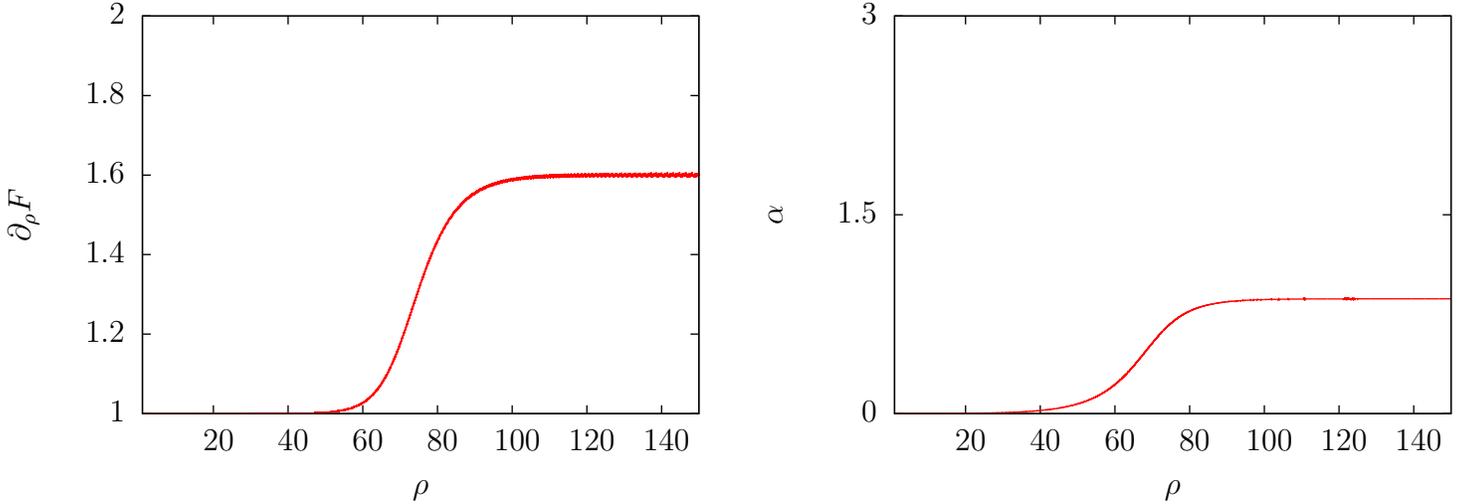
\begin{figure}[htb]
\centering\hbox{\hspace{-2cm}\input{MVM_Li_AdS_3.tex}}
\caption{Holographic RG flow in $d=3$ from a Lifshitz spacetime with $z=1.6$ in the UV to an AdS$_4$ spacetime in the IR. In our numerical analysis we use a radial coordinate $\rho=\ln r$. Note that $\partial_\rho F \to z$ as we approach one of the AdS or Lifshitz solutions.}
\label{fig:MVM_Li_AdS_flow_3}
\end{figure}

We have numerically constructed examples of such flows for $d=3,4,5$. An example with 
$z=1.6$ in $d=3$ is shown in figure \ref{fig:MVM_Li_AdS_flow_3}. The flows typically 
rapidly approach the Lifshitz solution at large $r$, although in the $z=2$ case it is much 
slower. This is to be expected, since in this case the direction we are approaching the 
Lifshitz point along is marginal at the linear level.

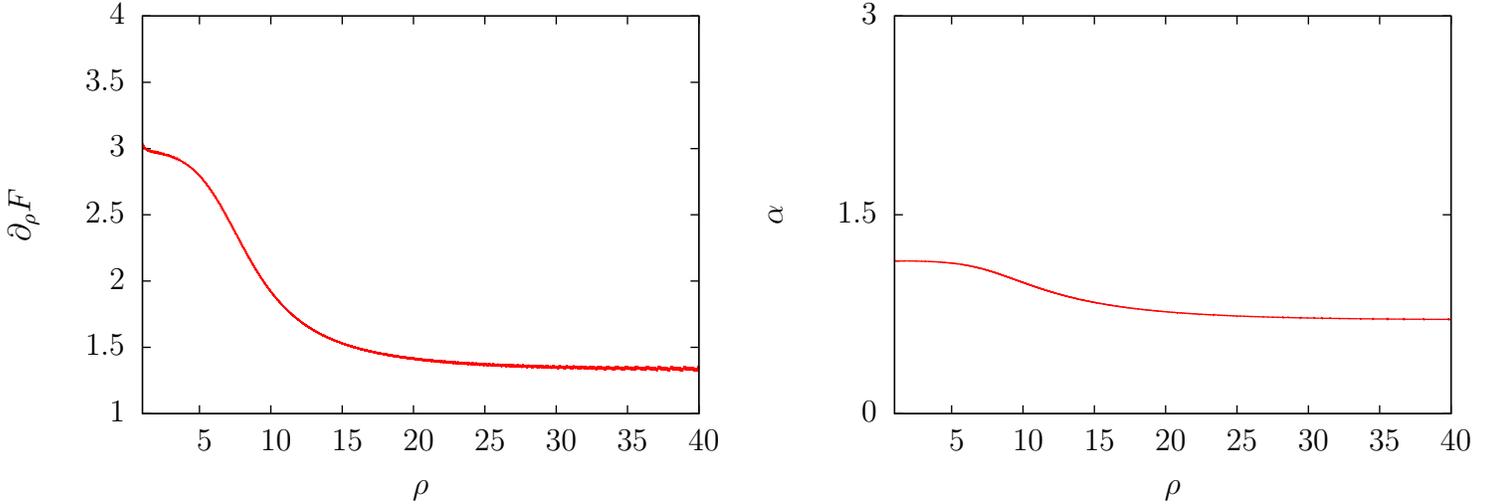
\begin{figure}[htb]
\centering\hbox{\hspace{-2cm}\input{MVM_Li_Li_1.tex}}
\caption{Holographic RG flow in $d=3$ from a Lifshitz spacetime with $z=1.333$ in the UV to one with $z=3$ in the IR. In our numerical analysis we use a radial coordinate $\rho=\ln r$. Note that $\partial_\rho F \to z$ as we approach one of the AdS or Lifshitz solutions.}
\label{fig:MVM_Li_Li_flow_1}
\end{figure}

The Lifshitz solution with larger $z$ has an irrelevant perturbation, so we would expect to 
be able to construct solutions which approach this solution at small $r$; from the analysis 
in figure \ref{fig:MVM_Li_AdS_flow_3}, we expect that for $\Lambda/m^2 > -d/2$ this will 
be a flow from the small $z$ Lifshitz solution at large $r$ (if we choose the sign of the 
perturbation at small $r$ to be in the direction of decreasing $\alpha$). These will 
therefore be examples of Lifshitz to Lifshitz flows. In figure  \ref{fig:MVM_Li_Li_flow_1} we 
show an example of such a flow for $d=3$. We have also constructed examples in 
$d=4,5$ which are qualitatively similar.

\begin{figure}[htb]
\centering\hbox{\hspace{-2cm}\input{MVM_AdS_Li_1.tex}}
\caption{Holographic RG flow in $d=3$ from an AdS$_4$ spacetime in the UV to a Lifshitz spacetime with $z=6$ in the IR. In our numerical analysis we use a radial coordinate $\rho=\ln r$. Note that $\partial_\rho F \to z$ as we approach one of the AdS or Lifshitz solutions.}
\label{fig:MVM_AdS_Li_flow_1}
\end{figure}
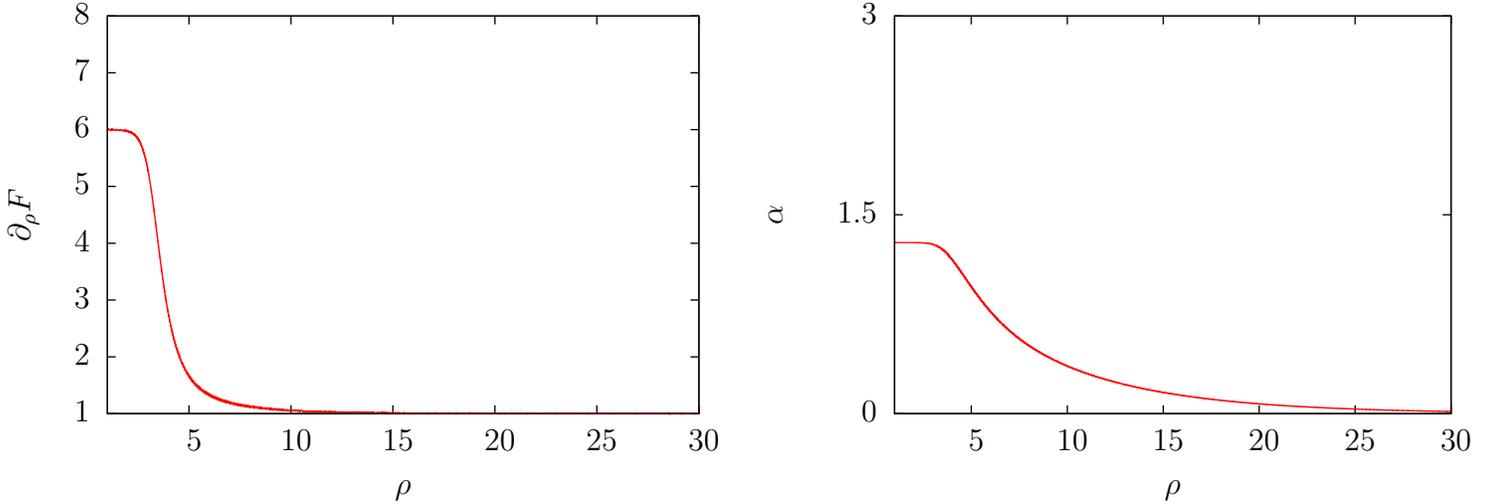

For  $\Lambda/m^2 < -d/2$, we expect the irrelevant deformation around the large $z$ 
Lifshitz solution to be associated with a flow from the AdS solution at large $r$ (again 
assuming we choose the sign of the perturbation at small $r$ to be in the direction of 
decreasing $\alpha$). We numerically constructed  examples in $d=3,4,5$; an example 
with $d=3$ is shown in figure \ref{fig:MVM_AdS_Li_flow_1}. 

Examples of flows from Lifshitz in the UV to AdS in the IR were previously constructed in 
\cite{Kachru:2008yh}, but the other two types of flow solutions we have constructed here, 
from Lifshitz to Lifshitz and from AdS to Lifshitz, are new. The latter are probably the most 
interesting. In the context of this simple massive vector model, these AdS to Lifshitz flows 
are interesting primarily for the potential to relate the study of the holographic dictionary 
in Lifshitz geometries to the better understood AdS case, by embedding asymptotically 
Lifshitz geometries in asymptotically AdS ones, and hence relating the calculation e.g. of 
correlation functions in the Lifshitz context to observables in the UV conformal field 
theory. The application to understanding the field theory dual to the Lifshitz geometry is 
hampered here by our lack of an understanding of the field theory dual to the AdS 
solutions in this massive vector model. This motivates us to turn in the next section to the 
consideration of Lifshitz solutions in a supergravity theory which can be embedded into 
string theory, where we can obtain a concrete interpretation of the geometries in terms of 
dual field theory. 

\section{$F(4)$ gauged supergravity}
\label{romans}

We now turn to the consideration of a more complicated theory, which can be embedded 
into string theory, the six-dimensional $\mathcal N=4$ $F(4)$ gauged supergravity of 
\cite{Romans:1985tw}. This theory can be obtained as a consistent truncation of a 
Kaluza-Klein reduction of massive type IIA supergravity \cite{Cvetic:1999un}, so solutions 
of this theory can be uplifted to solutions of string theory in a background including D8-
brane charge. 

The bosonic part of the action for this theory is
\begin{align} \label{6daction}
S =& \int d^6 x \sqrt{-g} \left[ \frac{1}{4} R - \frac{1}{2} \partial^\mu \phi \partial_\mu \phi - 
\frac{1}{4} e^{-\sqrt{2} \phi} (\mathcal H^{\mu\nu} \mathcal H_{\mu\nu} + F^{(i) \mu\nu} F^
{(i)}_{\mu\nu}) \right. \\ & \left. - \frac{1}{12} e^{2 \sqrt{2} \phi} G_{\mu\nu\rho}G^{\mu\nu
\rho} + \frac{1}{8} (g^2 e^{\sqrt{2} \phi} + 4gm e^{-\sqrt{2} \phi}- m^2 e^{-3 \sqrt{2} \phi} ) 
\right] \nonumber \\ \nonumber 
& + \frac{1}{8} \int d^6 x \left( B \wedge \mathcal F \wedge \mathcal F + m B \wedge B 
\wedge \mathcal F + \frac{m^2}{3} B \wedge B \wedge B + B \wedge F^{(i)} \wedge F^{(i)} 
\right). 
\end{align}
We follow the conventions of \cite{Romans:1985tw}; the bosonic fields are the metric $g_
{\mu\nu}$, the dilaton $\phi$, the two-form $B_{\mu\nu}$, an $SU(2)$ gauge field $A^{(i)}
_\mu$ and a $U(1)$ gauge field $\mathcal A_\mu$. The field strengths are 
\begin{equation}
\mathcal F = d \mathcal A, \quad F^{(i)} = dA^{(i)} + g \epsilon^{ijk} A^{(j)} \wedge A^{(k)}, 
\quad G = d B,
\end{equation}
and we write 
\begin{equation}
\mathcal H_{\mu\nu} = \mathcal F_{\mu\nu} + m B_{\mu\nu}.
\end{equation}
The Lagrangian involves two parameters, $g$ and $m$. We consider only $g > 0$, $m 
>0$, referred to as $\mathcal N = 4^+$ in the notation of  \cite{Romans:1985tw}. Note that 
as explained in \cite{Romans:1985tw}, there is a freedom to make field redefinitions 
which relates different theories; the inequivalent theories are labeled by the ratio $g/m$. 

We want to consider the AdS and Lifshitz solutions in this theory. The theory of course 
also has an AdS$_6$ solution, discussed in detail in \cite{Romans:1985tw}. We will very 
briefly review this solution, as we will be considering some flows involving asymptotically 
AdS$_6$ solutions, but we focus on describing the four-dimensional AdS and Lifshitz 
solutions obtained by considering a further compactification of this theory on a compact 
hyperbolic space. 

The AdS$_6$ solution has metric 
\begin{equation}
ds^2 = r^2 \eta_{\alpha\beta} dx^\alpha dx^\beta + L^2 \frac{dr^2}{r^2}, 
\end{equation}
the vector and two-form fields vanishing, and dilaton $\phi = \phi_0$ with $e^{-2\sqrt{2} \phi_0} = \frac{g}{m}$ or $e^{-2\sqrt{2} 
\phi_0} = \frac{g}{3m}$. The latter case is 
supersymmetric, and is dual to the conformal field theory obtained in the IR limit of the $
\mathcal N = 2$ $Sp(N)$ gauge theory on the worldvolume of D4-branes in the presence 
of D8-branes \cite{Ferrara:1998gv,Brandhuber:1999np}. 

The four-dimensional solutions are obtained by considering a compactification of this six-
dimensional theory on a compact hyperbolic space. To describe the AdS and Lifshitz 
solutions, we can take the metric to have the form
\begin{equation}
ds^2 = -e^{2F(r)} dt^2 + r^2 d\vec{x}^2 + e^{2d_0} \frac{dr^2}{r^2} + e^{2h_0} d
\Omega_2^2, 
\end{equation}
where $d_0, h_0$ are constants, and 
\begin{equation} \label{O2}
d\Omega_2^2 = \frac{1}{y_2^2} (dy_1^2 + dy_2^2)
\end{equation}
is the metric on $H^2$, a two-dimensional space of constant negative curvature. We take 
the global geometry of this two-dimensional space to be some compact quotient of 
$H^2$ by a discrete subgroup $\Gamma$ of its isometry group. For an AdS$_4$ solution, 
$F(r) = \ln r$, while for a Lifshitz solution, $F(r) = z \ln r$.
We take the dilaton to be a constant, $\phi = \phi_0$, and for the vector and two-form 
fields, we take 
\begin{equation}
F^{(3)} = \alpha_0 \frac{e^{F(r)+d_0}}{r} dt \wedge dr + \frac{\gamma}{y_2^2} dy_1 
\wedge dy_2, 
\end{equation}
and
\begin{equation}
\mathcal{H} = m B = m \frac{\bar \beta_0}{2} r^2 dx_1 \wedge dx_2,
\end{equation}
so we consider a flux of one component of the $SU(2)$ gauge field on the compact 
space. 

The equations of motion fix $\alpha_0 = \gamma \bar \beta_0 e^{\sqrt{2} \phi_0} e^
{-2h_0}$. Charge conservation implies $\gamma$ is fixed (in particular, in the 
interpolating solutions it will remain a constant), and from the four-dimensional point of 
view it corresponds to a parameter labeling the theory rather than a feature of a particular 
solution. Thus, in determining Lifshitz solutions, we should look to solve for the other 
parameters $z$, $d_0$, $h_0$, $\phi_0$ and $\bar \beta_0$ in terms of $g$, $m$ and $
\gamma$. As was observed in \cite{Gregory:2010gx}, in this ansatz there is a further  
freedom to rescale fields additional to the field redefinition of \cite{Romans:1985tw}; as a 
result both $g$ and $m$ just set an overall scale for fields. It will be convenient for us to 
define a slightly different set of rescaled variables to those considered in 
\cite{Gregory:2010gx}. We define 
\begin{equation} 
\varphi_0 = \sqrt{\frac{m}{g}} e^{-\sqrt{2} \phi_0}, \quad e^{-2H_0} = \frac{\gamma}{\sqrt
{gm}} e^{-2h_0}, \quad e^{2D_0} = \sqrt{g^3 m}e^{2d_0}, \quad \beta_0 = \sqrt{\frac{m}
{g}} \bar \beta_0. 
\end{equation}

\begin{figure}
\begin{center}
\includegraphics[width=6.2cm]{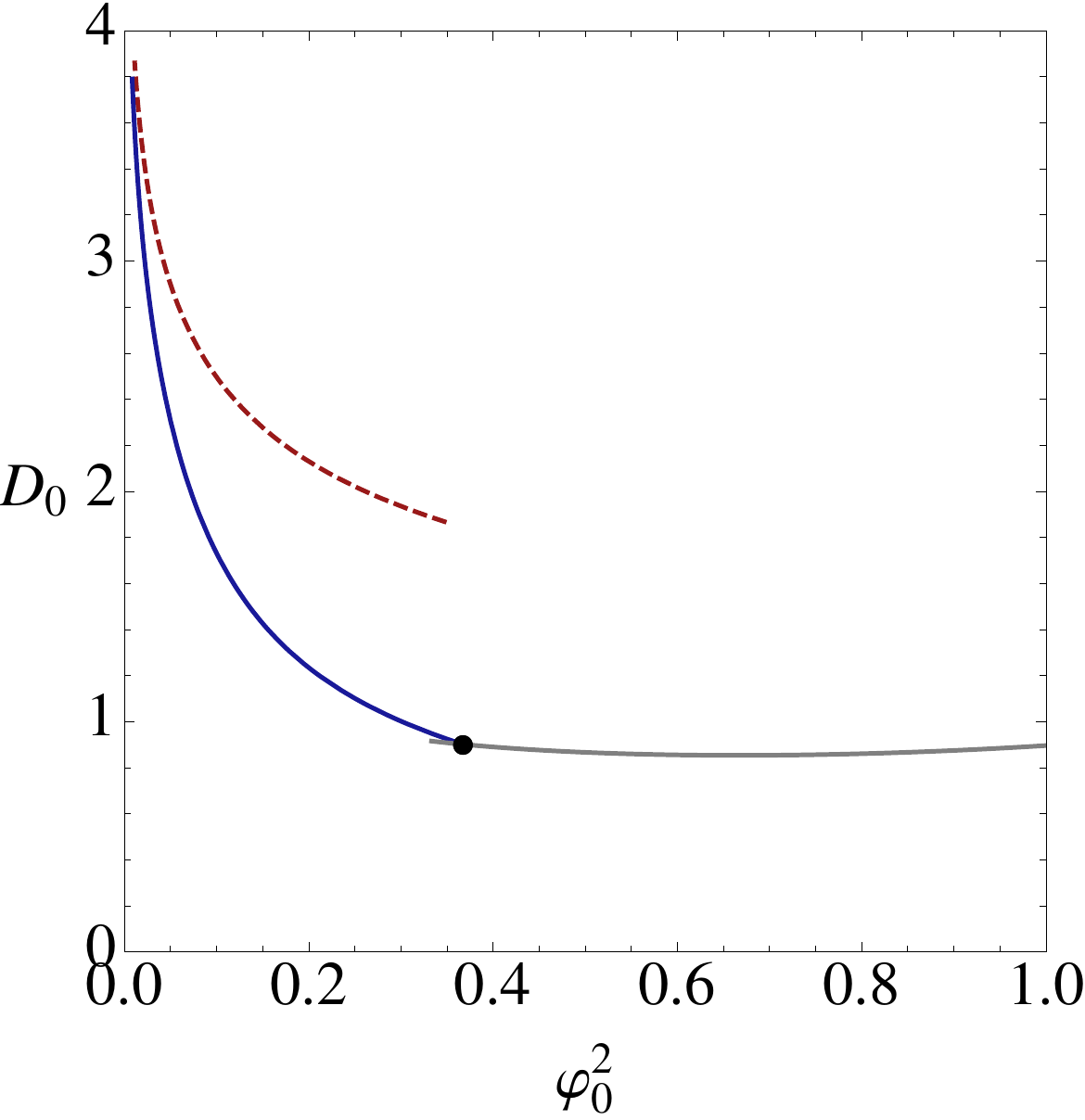}~~~~\includegraphics[width=7cm]{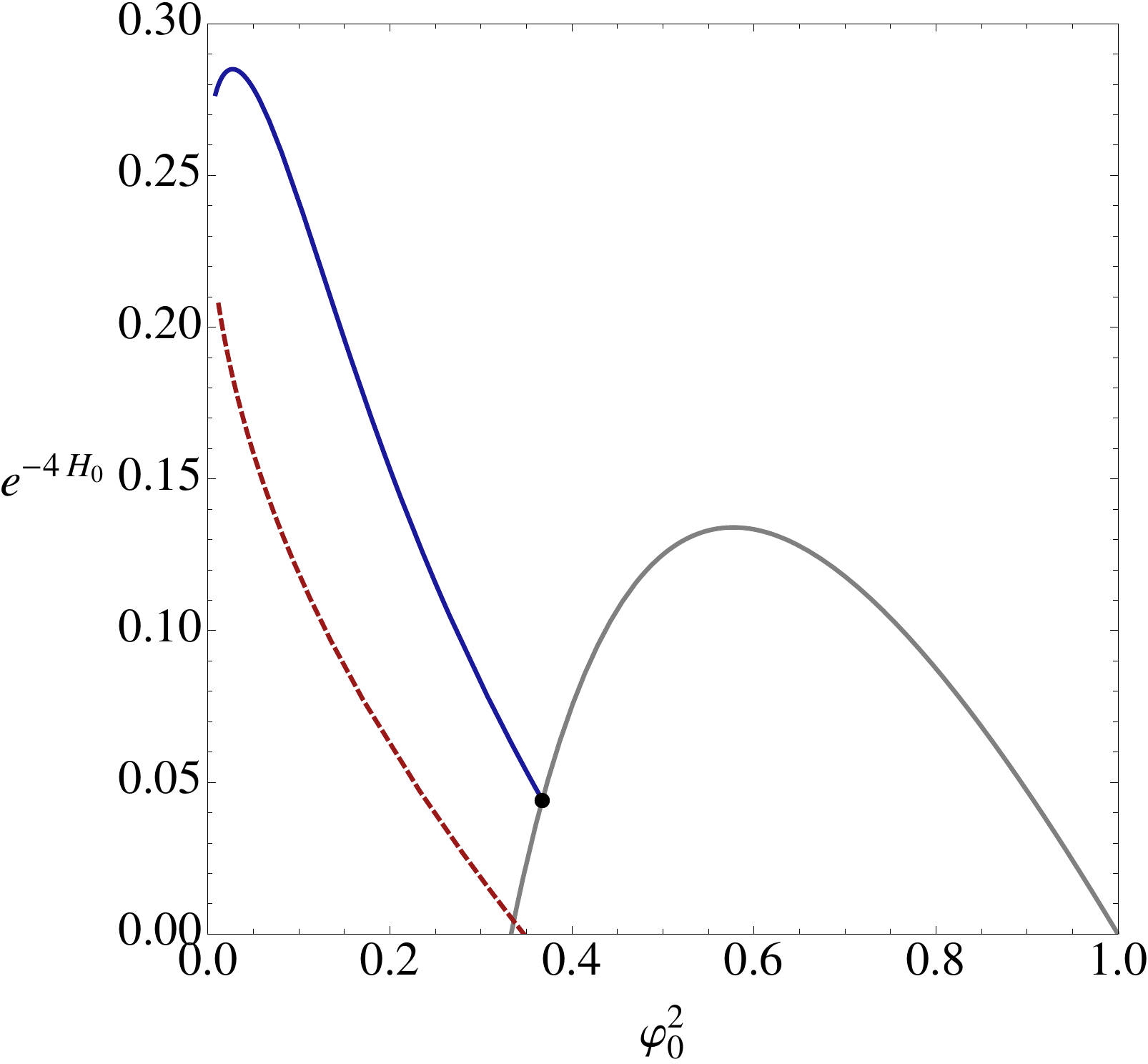}
\caption{Plots of $D_0$ and $H_0$ for the AdS and Lifshitz solutions as a function of
$\varphi_0^2$. The AdS solutions are shown in grey, the smaller Lifshitz in blue,
and the larger $z$ Lifshitz in dashed red.}
\label{DandH}
\end{center}
\end{figure}

AdS$_4$ solutions were discussed in \cite{Romans:1985tw} and more recently in \cite
{Nunez:2001pt}. They have $z=1$ and $\beta_0 = 0$. Solving the equations of motion 
then gives us a relation between the flux $\gamma$ and the constant values $\varphi_0$, 
$H_0$ and $D_0$. Surprisingly, solutions only exist for a certain range of values of $g^2 
\gamma^2$. Solving the equations of motion gives 
\begin{equation} \label{vphi}
g^2 \gamma^2 =  \frac{ (1 - \varphi_0^2)(3 \varphi_0^2-1)}{(1-2\varphi_0^2 + 2 
\varphi_0^4)^2}, 
\end{equation}
which has two solutions for $0<g^2 \gamma^2 < \frac{9-\sqrt{216}}{\sqrt{1536}-44} 
\approx 1.185$ (one either side of $\varphi_0^2 = 1-1/\sqrt{6} \approx 0.592$; note $
\varphi_0 >0$ for real $\phi_0$), and no solutions for larger $g^2 \gamma^2$. We will 
refer to these as the small $\varphi_0$ and large $\varphi_0$ solutions; the small $
\varphi_0$ solution has $\varphi_0^2 \in  \left(\frac{1}{3},1-\frac{1}{\sqrt{6}}\right)$ and the 
large $\varphi_0$ solution has $\varphi_0^2 \in \left(1-\frac{1}{\sqrt{6}},1\right)$. The 
values of the other fields at the AdS$_4$ solution are most conveniently written in terms 
of $\varphi_0$, 
\begin{equation} \label{c1}
e^{-2D_0} = \frac{\varphi_0 (2 - \varphi_0^2)}{6}, \quad e^{-2H_0} = \frac{ \sqrt{(1-
\varphi_0^2)(3 \varphi_0^2-1)}}{2 \varphi_0}.
\end{equation}
Note that these are both positive for all $\varphi_0^2 \in \left( \frac{1}{3},1 \right)$,
see figure \ref{DandH}.

For $\varphi_0^2 = \frac{1}{2}$, these solutions preserve half the supersymmetry of the 
original six-dimensional theory. In \cite{Nunez:2001pt}, these supersymmetric AdS$_4$ 
solutions were related to the conformal field theory obtained by taking the low-energy 
limit of a five-dimensional twisted field theory. We will describe the flows corresponding 
to this IR limit in the next section.

In \cite{Gregory:2010gx}, Lifshitz solutions with arbitrary $z$ were obtained in this ansatz. 
These Lifshitz solutions break all of the supersymmetry of the theory. Solving the 
equations of motion for $z \neq 1$ gives 
\begin{equation}
g^2 \gamma^2 = \frac{(z+4)[(z+2)(z-3) \pm 2 \sqrt{2z+8}]}{[3z+6\mp2 \sqrt{2z+8}]^2},
\end{equation}
which fixes $z$ for given $g^2 \gamma^2$, and the other parameters are given by
\begin{equation} \label{betal}
\varphi_0^2 = \frac{ z^2(z+4)}{6+z \mp 2 \sqrt{2(z+4)}}, \quad \beta_0 = \varphi_0 \sqrt
{z-1}, \end{equation}
and
\begin{equation}
e^{4D_0} = 4 z \sqrt{ (4+z)^3 [6+z \mp 2 \sqrt{2(z+4)}]}, \quad e^{-2H_0} = g \gamma e^{-2 
D_0} ( 6 + 3z \mp 2 \sqrt{2(z+4)}).
\end{equation}
These different solutions are shown in figure \ref{DandH}.
There is a single sign choice here; we can choose either the upper or the lower sign in all 
expressions to obtain a solution. Thus, for given $g^2 \gamma^2$, there are two possible 
values for $z$, and the other fields are then uniquely specified once one of these two 
values is chosen. As the lower sign gives larger values of $z$, we refer to this as the 
larger $z$ branch of solutions, and the upper sign as the smaller $z$ branch.  

We are restricted to solutions with $z \geq 1$ by \eqref{betal}. There are then solutions on 
the larger $z$ branch for all values of $g^2 \gamma^2$; $z$ increases monotonically from 
$z=4.29$ at $g^2 \gamma^2 = 0$. The solutions on the smaller $z$ branch have $z =1$ at 
$g^2 \gamma^2 = \frac{5 (-6+2 \sqrt{10})}{(9+2\sqrt{10})^2} \approx 0.23$ and again 
have $z$ increasing monotonically with $g^2 \gamma^2$. 

\begin{figure}
\begin{center} \includegraphics[width=12cm]{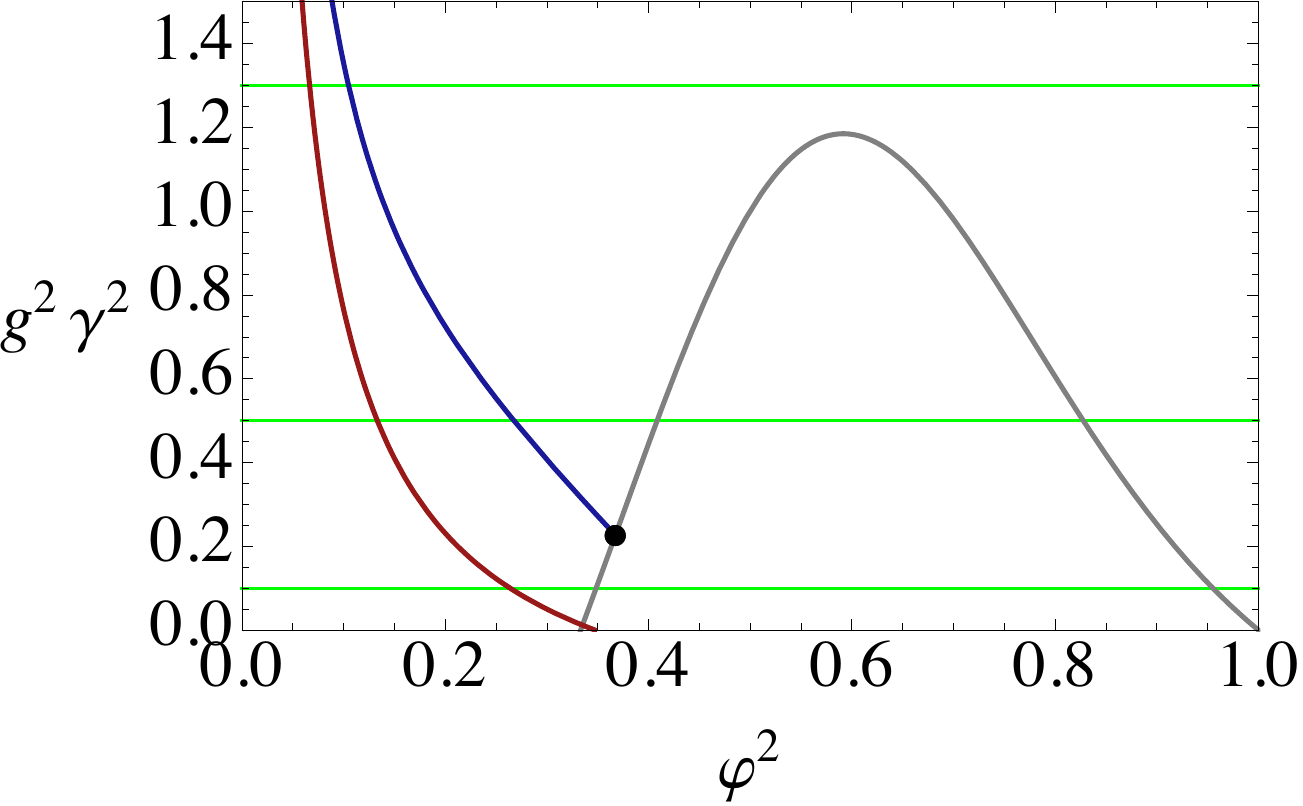}
\end{center}
\caption{The possible AdS and Lifshitz solutions for the gauged supergravity theory as a function of the flux $g^2 \gamma^2$. Reading from left to right, the curves represent the large $z$ Lifshitz solution, the small $z$ Lifshitz solution and the AdS solutions. Flows interpolating between these solutions will be at fixed $g^2 \gamma^2$, so the horizontal lines indicate possible flows. As discussed in the text, the structure of the expected flows is similar to that shown in figure \ref{fig:MVM_flows} for the massive vector case; the main difference is the appearance of two AdS solutions here.}
\label{lifsolns}
\end{figure}

In summary, there are two AdS solutions for $g^2 \gamma^2 < 1.18$, and none for larger 
values. There is one Lifshitz solution for $g^2 \gamma^2 < 0.23$, with $z > 4.29$; there 
are two Lifshitz solutions for $g^2 \gamma^2 > 0.23$, with the second solution starting 
from $z=1$ at $g^2 \gamma^2 = 0.23$, where it coincides with one of the two AdS 
solutions. This structure is reminiscent of what we saw for the massive vector model 
earlier, although these Lifshitz solutions never meet. The solutions are plotted in figure \ref{lifsolns}.

\subsection{Consistent truncation}

We are interested in studying solutions which interpolate between the fixed point AdS 
and Lifshitz solutions identified above. It is straightforward to study these solutions in 
terms of the six-dimensional theory, and retaining this point of view will prove useful for 
understanding the field theory interpretation of the solutions later. However, examining 
the solutions from the point of view of a Kaluza-Klein reduction provides a 
complementary viewpoint which also provides useful insight. Since all the solutions we 
are considering excite only the overall volume of the internal space, this reduction is fairly 
simple, and we give here a consistent truncation of the reduced equations of motion 
which will include the solutions of interest. 

We consider a metric ansatz
\begin{equation}
ds_6^2 = e^{-2h} g_{\mu\nu} dx^\mu dx^\nu + e^{2h} d\Omega_2^2,
\end{equation}
where $d\Omega_2^2$ is given in \eqref{O2}, $g_{\mu\nu}$ is an arbitrary four-
dimensional metric, and $h$ is a function of the $x^\mu$. For the matter fields, we take $
\mathcal F = 0$, 
\begin{equation}
F_6^{(3)} = F + \gamma \epsilon_2, \quad B_6 = B + B_0 \epsilon_2.
\end{equation}
The dilaton $\phi$ is a function of the $x^\mu$. Here $\epsilon_2$ is the volume form 
on the internal space, and $F$, $B$ are two-forms on the four-dimensional spacetime. 
The equations of motion imply $\gamma$ is a constant, but $B_0$ is a function of the $x^
\mu$. 

This ansatz will satisfy the six-dimensional equations of motion following from the action 
\eqref{6daction} if $g_{\mu\nu}, h, \phi, B_0, B_{\mu\nu}$ and $F_{\mu\nu}$ satisfy the 
equations of motion of the action 
\begin{align} \label{4daction}
S =& \int d^4 x \sqrt{-g} \left[ \frac{1}{4} R  - \frac{1}{12} e^{2 \sqrt{2} \phi + 4h} G_{\mu\nu
\rho} G^{\mu\nu\rho} - \frac{1}{4} e^{-\sqrt{2} \phi + 2h} (F_{\mu\nu} F^{\mu\nu} + m^2 B_
{\mu\nu} B^{\mu\nu}) \right. \nonumber \\ & - (\nabla h)^2 - \frac{1}{2} (\nabla \phi)^2 - \frac
{1}{2} e^{2 \sqrt{2} \phi - 4h} (\nabla B_0)^2 + \frac{1}{2} e^{-4h} + e^{-2h} P - \frac{1}{2} 
m^2 e^{-\sqrt{2} \phi - 6h} B_0^2 \\ &\nonumber \left. - \frac{1}{2} \gamma^2 e^{-\sqrt{2} 
\phi - 6h} \right] + \frac{1}{8} \int d^4 x ( m^2 B_0 B \wedge B + B_0 F \wedge F + \gamma 
B \wedge F ),
\end{align}
where $G= dB$ and the dilaton potential $P = \frac{1}{8} (g^2 e^{\sqrt 2 \phi} + 4gm e^{-
\sqrt 2 \phi} -m^2 e^{-3 \sqrt 2 \phi})$. 

One advantage of the consistent truncation is that it makes the relation between 
seemingly different solutions evident. In \cite{Romans:1985tw}, it was noted that theories 
with seemingly different values of $g$ and $m$ were in fact related by a simple scaling of 
the fields, so that for $g >0$, $m>0$ the inequivalent theories were labelled by $\frac{g}
{m}$. In our truncated theory, there are apparently three parameters, $g$, $m$ and $
\gamma$; however, as one might expect from the preceding analysis of the fixed points, 
the inequivalent theories are labelled by a single invariant combination, $g^2 
\gamma^2$. This can be seen explicitly by starting from the theory with any $g>0$, $m 
>0$, setting $\gamma = \frac{\Gamma}{g}$, and making the field redefinitions
\begin{equation}
e^{-\sqrt{2} \phi} = \sqrt{\frac{g}{m}} e^{-\sqrt{2} \phi'}, \quad e^{-2h} = \sqrt{g^3 m}e^{-2h'}, 
\quad B_0 = \frac{1}{gm} B_0',  
\end{equation}
\begin{equation}
g_{\mu\nu} = \frac{1}{g^3 m} g'_{\mu\nu}, \quad F^{(3)} = \frac{1}{g} F^{(3)'}, \quad B = \frac
{1}{mg} B'.
\end{equation}
The action \eqref{4daction} then reduces to an overall factor of $\frac{1}{g^3 m}$ times 
the same action for the primed fields with $g=m=1$, depending only on $\Gamma$. 
Thus, the inequivalent theories are labelled by $\Gamma^2 = g^2 \gamma^2$. 

Having this four-dimensional action also makes it easier to compare to the 
phenomenological models which have previously been studied. We can see that the 
action is qualitatively similar to the theory considered in \cite{Kachru:2008yh}. The most 
significant difference is perhaps the presence of  a mass term for the two-form $B$, 
although there are also additional scalar fields, and some more complicated couplings.

\section{Flows in the $F(4)$ gauged supergravity}
\label{romansflows}

We wish to construct domain wall solutions which interpolate between these solutions. 
Assuming the fields are only functions of the radial coordinate and that the interpolating 
solutions preserve the rotational symmetry in the spatial directions, we can by choice of 
gauge write the most general such solution as
\begin{equation}
ds^2 = - e^{2F(r)} dt^2 + r^2 d\vec{x}^2 + e^{2 d(r)} \frac{dr^2}{r^2} + e^{2 h(r)} d
\Omega_2^2,
\end{equation}
where $d\Omega_2^2 = \frac{1}{y_2^2}(dy_1^2 + dy_2^2)$, with the matter fields $\phi(r)
$, 
\begin{equation}
F^{(3)} = \alpha(r) \frac{e^{F(r)+d(r)}}{r} dt \wedge dr + \frac{\gamma}{y_2^2} dy_1 \wedge 
dy_2, 
\end{equation}
and
\begin{equation}
\mathcal{H} = m B = m \frac{\bar \beta(r)}{2} r^2 dx_1 \wedge dx_2.
\end{equation}
Note $dF^{(3)} = 0$ implies $\gamma$ is a constant throughout the flow.  

The matter equations of motion are
\begin{align} \label{phi}
\frac{\sqrt 2}{r e^{F+d + 2h}} \partial_r \left(r^3 e^{F-B+2h} \partial_r \phi \right) + \frac{1}{4} 
(g^2 e^{\sqrt{2} \phi} - 4 gm e^{-\sqrt{2} \phi} + 3m^2 e^{-3\sqrt{2} \phi}) & \nonumber  \\ - 
\frac{1}{2r^2 e^{2d}} e^{2 \sqrt 2 \phi} (\partial_r (r^2 \bar \beta))^2 + \frac{m^2}{4} e^{-\sqrt
{2} \phi} \bar \beta^2 - e^{-\sqrt 2 \phi} \alpha^2 + e^{-\sqrt 2 \phi} \gamma^2 e^{-4h}&=0,
\end{align}
\begin{equation} \label{beta}
\frac{r}{e^{F+d + 2h} } \partial_r \left( \frac{e^{2 \sqrt 2 \phi+ F+ 2 h- d}}{2r} \partial_r (r^2 
\bar\beta) \right) = \frac{m^2}{2} e^{-\sqrt{2} \phi} \bar \beta + 2 \alpha \gamma e^{-2h}, 
\end{equation}
and
\begin{equation} \label{alpha}
\frac{1}{ r e^{d + 2h} } \partial_r \left( e^{-\sqrt 2 \phi} r^2 e^{2h} \alpha \right) = \frac
{\gamma}{r e^{d+2h}} \partial_r(r^2 \bar \beta).
\end{equation}
This last equation can be integrated to obtain
\begin{equation} \label{alpha2}
\alpha = \gamma \bar \beta e^{\sqrt 2 \phi}  e^{-2h},
\end{equation}
where we have set a constant of integration to zero because it vanishes in the solutions 
we want to consider interpolating between.

Einstein's equations give
\bea
\frac{1}{r e^{F+d+2h}} \partial_r \left( r^3 e^{-d+2h} \partial_r e^F \right) &=& P + \frac{1}{4r^2 
e^{2d}} e^{2 \sqrt 2 \phi}  (\partial_r (r^2 \bar \beta))^2 \nonumber \\
&& + e^{-\sqrt 2 \phi} \left( \frac{m^2 
\bar \beta^2}{8} + \frac{3 \alpha^2}{2} + \frac{ \gamma^2}{2} e^{-4h} \right),
\label{d}\\
\frac{1}{r e^{F+d+2h}} \partial_r( r^2 e^{F-d+2h}) &=& P - \frac{1}{4r^2 e^{2d}} e^{2 \sqrt 2 
\phi} (\partial_r (r^2 \bar \beta))^2 \nonumber \\
&& + e^{-\sqrt 2 \phi} \left( - \frac{3 m^2 \bar \beta^2}{8} - 
\frac{\alpha^2}{2} + \frac{\gamma^2}{2} e^{-4h} \right),
\label{f}\\
e^{-2h} + \frac{1}{2 r e^{F+d+2h}} \partial_r( r^3  e^{F-d} \partial_r e^{2h}) &=& P + \frac{1}
{4r^2 e^{2d}} e^{2 \sqrt{2} \phi} (\partial_r (r^2 \bar \beta))^2 \nonumber \\
&& + e^{-\sqrt{2} \phi} \left( \frac
{m^2 \bar \beta^2}{8} - \frac{\alpha^2}{2} - \frac{3 \gamma^2}{2} e^{-4h} \right),
\label{h}
\eea
and
\bea
\label{o1}
&&e^{-2h} + \frac{1}{f e^{2d}} \left(f + f r^2 (\partial_r h)^2 + 2 r^2 \partial_r h \partial_r f + 2r 
\partial_r f + 4 r f \partial_r h    \right) \\ 
&&=  r^2 e^{-2d} (\partial_r \phi)^2 + 2 P + 
\frac{1}{4r^2 e^{2d}} e^{2 \sqrt 2 \phi} (\partial_r (r^2 \bar \beta))^2 + e^{-\sqrt{2} \phi} \left
( - \frac{ m^2 \bar \beta^2}{4} - \alpha^2 - \gamma^2 e^{-4h} \right),\nonumber 
\eea
where as before $P = \frac{1}{8} (g^2 e^{\sqrt 2 \phi} + 4gm e^{-\sqrt 2 \phi} -m^2 e^{-3 
\sqrt 2 \phi})$. The final equation comes from the $rr$ component of the Einstein tensor 
rather than the Ricci tensor.

There are seven equations and only six unknown functions, but one of the equations is 
redundant because of the Bianchi identity, which involves the $r$ derivative of the last 
equation.

As explained in the previous section, for the Lifshitz and AdS solutions, the parameters 
$g, m$ in the Lagrangian for our supergravity theory only affect the overall scale of the 
fields; the value of $z$ is determined entirely by $g^2 \gamma^2$. In the four-
dimensional truncated theory, this was manifest at the level of the action. We will use a 
similar scaling of the fields here to simplify the  equations of motion. However, in this 
context we find it convenient to include a factor of $\gamma$ in the scaling of $h$. We set
\begin{equation}
\varphi = \sqrt{\frac{m}{g}} e^{-\sqrt{2} \phi}, \quad e^{-2H} = \frac{\gamma}{\sqrt{gm}} e^
{-2h}, \quad e^{-2D} = \frac{1}{\sqrt{g^3 m}} e^{-2d}, \quad \beta = \sqrt{\frac{m}{g}} \bar 
\beta.
\end{equation}
Then writing the radial coordinate as $r = e^\rho$, we have four second-order equations,
\begin{align} \label{em1}
\partial_{\rho} \partial_{\rho} \beta = & -2\partial_{\rho}\beta - \left(2\beta + \partial_{\rho}
\beta\right)\left(\partial_{\rho} F - \partial_\rho D + 2 \partial_{\rho} H  - 2 \varphi^{-1} 
\partial_{\rho} \varphi \right) \nonumber \\
&+ e^{2D} \varphi \left(\varphi^2 + e^{-4H}\right)\beta \\
\label{em2} \partial_{\rho} \partial_{\rho} \varphi = & -\partial_{\rho} \varphi \left(2 + 
\partial_{\rho}F - \partial_\rho D + 2 \partial_\rho H - \varphi^{-1}\partial_{\rho}\varphi 
\right)  - \frac{1}{2\varphi} \left(\partial_{\rho}
\beta + 2\beta\right)^2 \nonumber \\ 
&+ \frac{e^{2D}}{4} \left(1 - 4\varphi^2 + 3\varphi^4 + \left(\varphi^2 - 
4e^{-4H}\right)\beta^2 + 4\varphi^2 e^{-4H}\right)\\
\label{em3} \partial_{\rho} \partial_{\rho} F = & \partial_\rho F \left(2 + \partial_\rho F - 
\partial_\rho D + 2 \partial_\rho H \right) + \frac{1}{4\varphi}\left(\partial_{\rho}\beta + 
2\beta\right)^2 \nonumber \\
&+ \frac{e^{2D}}{8\varphi}\left(1 + 4\varphi^2 - \varphi^4 + \left(\varphi^2 + 12e^{-4H}
\right)\beta^2 + 4\varphi^2e^{-4H}\right) \\
\partial_{\rho} \partial_{\rho} H = &  \partial_\rho H (2 + \partial_\rho F + \partial_\rho D - 
\partial_\rho H) + 2 \partial_\rho H + 2 \partial_\rho F +1 \nonumber \\ &+
\frac{e^{2D}}{8\varphi}\left(1 + 4\varphi^2 - \varphi^4 + \left(3 \varphi^2 + 7 e^{-4H}\right)
\beta^2 \right) \label{em4} \end{align}
and two further equations
\begin{equation} \label{em5}
2 + 2 \partial_\rho H + \partial_\rho F - \partial_\rho D + \frac{1}{4 \varphi} ( \partial_\rho 
\beta + 2 \beta)^2 = \frac{e^{2D}}{8\varphi} ( 1 + 4 \varphi^2 - \varphi^4 - (3 \varphi^2 + 4 
e^{-4H}) \beta^2 + 4 \varphi^2 e^{-4H}),
\end{equation}
and
\begin{align} \label{cons}
&\frac{1}{g^2 \gamma^2} = - \frac{e^{2H}}{4 \varphi}(\varphi^2 + 4 e^{-4H}) \beta^2 - 
\varphi e^{-2 H}\\
&- e^{2H-2D}\left(1 + (\partial_\rho H)^2 + 2 \partial_\rho H 
\partial_\rho F + 2 \partial_\rho F + 4 \partial_\rho H - 
\frac{(\partial_\rho \varphi)^2}{2\varphi^2} 
+ \frac{ \left(\partial_{\rho}\beta + 2\beta\right)^2 }{4 \varphi^2}\right) .
\nonumber 
\end{align}
Note that this last equation involves only $g^2 \gamma^2$, while the other equations do 
not involve the parameters in the theory at all. Note also that the equations involve only $
\partial_\rho F$, and not $F$. 

We can look at this system in two different ways. If we consider it as a dynamical system, 
it is convenient to view (\ref{em1}-\ref{em5}) as a system of nine coupled first-order 
equations in the variables $e^{H}, e^{D}$, $\varphi$, $\beta$, $\partial_\rho H$, $\partial_
\rho D$, $\partial_\rho \varphi$, $\partial_\rho \beta$ and $\partial_\rho F$. We can solve 
\eqref{em5} algebraically for one of the variables. The system is parameter-free, and the 
remaining set of first-order ODEs defines an eight-dimensional autonomous dynamical 
system. The equation \eqref{cons} then specifies a subspace in this dynamical system 
determined by the value of $g^2 \gamma^2$. The fact that the equations are compatible 
implies that this is an invariant subspace; flows starting in a space of a given value of 
$g^2 \gamma^2$ will remain in that space. That is, the right-hand side of \eqref{cons} is 
constant by virtue of (\ref{em1}-\ref{em5}), as required physically for conservation of the 
flux on the compact space.  This is a convenient way to describe the equations of motion 
which makes the abstract structure clear, but the high dimension unfortunately makes 
any detailed analysis of the structure of the flows difficult. 

Alternatively, when we solve these equations explicitly to find the interpolating solutions 
of interest, we will specify a value of $g^2 \gamma^2$ and explicitly solve (\ref{em1}-\ref
{em4}) and \eqref{cons}. The remaining equation \eqref{em5} is then redundant; it follows 
from  (\ref{em1}-\ref{em4}) and the derivative of \eqref{cons}. 

\subsection{Linearised equations}

To see which directions we would expect flows in, we consider the linearisation of the 
equations of motion about each of the background solutions. We first consider the 
linearization about the AdS$_4$ solutions, as this can be done analytically; we then 
discuss the numerical results from linearisation around the Lifshitz solutions. In the 
linearised equations, we find that there are some linearised modes which violate the 
Breitenlohner-Freedman bound \cite{Breitenlohner:1982bm}, implying that some of these 
solutions are unstable. 

The AdS$_4$ solutions have $F= \ln r$, $\beta = 0$, and $\varphi$, $D$ and $H$ taking 
the constant values given in (\ref{vphi},\ref{c1}). We linearise by writing
\begin{equation}
F = \ln r + \delta F, \quad D = D_0 + \delta D, \quad H= H_0 + \delta H, \quad \varphi = 
\varphi_0 (1 + \delta \varphi), 
\end{equation}
and noting that $\beta$ is itself a linear perturbation. This last
fact makes the linearization simple, as some equations decouple. 

There are two decoupled equations in the AdS$_4$ case. First a combination of \eqref
{em3} and \eqref{em5} gives a simple decoupled equation for $\delta F$, 
\begin{equation}
\partial_\rho \partial_\rho  \delta F + 3 \partial_\rho \delta F = 0, 
\end{equation}
with solution $\delta F = f_0 + f_1 e^{-3 \rho}$. As in the massive vector case, $f_0$ 
corresponds to a deformation of the timelike component of the background metric, which 
acts as a source for the energy density. The mode $f_1$ will then correspond to the 
vacuum expectation value of the energy density. The equation of motion \eqref{cons} 
implies that the mode $f_1$ also appears in $\delta D$, corresponding to the pressure 
required by tracelessness of the stress tensor. Since the energy density is a precisely 
marginal operator, deforming by its source by turning on non-zero $f_0$ does not 
generate a flow. In fact, this is just a diffeomorphism of the bulk spacetime, rescaling the 
time coordinate. 

We also get a simple decoupled linear equation for $\beta$ from \eqref{em1}, 
\begin{equation} \label{decb}
\partial_\rho \partial_\rho \beta = -3 \partial_\rho \beta - 2 \beta + e^{2D_0} \varphi_0 
(\varphi_0^2 + e^{-4 H_0}) \beta. 
\end{equation}
If we look for solutions of the form $\beta = \beta_1 e^{-\Delta \rho}$, this implies 
\begin{equation} \label{D1}
(\Delta-2)(\Delta-1) = -6 \frac{2 \varphi_0^4 - 4 \varphi_0^2 +1}{\varphi_0^2 (2- 
\varphi_0^2)},
\end{equation}
where $\varphi_0$ takes one of the two possible values given by solving \eqref{vphi}. 
Since $\beta$ appears quadratically in the other equations of motion, this solution will not 
source other fields at linear order. 

There are two solutions $\Delta_1 > \Delta_2$ such that $\Delta_1 + \Delta_2 =3$. With 
standard boundary conditions, these solutions correspond to the expectation value of 
and source for a dual operator of conformal dimension $\Delta = \Delta_1$. When the 
dual operator is relevant, turning on the source will deform the solution, generating a flow 
starting from this solution in the UV.

The operator is relevant, $\Delta < 3$, if $\varphi_0^2 < 0.3675$. This is in the range 
corresponding to the small $\varphi_0$ solution, and corresponds to $g^2 \gamma^2 < 
0.23$, when there is a single Lifshitz solution. Since turning on this mode corresponds to 
deforming the solution by exciting the two-form  which is present in the Lifshitz solution 
but absent in the AdS solution, the natural guess is that this will lead to an RG flow from 
the small $\varphi_0$ AdS solution in the UV to the Lifshitz solution in the IR. When $
\varphi_0^2 > 0.3675$, the operator is irrelevant, and we would expect to have flows that 
approach the AdS solution in the IR along this direction. We will show that flows from the 
smaller $z$ Lifshitz solution in the UV can indeed reach these solutions in the IR. 

As noted above, our system of equations can be reduced to an eight-dimensional 
dynamical system. There will therefore be four more linearly independent solutions of the 
linearised equations of motion. We can obtain the associated powers by considering 
\eqref{em2} and \eqref{em4}, which give 
\begin{equation}
\frac{2}{3} \varphi_0^2 (2-\varphi_0^2) \left( \partial_\rho \partial_\rho \delta \varphi + 3 
\partial_\rho \delta \varphi \right) =  -2 (1 - 3 \varphi_0^4) \delta \varphi + 4 
(1-3\varphi_0^2)(1-\varphi_0^2) \delta H
\end{equation}
and
\begin{equation}
\frac{4}{3} \varphi_0^2 (2- \varphi_0^2) \left( \partial_\rho \partial_\rho \delta H + 3 
\partial_\rho \delta H \right) = 2 (1- 3 \varphi_0^2)(1- \varphi_0^2) \delta \varphi + 4(-1 + 8 
\varphi_0^2 - 5 \varphi_0^4) \delta H. 
\end{equation}
These linear perturbations are dual to a pair of scalar operators in the dual CFT. As the 
equations are coupled, we should perform a field redefinition to diagonalise this system 
to obtain the bulk fields dual to the individual operators. But as all we are mainly 
interested in is finding the dimensions of the operators, we can proceed by considering a 
solution of the form $\delta \varphi = \varphi_1 e^{-\Delta \rho}$, $\delta H = H_1 e^{-
\Delta \rho}$, which gives
\begin{equation}
[ \varphi_0^2 (2-\varphi_0^2) \Delta (\Delta-3) -3(3 \varphi_0^4 -1)]\varphi_1  = 6 (1-4 
\varphi_0^2 + 3 \varphi_0^4) H_1, 
\end{equation}
and 
\begin{equation}
[ \varphi_0^2 (2-\varphi_0^2) \Delta (\Delta-3) +3(1 - 8 \varphi_0^2 + 5 \varphi_0^4)]H_1 
= \frac{3}{2} (1-4 \varphi_0^2 + 3 \varphi_0^4) \phi_1. 
\end{equation}
Solving for $\Delta$, 
\begin{equation} \label{scD}
\varphi_0^2 (2- \varphi_0^2) \Delta (\Delta -3) = -(6 \varphi_0^4 - 15 \varphi_0^2 + \frac{9}
{2}) \pm \frac{3}{2} (\varphi_0^2-1) \sqrt{25 \varphi_0^4 - 6 \varphi_0^2 + 1}.
\end{equation}

As before, we get solutions in pairs with $\Delta_1 > \Delta_2$ such that $\Delta_1 + 
\Delta_2 =3$. With standard boundary conditions, these correspond to the expectation 
value of and source for a dual scalar operator of conformal dimension $\Delta = 
\Delta_1$. The dimensions of the two scalar operators can thus be obtained by taking the 
larger solution for $\Delta$ in \eqref{scD} for each choice of sign. 

For the upper sign, the solutions of \eqref{scD} are real and the operator has $\Delta > 3$ 
for all $\varphi_0^2 \in (\frac{1}{3}, 1)$. This thus corresponds to an irrelevant operator 
around the AdS solution, and we would expect it to correspond to a direction along which 
we can approach either of these AdS solutions. We will see below that this direction can 
be reached by flows from the asymptotically AdS$_6$ solution in the UV (we construct 
the flow to the smaller $\varphi_0$ solution, but we expect such a flow to exist also for the  
larger $\varphi_0$ solution).

For the lower sign, the solutions of \eqref{scD} are complex for $\varphi_0^2 < 0.354$. 
This indicates that the linearised scalar in the bulk has a mass violating the 
Breitenlohner-Freedman bound. Here we have restricted to an ansatz where the fields 
have only radial dependence, but one can easily extend this to analyse general 
linearised perturbations around the AdS solution in the context of the four-dimensional 
action \eqref{4daction} to see this explicitly. There is a decoupled pair of equations 
involving the perturbations $\delta h(x^\mu)$ and $\delta \phi(x^\mu)$. Diagonalizing 
these equations gives two massive scalars on the AdS background, with masses 
\begin{equation}
m^2 L^2 = -\frac{3}{\varphi_0^2 (2- \varphi_0^2)} [(1-4\varphi_0^2 + \varphi_0^4) \pm (1-
\varphi_0^2) \sqrt{25 \varphi_0^4 - 6 \varphi_0^2 +1}],
\end{equation}
where $L^2$ is the background AdS scale. The Breitenlohner-Freedman bound is 
violated if $m^2 L^2 < - \frac{9}{4}$, which happens for the upper sign for sufficiently 
small $\varphi_0$. Thus, this mode indicates a dynamical stability of the AdS$_4$ 
spacetime to exponentially growing modes for this scalar. At $\varphi_0 = \frac{1}{3}$, $
\gamma=0$, and this mode is simply an excitation of $\delta h$, the overall volume of the 
compact space. As we increase the flux, moving away from $\varphi_0 = \frac{1}{3}$, the 
unstable mode involves excitation of the dilaton as well.

Setting aside this issue of instability, we can look at the issue of relevant operators. The 
operator has $\Delta < 3$ for $\varphi_0^2 < 0.59$. This is the mid-point corresponding 
to $g^2 \gamma^2 = 1.18$, so the AdS solution with smaller $\varphi_0$ has a relevant 
mode which excites the scalars but not the two-form. The natural interpretation is that a 
deformation by this operator will lead to an RG flow from the AdS solution with smaller $
\varphi_0$ in the UV to the AdS solution with larger $\varphi_0$ in the IR. We will 
construct such solutions explicitly in the next section. 

\begin{figure}[htb]
\centering
\input{AdS_evals.tex}
\caption{Operator dimensions for the linearised perturbations about the AdS solutions, determined from the asymptotic scaling of bulk fields. $\Delta_1$ is the energy density, $\Delta_2$ is the mode which excites the two-form, and $\Delta_3$, $\Delta_4$ are the modes which excite the scalars $h$, $\phi$. We plot the real parts, but note that $\Delta_4$ is complex where the real part is $\frac{3}{2}$, and it will then not correspond to the dimension of an operator.}
\label{fig:AdS_evals}
\end{figure}
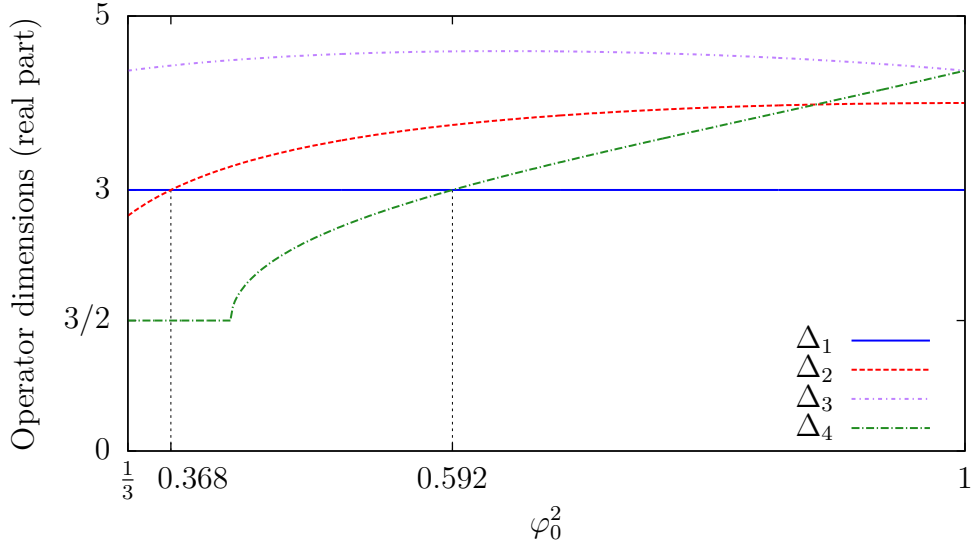

In summary, there are four operators which appear in our analysis around AdS; the field 
theory energy density, which is marginal; the operator dual to the excitation $\beta$ of the 
two-form, which is relevant for $\varphi_0^2 < 0.3675$, and two scalar operators, one of 
which is always irrelevant, the other of which is relevant for $\varphi_0^2 < 0.59$. The 
dimensions of these operators as a function of $\varphi_0$ are shown in figure \ref
{fig:AdS_evals}. 

\begin{figure}[htb]
\centering
\input{lower_Li_evals.tex}
\caption{Real part of the scaling of the linearised perturbations about the larger $z$ Lifshitz solution. The values are normalized to $z+2$. When the values are real, they should be interpreted as the dimension of the corresponding operator, assuming the usual holographic dictionary. In particular, $\Delta_1$ corresponds to the energy density. Note that $\Delta_4$ is complex in the region where the real part is $\frac{z+2}{2}$. }
\label{fig:lower_Li_evals}
\end{figure}
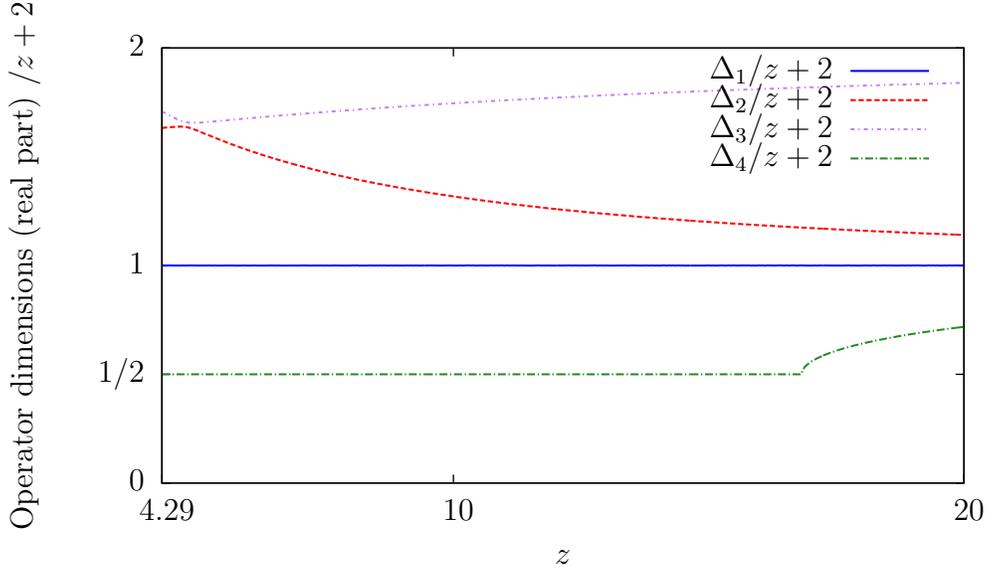

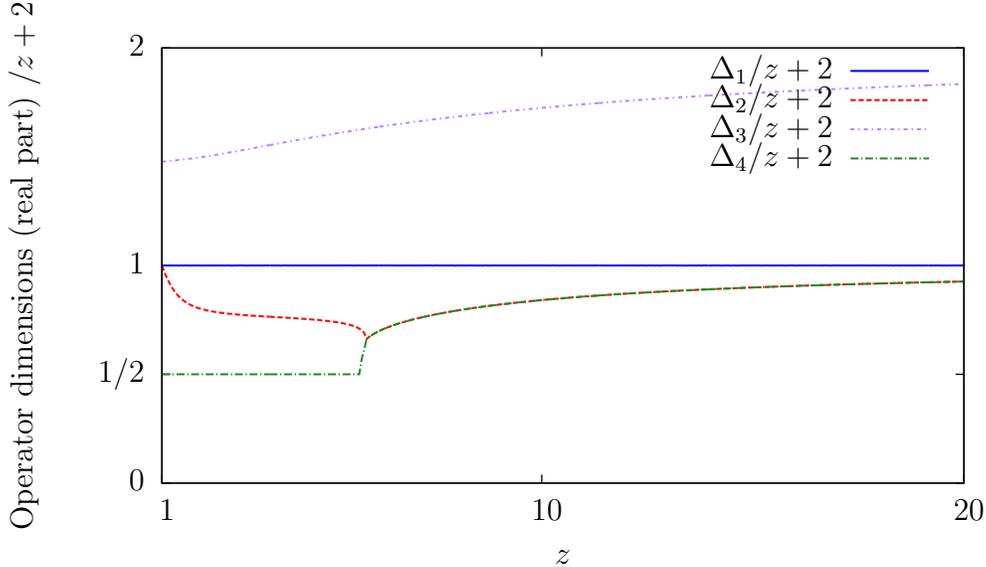
\begin{figure}[htb]
\centering
\input{upper_Li_evals.tex}
\caption{Real part of the scaling for the linearised perturbations about the smaller $z$ Lifshitz solution. The values are normalized to $z+2$. When the values are real, they should be interpreted as the dimension of the corresponding operator, assuming the usual holographic dictionary. In particular, $\Delta_1$ corresponds to the energy density. Note that $\Delta_4$ is complex in the region where the real part is $\frac{z+2}{2}$, and that $\Delta_2$ and $\Delta_4$ are both complex after their real parts merge.}
\label{fig:upper_Li_evals}
\end{figure}

We now turn to the linearisation around the Lifshitz solutions. These are more 
complicated, as nothing obviously decouples. We can however rewrite the problem as a 
simple linear analysis problem by writing the linearised system of equations as $r\frac{d}
{dr} \delta\mathbf{x} = A \delta\mathbf{x}$, where $\delta \mathbf{x}$ collectively denotes 
the linearised perturbations, and $A$ is a matrix depending on the background field 
values. The problem of finding solutions then reduces to finding the eigenvectors and 
eigenvalues of the matrix $A$. If $A$ has eigenvectors $\mathbf{v}_i$ with eigenvalues $
\Delta_i$, the linearised system has solutions $\delta \mathbf{x} = \sum_i \mathbf v_i e^{-
\Delta_i \rho}$. The eigenvalues were calculated numerically.

In the larger $z$ solutions, there is a pair of complex eigenvalues with real part $\frac{z
+2}{2}$ for $z<16.82$. Extrapolation from the AdS case would lead us to expect that this 
is associated with a dynamical instability of the Lifshitz solution, although constructing 
such an instability explicitly requires work outside of our ansatz. For the smaller $z$ 
solutions, the situation is a little more complicated. There is a similar pair of complex 
eigenvalues for $z < 5.69$, then there is a small window where all eigenvalues are real 
up to $z=5.83$, and then a set of four complex eigenvalues but whose real parts are not 
$\frac{z+2}{2}$. We would again expect that, at least for $z < 5.69$, the complex 
eigenvalues signal a mode which is violating the Breitenlohner-Freedman bound. In the 
second region of complex eigenvalues, the interpretation is less clear, but it is certainly 
problematic to interpret these eigenvalues in terms of operator dimensions in a dual field 
theory.

This instability is clearly an important aspect of the physics of these Lifshitz solutions, 
particularly as it is appearing for phenomenologically interesting values like  $z=2$ 
which we would most like to understand. An important task for the future will be to 
perform a more general linearised analysis to exhibit the instability explicitly and 
understand its character.

For the present, we will leave this instability to one side and return to analysing the flows. 
We have not considered in detail the identifaction of these linearised solutions with dual operators, so the 
interpretation from the field theory point of view is somewhat hueristic, but we see the 
structures we would expect.\footnote{Extending \cite{ALL} to construct the holographic dictionary for this theory should in principle be straightforward.} As in the AdS case, the solutions 
come in pairs  with $\Delta_i + \Delta_{i+1} = z
+2$, which should correspond to the source for and expectation value of dual operators respectively. When $
\Delta_i > z+2$, $\Delta_{i+1} < 0$, so the effect of this ``source mode'' is large at large $r
$, and corresponds to a deformation which grows in the UV; conversely if $\Delta_i < z
+2$, the effect is important at small $r$, corresponding to a deformation which grows in 
the IR. The operator dimensions (the larger eigenvalue in each pair)  are plotted in 
figures \ref{fig:lower_Li_evals} and \ref{fig:upper_Li_evals}. 

To construct interpolating solutions dual to renormalization group flows, we are therefore 
interested in the perturbation of the Lifshitz solution by these ``source modes''. Since our 
explicit construction works out from the IR by considering a perturbation along an 
irrelevant direction, this construction is insensitive to the existence of the complex 
eigenvalues, at least at linear order. Thus, we construct interpolating solutions dual to 
flows without considering whether the solutions we are interpolating between are stable.  

Of the four deformations included in our ansatz, there is one which is exactly marginal; 
this corresponds again to the energy density in the dual field theory. There is one 
irrelevant operator in the smaller $z$ Lifshitz solution, and two in the larger $z$ Lifshitz 
solution. 

In summary, the expected flows here are very similar to in the massive vector model. 
There should be a flow from the small $\varphi_0$ AdS solution to the large $\varphi_0$ 
AdS solution. In the regime where there is a single Lifshitz solution, there should be flow 
from the small $\varphi_0$ AdS solution in the UV to the larger $z$ Lifshitz solution in the 
IR. Once the smaller $z$ Lifshitz solution appears, this will be replaced by two flows 
starting from the smaller $z$ Lifshitz solution in the UV, and running to the larger $z$ Lifshitz 
solution or an AdS solution in the IR. In addition to the ones required by these expected 
flows, there is an additional irrelevant direction about each solution; we will see in the 
next subsection that this is associated with a flow from an asymptotically AdS$_6$ 
solution.

\FloatBarrier

\subsection{Flows}

We can now turn to the discussion of the numerical solutions interpolating between the 
Lifshitz and AdS solutions. As in the massive vector model, solutions are obtained by 
starting from a candidate IR fixed point at small $r$, perturbing along one of the 
eigenvectors associated with an irrelevant direction, and integrating out to large $r$ to 
identify the UV fixed point at the source of the renormalization group flow. The analysis is 
made more complicated in the present context because of the existence of more than 
one irrelevant direction in many cases, which implies that to reach the desired UV fixed 
point we have to search for an appropriate direction for the perturbation at small $r$. 
However, with the exception of the large $\varphi_0$ AdS solutions, we have only one or 
two irrelevant directions, so this search can be simply carried out by interval bisection in 
the space of possible directions on the two-dimensional plane spanned by the two 
irrelevant eigenvectors.

There are a large number of cases to consider, so we have relegated the plots of 
numerical solutions to appendix \ref{plots}, and here give a description of the results and 
their interpretation. 

\subsubsection{Flows from 6D AdS}
\label{subsubsec:6DUVflows}

The simplest case to consider is the generic perturbation from the AdS  or Lifshitz 
solutions, perturbing along the most irrelevant direction. Here we have only a choice of 
sign in the perturbation. These solutions are particularly interesting as they enable us to 
give a description of the field theory dual to the Lifshitz solutions in the context of this 
gauged supergravity theory. 

These flows do not approach any of the fixed points we discussed previously in the UV; 
they have $e^{2H}$ scaling like $r^2$ in the UV, and $e^{2D}$ tending to some finite, 
non-zero value (for one choice of sign). That is, the compact hyperbolic space has a 
proper size growing like $r^2$ at large $r$. These can therefore be identified as flows 
from an asymptotically AdS$_6$ geometry, 
\begin{equation} \label{6D_space} ds^2 \approx -r^2 dt^2 + r^2 \left(dx_1^2 + 
dx_2^2\right) +L^2 \frac{dr^2}{r^2} + C \frac{r^2}{y_2^2} \left(dy_1^2 + dy_2^2\right).
\end{equation}

When the solution in the IR is AdS$_4$, this type of flow was previously discussed in \cite
{Nunez:2001pt}, building on the work of \cite{Maldacena:2000mw}.  The asymptotics 
correspond to considering a five-dimensional field theory on $\mathbb{R}^{1,2} \times 
H^2$. Specifically, the field theory is the five-dimensional $\mathcal N=2$ conformal field 
theory obtained from the IR limit of the $Sp(N)$ D4-D8 theory. The conformal symmetry in 
the UV is broken by introducing curvature in the background spacetime, giving an 
asymptotically AdS$_6$ geometry.\footnote{This is thus technically slightly different from 
the flows constructed by deforming the field theory Lagrangian we are otherwise 
considering, but the construction of the bulk solution is essentially the same.} In the UV, 
the presence of the flux $\gamma$ on the compact space implies that the UV field theory 
is a twisted field theory, as in \cite{Maldacena:2000mw}. The interpolating geometries 
with AdS$_4$ in the IR then describe the flow from the a five-dimensional field theory in 
the UV to a three-dimensional conformal theory in the IR.  If we choose $\varphi_0^2 = 
\frac{1}{2}$, which corresponds to $g^2 \gamma^2 = 1$, the AdS$_4$ solution preserves 
half the supersymmetry. This supersymmetry is in fact preserved along the flow. It is these 
supersymmetry-preserving flows which are explicitly considered in  \cite{Nunez:2001pt}. 

To understand the flows to the Lifshitz theories in the IR, we just need to consider a more 
general deformation of the asymptotically AdS$_6$ solution where we turn on the two-
form by considering a perturbation involving $\beta$. The field theory dual to the Lifshitz 
solutions can then be defined as the result of considering the IR limit of the $\mathcal 
N=2$ twisted field theory on $\mathbb{R}^{1,2} \times H^2$ with this further deformation. 
In principle, this gives a constructive definition of the field theory duals of the Lifshitz 
solutions of the gauged supergravity theory, in terms of a controllable deformation of an 
explicit supersymmetric field theory. It would be interesting to understand this description 
further.  Examples of such flows with either an AdS or a Lifshitz solution in the IR are 
shown in figures \ref{fig:6D_AdS_flow_2} and \ref{fig:6D_us_Li_flow_1}.

\subsubsection{Asymptotically AdS flows}
\label{ss:AdS}

Next we consider flows from an AdS$_4$ solution in the UV. The simplest case is flows 
between two AdS solutions; the linearised analysis led us to expect a solution 
interpolating between the small $\varphi_0$ and large $\varphi_0$ AdS solutions. This 
flow will have $\beta=0$, so we can look for it by starting from the large $\varphi_0$ AdS 
solution at small $r$ perturbed by some linear combination of the two irrelevant 
directions associated with the scalar operators, which keep us in this subspace.

To find a flow to the small $\varphi_0$ solution, we scan across the possible linear 
combinations. We find that the flow that hits the other AdS point is very close to the $
\Delta_4$ direction (perturbations that involve the $\Delta_3$ direction will lead to the 
asymptotically AdS$_6$ solution considered above). A typical example of such a flow, 
from  $\varphi_0^2=0.425$ to $\varphi_0^2=0.8$, is shown in figure \ref
{fig:AdS_AdS_flow}.

In \cite{Copsey:2010ya}, it was pointed out that interpolating solutions which approach a 
Lifshitz solution in the UV and an AdS solution in the IR often have a mild singularity in 
the IR. In fact, the singularity pointed out there is a purely IR feature, which will appear 
whenever an interpolating solution approaches AdS in the interior, and the associated 
irrelevant direction in the field theory is close to marginal. We discuss this here as the 
AdS to AdS flows we are considering here provide a simple example of this point. 

Consider the behaviour of the flow near the IR fixed point. This is controlled by the 
leading irrelevant direction, which for the flows considered here will be the one 
associated with the eigenvalue $\Delta_4$. The IR fields are therefore of the general 
form $\delta \mathbf x \sim r^{\Delta_4 -3}$.  As $r \to 0$, these perturbations decay to 
zero. However, as $\Delta_4 < 5$ in all the large $\varphi_0$ solutions, the second 
derivatives $\partial_r^2 \delta \mathbf x$ will blow up at small $r$. As noted in  \cite
{Copsey:2010ya}, this is reflected in a divergence of the components of the Riemann 
tensor in a parallely propagated orthonormal frame as $r \to 0$. This divergence should 
signal some pathology in the behaviour of the field theory. Further analysing these 
divergences is an interesting problem for the future; these AdS to AdS flows provide a 
useful laboratory to do so, as they are simple deformations of a relativistic conformal field 
theory, although they do not preserve supersymmetry, as the large $\varphi_0$ AdS 
solution is never supersymmetric. 

Flows from an AdS solution in the UV to a Lifshitz solution in the IR are only expected to 
be possible for the small $\varphi_0$ AdS solution for $g^2\gamma^2 \lesssim 0.227$, 
as this is when the operator dual to $\beta$ is relevant. We expect the IR end of this flow 
to be the larger $z$ Lifshitz solution, as it has an additional irrelevant direction (in 
addition to the one associated with flows from AdS$_6$). We numerically found such 
flows for a range of values of $z$. The flow with $z=5$ in the IR is shown in figure \ref
{fig:AdS_Li_flow_1}. 

\subsubsection{Asymptotically Lifshitz flows}
\label{ss:Li} 

We will also have flows with a Lifshitz field theory in the UV; as discussed previously, 
when $g^2\gamma^2 >0.227$, the AdS solution can no longer be the UV fixed point 
associated with the larger $z$ Lifshitz solution in the IR.  Since this is the value at which 
the smaller $z$ Lifshitz solution appears, it is natural to assume that the flow is replaced 
by new flows involving this solution. Indeed, shooting from the IR will now produce flows 
from the smaller $z$ Lifshitz solution in the UV to the larger $z$ Lifshitz solution in the IR. 
The flow from $z=1.248$ to $z=9$ is shown as figure \ref{fig:Li_Li_flow_2}. 

We would also expect there to be a Lifshitz to AdS flow in this range of parameters, 
corresponding roughly speaking to deforming the smaller $z$ Lifshitz solution in the 
opposite direction, decreasing $\beta$. We should be able to obtain the flows 
numerically by starting from the small $\varphi_0$ AdS solution and considering the 
irrelevant perturbation along the $\Delta_2$ direction. Such flows could also exist for the 
large $\varphi_0$ AdS solutions, but they might be more difficult to find because of the 
additional irrelevant direction there. In fact, it proved possible to find flows with both AdS 
solutions in the IR using a simple shooting algorithm. Examples are shown in figures \ref
{fig:Li_AdS_flow_2} and \ref{fig:Li_AdS_flow_8} respectively. The differences between 
these two flows is typical of the difference between flows from the same Lifshitz space to 
AdS spaces on the two different branches.

There is an interesting special case of the flow from smaller $z$ Lifshitz in the UV to large $
\varphi_0$ AdS in the IR. Since there is also a flow from the Lifshitz solution to the small $
\varphi_0$ AdS solution, it should be possible to tune the deformation so that the flow 
from smaller $z$ Lifshitz in the UV to large $\varphi_0$ AdS passes near the fixed point 
associated with the small $\varphi_0$ AdS solution at intermediate energy scales.  These 
flows provide an interesting illustration of the general shooting technique, showing how 
starting from a more generic flow we can tune in to a different flow by varying the 
irrelevant direction of perturbation at the IR end of the flow.  An illustrative flow geometry 
is shown in  figure \ref{fig:Li_AdS_AdS_flow_1}. Note that such solutions are only 
possible for the region of parameter space in which $\Delta_3$ is positive for both AdS 
solutions, namely $0.227 \lesssim g^2 \gamma^2 \lesssim 1.185 $.  It would also be 
interesting to understand what happens to the flows from the Lifshitz solution for $ g^2 
\gamma^2> 1.185 $,  when the AdS solutions no longer exist. 

Since the smaller $z$ Lifshitz branch emanates from the small $\varphi_0$ AdS solution at 
$g^2 \gamma^2 \approx 0.23$, flows from a smaller $z$ Lifshitz fixed point in the UV to the 
small $\varphi_0$ AdS solution in the IR for $z$ near 1 involve only a small change from 
the initial solution. We therefore thought it might be interesting to see if such flows could 
be analysed perturbatively. However, it is easy to see that they cannot be analysed in a 
purely linear approximation: starting from the IR AdS solution, the decoupled equation 
\eqref{decb} tells us that $\beta = \beta_1 e^{-\Delta \rho}$ at the linearised level, and the 
linear approximation must therefore break down at sufficiently large $r$, with the 
quadratic or higher corrections causing the asymptotic value of $\beta$ to approach the 
constant value associated with the Lifshitz solution. Thus, even though the asymptotic 
value of $\beta$ is small for $z$ near 1, a simple linearised analysis of the flow is not 
possible.

\section{Conclusions}
\label{concl}

Our main goal in this paper was to explore the renormalization group flows between 
theories with isotropic or anisotropic scaling symmetries from the dual holographic 
viewpoint. This analysis sheds further light on the field theory interpretation of the Lifshitz 
geometries; several motivations and possible applications were discussed in the 
introduction.

We studied the flows in the context of the simple massive vector model of \cite{Taylor:2008tg} 
and in the gauged supergravity theory considered in \cite{Gregory:2010gx}. We 
found that the two theories had a surprisingly similar structure of flows. There were a 
range of different possibilities: flows between different AdS solutions, AdS to Lifshitz 
flows, Lifshitz to AdS flows and Lifshitz to Lifshitz flows. In the gauged supergravity 
model, there were also flows from a six-dimensional asymptotically AdS solution to four-
dimensional AdS or Lifshitz solutions. These last cases are particularly interesting as 
they offer a description of the field theory dual to the Lifshitz solutions, in terms of a flow 
from a five-dimensional $Sp(N)$ theory compactified on $H^2/\Gamma$. Exploring the consequences for the structure of the field theory dual to the Lifshitz solution is an interesting direction for future work.

In analysing the linearised perturbations, we noted that some of the AdS and Lifshitz 
solutions of the gauged supergravity theory appear to be unstable. This contrasts with the 
massive vector model, where no instability appeared within our ansatz. The appearance 
of instabilities for non-supersymmetric AdS solutions is not a surprise; \cite{Bobev:2010ib} 
has argued that this will be generic for non-supersymmetric AdS solutions. 
Further analysis and characterisation of these instabilities is probably the most important 
open direction in our work.

Another direction for further development will be to consider a similar analysis for other 
theories, notably the similar construction of three-dimensional Lifshitz geometries in type 
IIB supergravity in \cite{Gregory:2010gx}. In the context of the circle reductions which give $z=2$ Lifshitz solutions, it would be interesting to study the flows in the model of \cite{Cassani:2011sv}, which has both Lifshitz and AdS solutions. One could also look for flows 
between higher-dimensional AdS solutions and Lifshitz solutions in all these circle reductions. It would also be 
interesting to further analyse the curvature singularity in the IR region of the flows 
signalled in \cite{Copsey:2010ya} and to understand its interpretation in the field theory. 

\section*{Acknowledgements}

We acknowledge helpful discussions with Luke Barclay. 
This work was supported in part by STFC under the rolling 
grant ST/G000433/1. RG would like to acknowledge the Aspen Center 
for Physics, NSF grant 1066293, for hospitality while this work was
being completed. SFR thanks the Centre de Ciencias de Benasque for hospitality while this work was being completed.

\appendix

\section{Numerical plots for $F(4)$ theory}
\label{plots}

\begin{figure}[htb]
\centering\hbox{\hspace{-2cm}\input{6D_AdS_2.tex}}
\caption{Holographic RG flow from the 6D spacetime \eqref{6D_space} to the $\varphi_0^2=0.45$ AdS solution. $F = 1$, $\beta=0$ throughout this flow.}
\label{fig:6D_AdS_flow_2}
\end{figure}
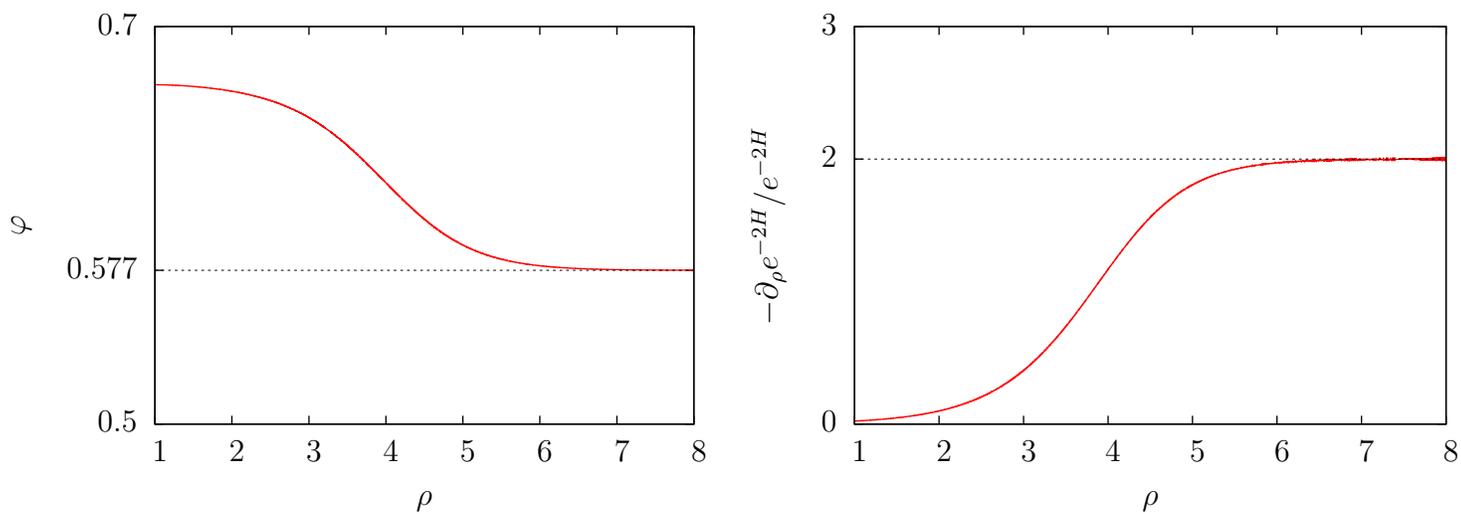

\begin{figure}[htb]
\centering\hbox{\hspace{-2cm}\input{6D_us_Li_1.tex}}
\caption{Holographic RG flow from the 6D spacetime \eqref{6D_space} to the $z=2$ Lifshitz solution on the smaller $z$ branch of Lifshitz solutions.}
\label{fig:6D_us_Li_flow_1}
\end{figure}
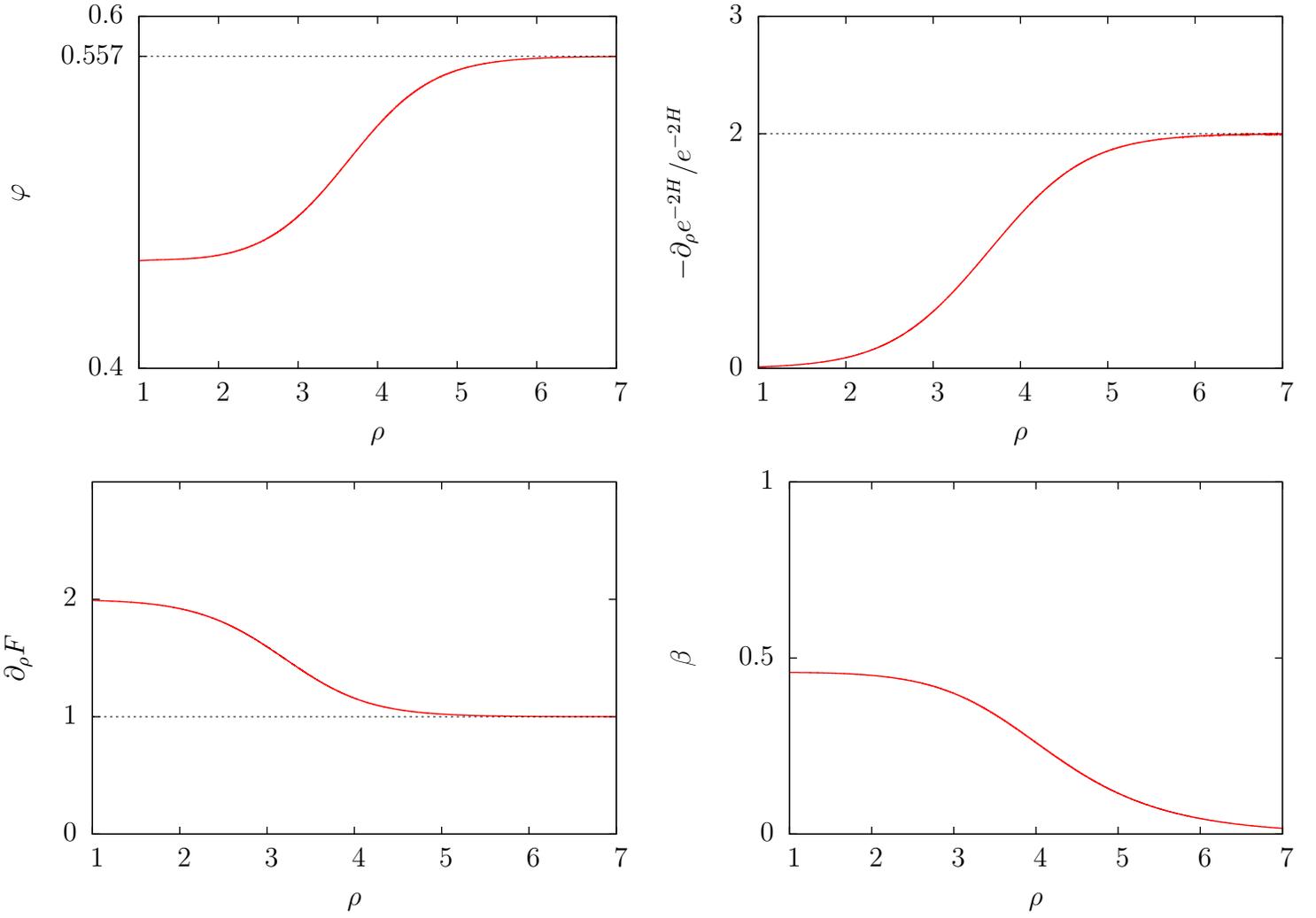

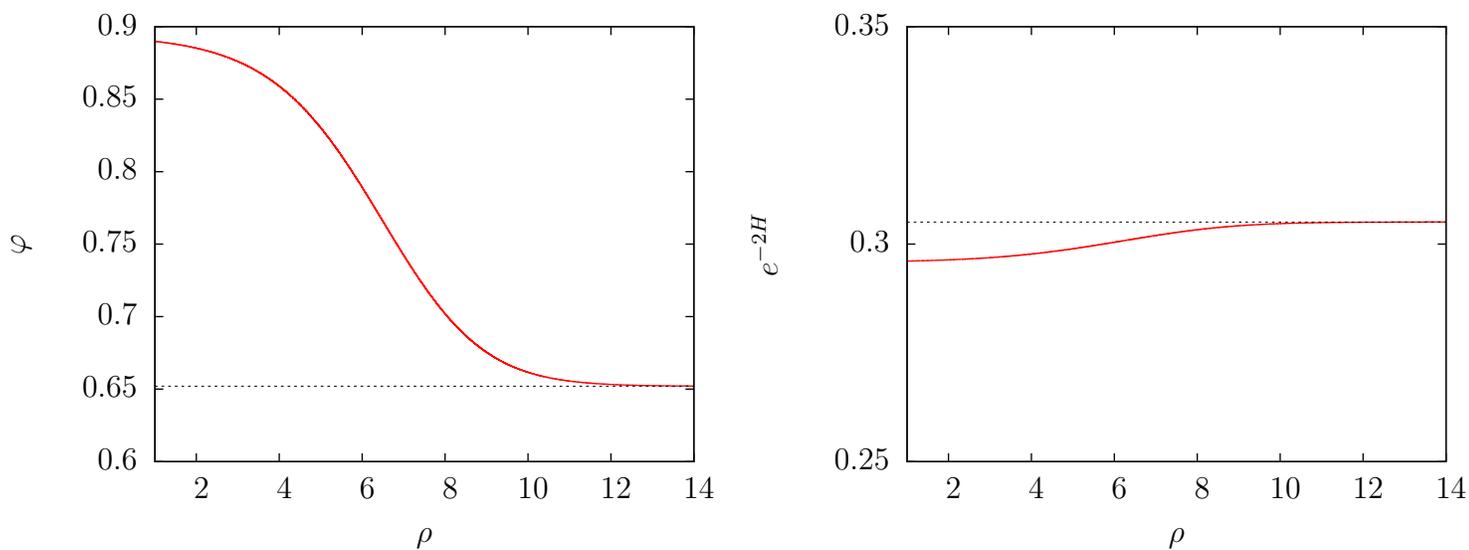
\begin{figure}[htb]
\centering\hbox{\hspace{-2cm}\input{AdS_AdS_1.tex}}
\caption{Holographic RG flow from an AdS space with $\varphi_0^2=0.425$ to an AdS space with $\varphi_0^2=0.8$. The dashed lines show the exact values of $\varphi_0$ and $e^{-2H}$ of the small $\varphi_0$ AdS space we expected to hit. $f=r$ and $\beta=0$ throughout the flow, as expected.}
\label{fig:AdS_AdS_flow}
\end{figure}

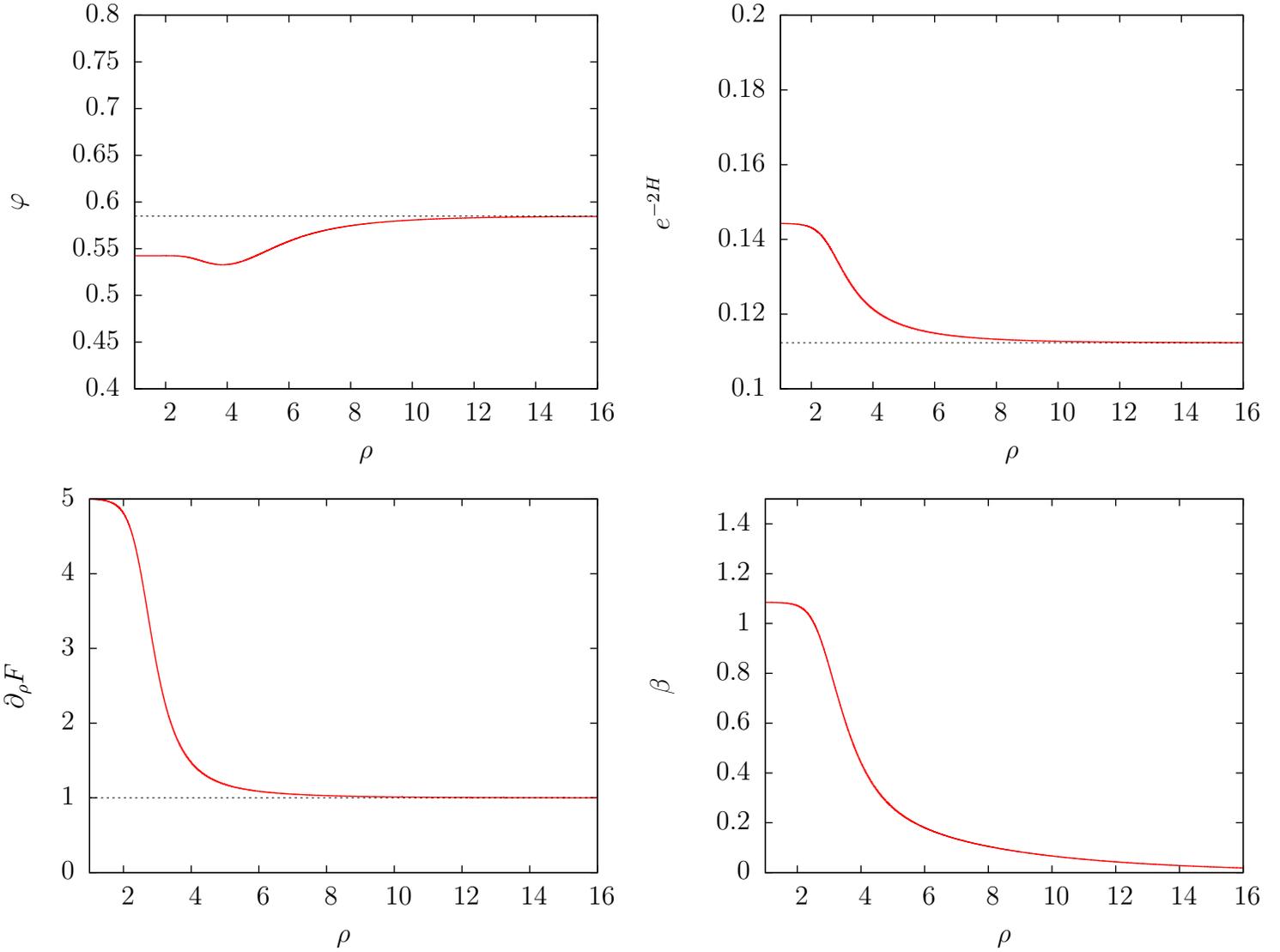
\begin{figure}[htb]
\centering\hbox{\hspace{-2cm}\input{AdS_Li_1.tex}}
\caption{Holographic RG flow from an AdS space with $\varphi_0^2=0.342$ to a Lifshitz space on the lower sign branch with $z=5$. The dashed lines show the exact field values of the UV AdS solution. Note that $F$ provides an estimate of $z$.}
\label{fig:AdS_Li_flow_1}
\end{figure}

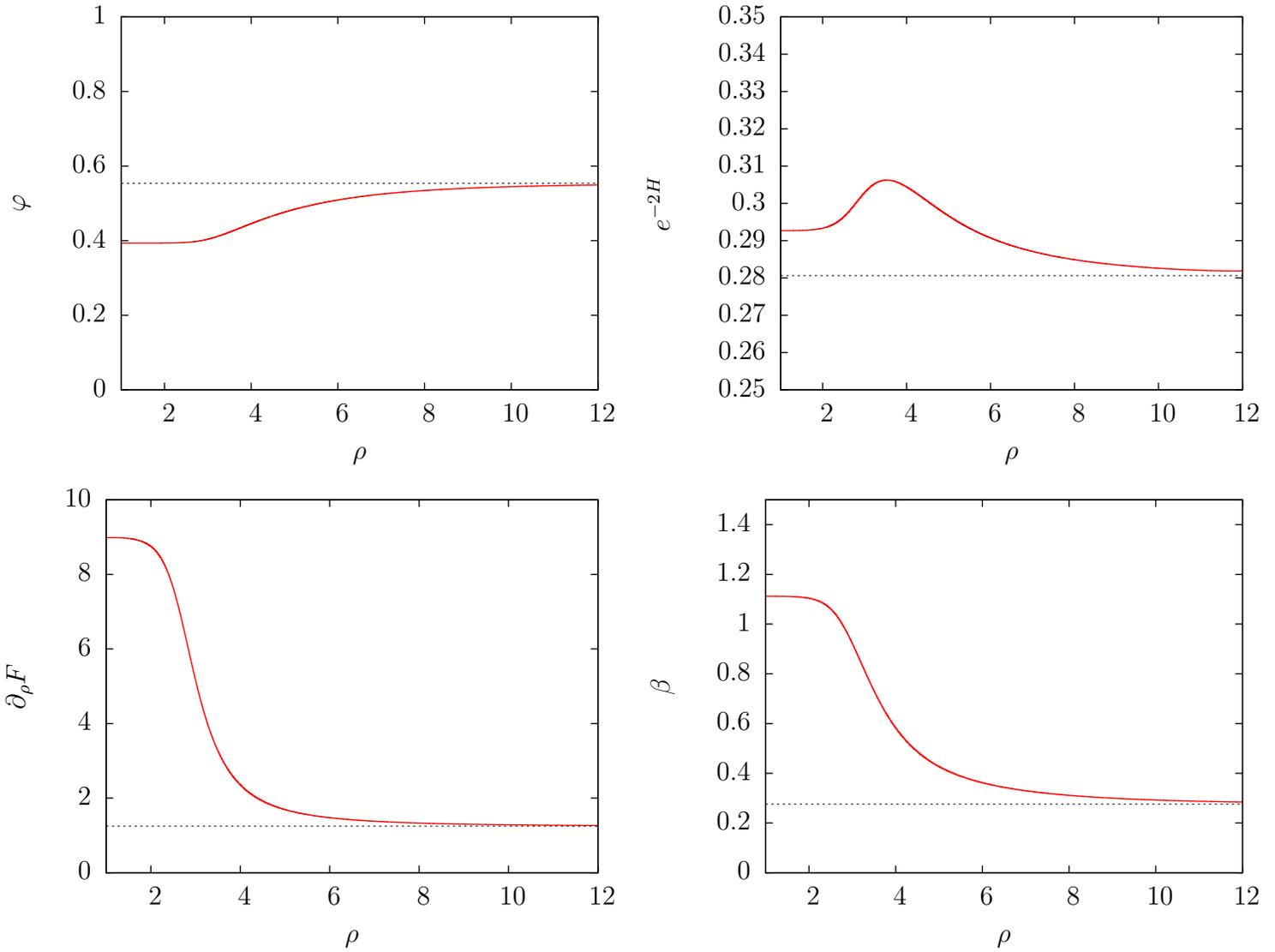
\begin{figure}[htb]
\centering\hbox{\hspace{-2cm}\input{Li_Li_2.tex}}
\caption{Holographic RG flow from an smaller $z$ Lifshitz space with $z=1.248$ to a larger $z$ branch Lifshitz solutions with $z=9$. The dashed lines show the exact field values of the Lifshitz space in the UV. Note that $F$ provides an estimate of $z$.}
\label{fig:Li_Li_flow_2}
\end{figure}

\begin{figure}[htb]
\centering\hbox{\hspace{-2cm}\input{Li_AdS_2.tex}}
\caption{Holographic RG flow from an smaller $z$ Lifshitz space with $z=2.258$ to an AdS space on the small $\varphi_0$ branch with $\varphi_0^2=0.45$. The dashed lines show the exact field values of the Lifshitz space.}
\label{fig:Li_AdS_flow_2}
\end{figure}

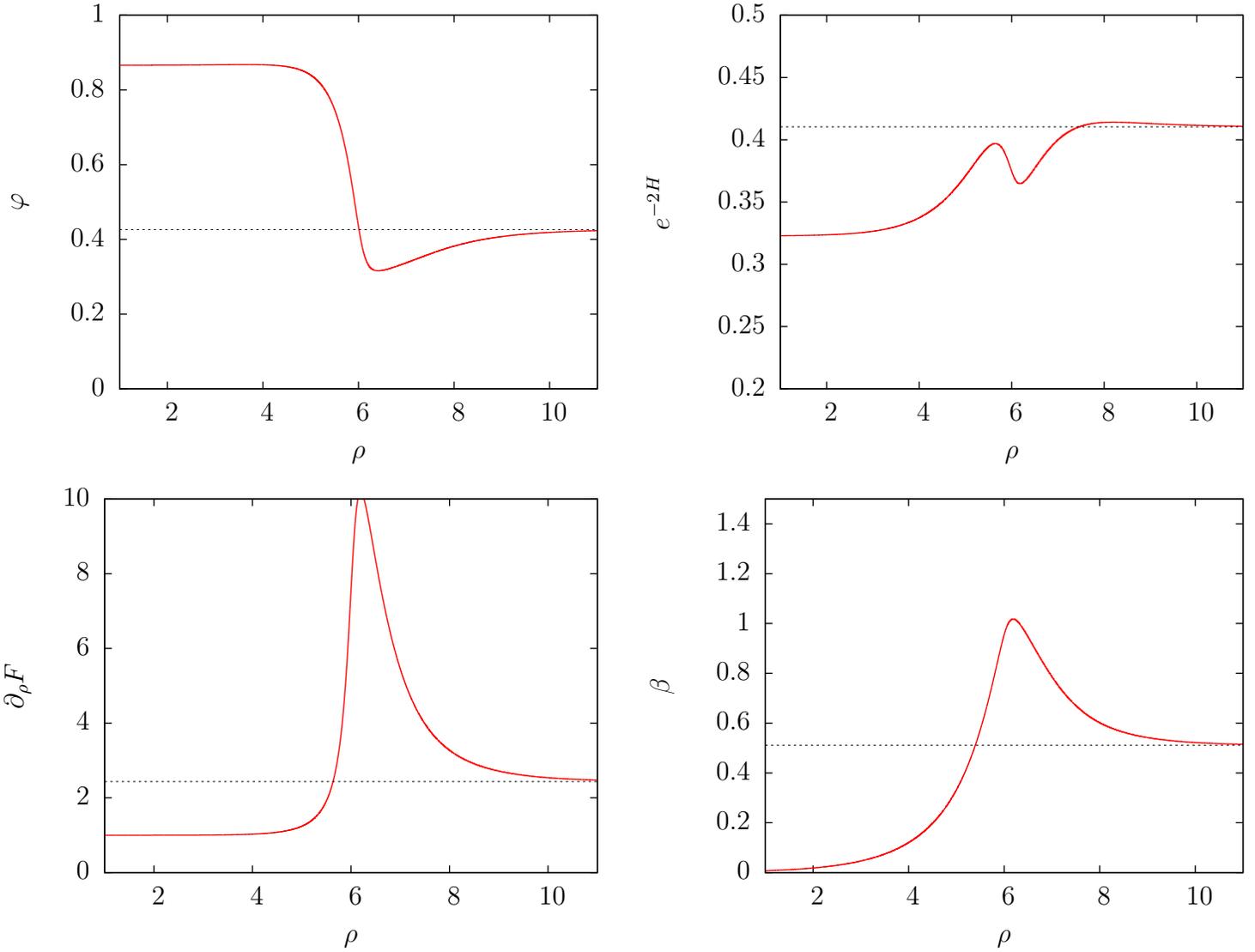
\begin{figure}[htb]
\centering\hbox{\hspace{-2cm}\input{Li_AdS_8.tex}}
\caption{Holographic RG flow from a small $z$ Lifshitz space with $z=2.437$ to an AdS space on the large $\varphi_0$ branch with $\varphi_0^2=0.75$. The dashed lines show the exact field values of the Lifshitz space.}
\label{fig:Li_AdS_flow_8}
\end{figure}

\begin{figure}[htb]
\centering\hbox{\hspace{-2cm}\input{Li_AdS_AdS_1.tex}}
\caption{Holographic RG flow from an small $z$ Lifshitz space with $z=1.756$ to \emph{a point close to} the AdS space on the small $\varphi_0$ branch with $\varphi_0=0.234$, and finally to the AdS space on the large $\varphi_0$ branch with $\varphi_0=0.8$. The dashed lines show the exact field values of the Lifshitz space.}
\label{fig:Li_AdS_AdS_flow_1}
\end{figure}
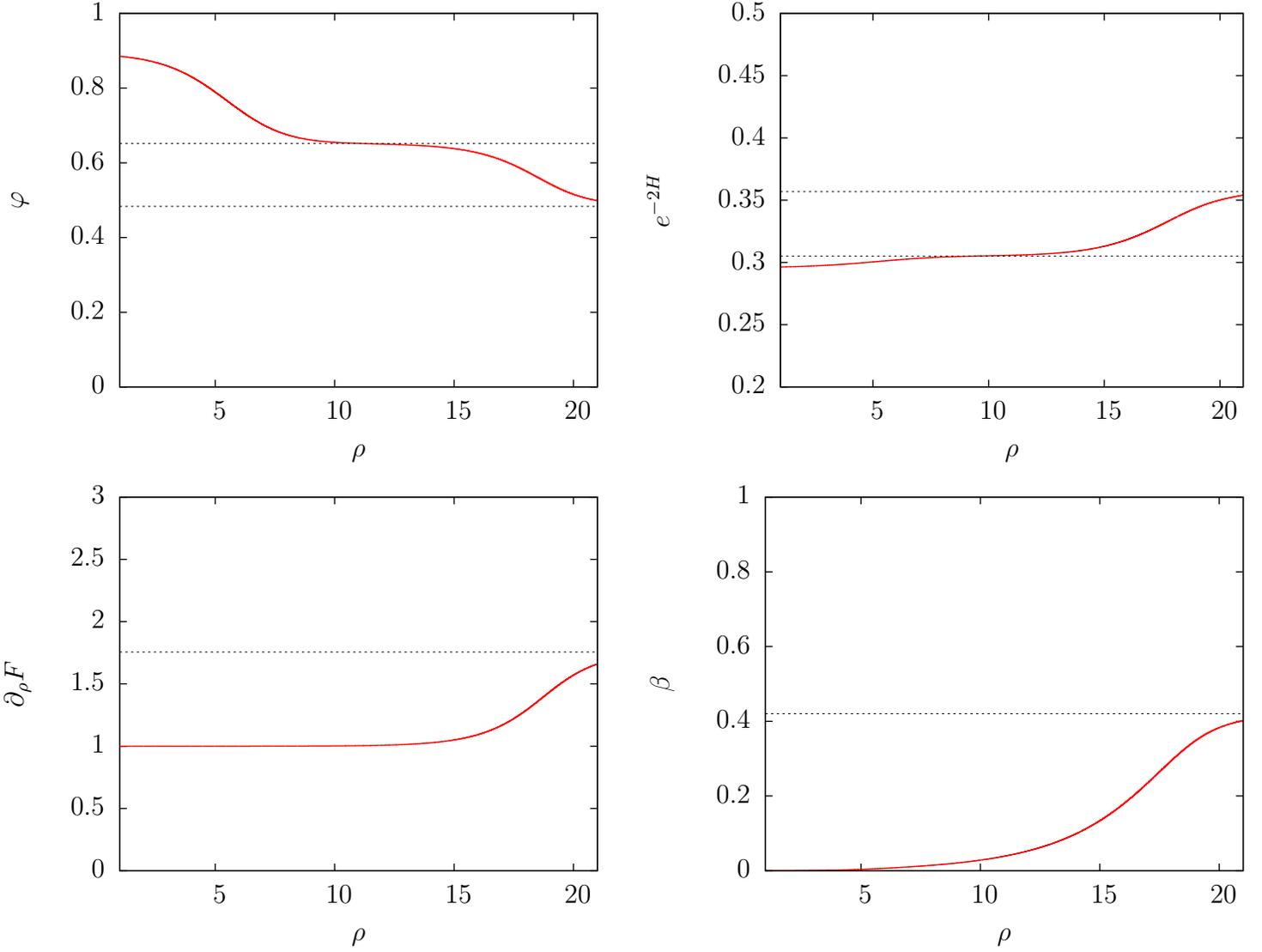

\FloatBarrier

\end{document}

%% file: MVM_vacua.tex
\begingroup
  \makeatletter
  \providecommand\color[2][]{%
    \GenericError{(gnuplot) \space\space\space\@spaces}{%
      Package color not loaded in conjunction with
      terminal option `colourtext'%
    }{See the gnuplot documentation for explanation.%
    }{Either use 'blacktext' in gnuplot or load the package
      color.sty in LaTeX.}%
    \renewcommand\color[2][]{}%
  }%
  \providecommand\includegraphics[2][]{%
    \GenericError{(gnuplot) \space\space\space\@spaces}{%
      Package graphicx or graphics not loaded%
    }{See the gnuplot documentation for explanation.%
    }{The gnuplot epslatex terminal needs graphicx.sty or graphics.sty.}%
    \renewcommand\includegraphics[2][]{}%
  }%
  \providecommand\rotatebox[2]{#2}%
  \@ifundefined{ifGPcolor}{%
    \newif\ifGPcolor
    \GPcolortrue
  }{}%
  \@ifundefined{ifGPblacktext}{%
    \newif\ifGPblacktext
    \GPblacktexttrue
  }{}%
  \let\gplgaddtomacro\g@addto@macro
  \gdef\gplbacktext{}%
  \gdef\gplfronttext{}%
  \makeatother
  \ifGPblacktext
    \def\colorrgb#1{}%
    \def\colorgray#1{}%
  \else
    \ifGPcolor
      \def\colorrgb#1{\color[rgb]{#1}}%
      \def\colorgray#1{\color[gray]{#1}}%
      \expandafter\def\csname LTw\endcsname{\color{white}}%
      \expandafter\def\csname LTb\endcsname{\color{black}}%
      \expandafter\def\csname LTa\endcsname{\color{black}}%
      \expandafter\def\csname LT0\endcsname{\color[rgb]{1,0,0}}%
      \expandafter\def\csname LT1\endcsname{\color[rgb]{0,1,0}}%
      \expandafter\def\csname LT2\endcsname{\color[rgb]{0,0,1}}%
      \expandafter\def\csname LT3\endcsname{\color[rgb]{1,0,1}}%
      \expandafter\def\csname LT4\endcsname{\color[rgb]{0,1,1}}%
      \expandafter\def\csname LT5\endcsname{\color[rgb]{1,1,0}}%
      \expandafter\def\csname LT6\endcsname{\color[rgb]{0,0,0}}%
      \expandafter\def\csname LT7\endcsname{\color[rgb]{1,0.3,0}}%
      \expandafter\def\csname LT8\endcsname{\color[rgb]{0.5,0.5,0.5}}%
    \else
      \def\colorrgb#1{\color{black}}%
      \def\colorgray#1{\color[gray]{#1}}%
      \expandafter\def\csname LTw\endcsname{\color{white}}%
      \expandafter\def\csname LTb\endcsname{\color{black}}%
      \expandafter\def\csname LTa\endcsname{\color{black}}%
      \expandafter\def\csname LT0\endcsname{\color{black}}%
      \expandafter\def\csname LT1\endcsname{\color{black}}%
      \expandafter\def\csname LT2\endcsname{\color{black}}%
      \expandafter\def\csname LT3\endcsname{\color{black}}%
      \expandafter\def\csname LT4\endcsname{\color{black}}%
      \expandafter\def\csname LT5\endcsname{\color{black}}%
      \expandafter\def\csname LT6\endcsname{\color{black}}%
      \expandafter\def\csname LT7\endcsname{\color{black}}%
      \expandafter\def\csname LT8\endcsname{\color{black}}%
    \fi
  \fi
  \setlength{\unitlength}{0.0500bp}%
  \begin{picture}(7652.00,4250.00)%
    \gplgaddtomacro\gplbacktext{%
      \csname LTb\endcsname%
      \put(1386,704){\makebox(0,0)[r]{\strut{}0}}%
      \put(1386,1032){\makebox(0,0)[r]{\strut{}1}}%
      \put(1386,1360){\makebox(0,0)[r]{\strut{}$(d-1)$}}%
      \put(1386,2016){\makebox(0,0)[r]{\strut{}$(d-1)^2$}}%
      \put(1386,3985){\makebox(0,0)[r]{\strut{}10}}%
      \put(1518,484){\makebox(0,0){\strut{}-2.5}}%
      \put(5115,484){\makebox(0,0){\strut{}$-\frac{d}{2}$}}%
      \put(6015,484){\makebox(0,0){\strut{}$-\frac{(3d-4)}{2(d-1)}$}}%
      \put(6914,484){\makebox(0,0){\strut{}-1}}%
      \put(4216,154){\makebox(0,0){\strut{}$\Lambda/m_0^2$}}%
      \put(1140,2837){\makebox(0,0)[l]{\strut{}$z$}}%
    }%
    \gplgaddtomacro\gplfronttext{%
      \csname LTb\endcsname%
      \put(5927,3812){\makebox(0,0)[r]{\strut{}With irrelevant directions}}%
      \csname LTb\endcsname%
      \put(5927,3592){\makebox(0,0)[r]{\strut{}Without irrelevant directions}}%
    }%
    \gplbacktext
    \put(0,0){\includegraphics{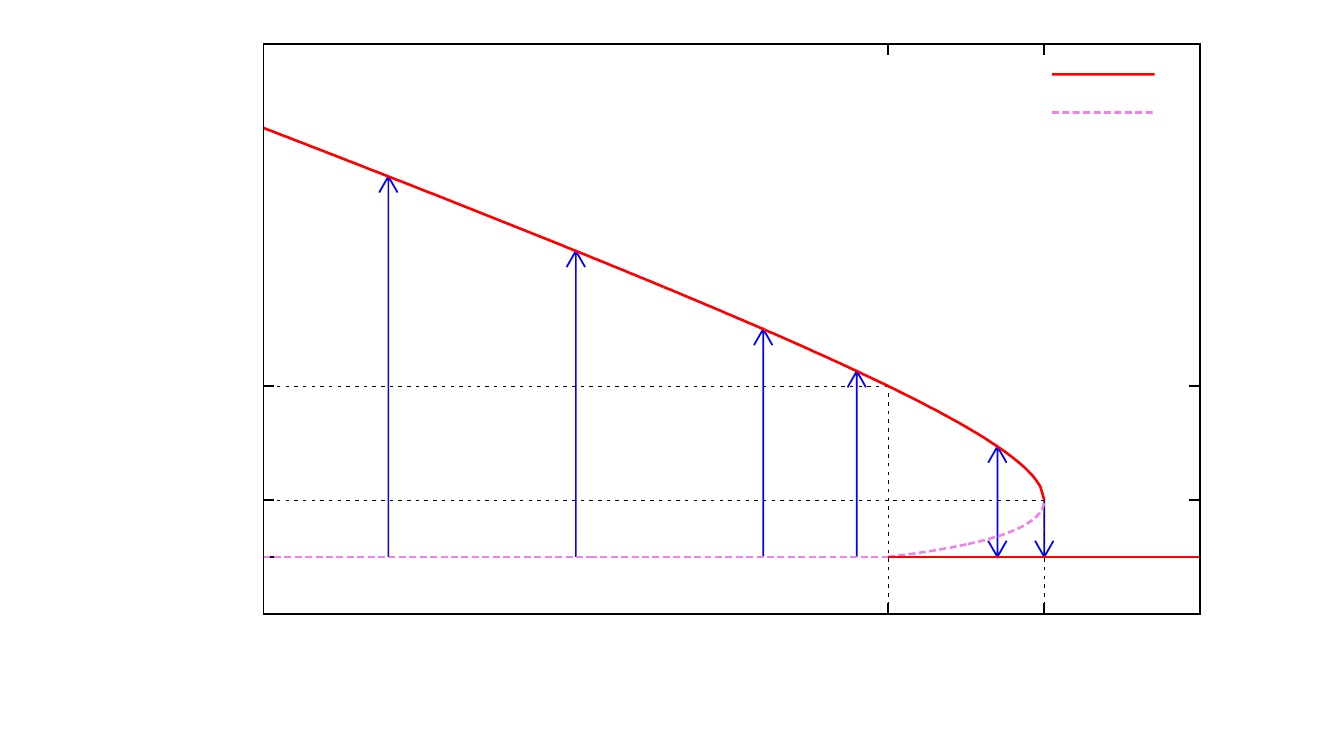}}%
    \gplfronttext
  \end{picture}%
\endgroup

%% file: MVM_Li_AdS_3.tex
\begingroup
  \makeatletter
  \providecommand\color[2][]{%
    \GenericError{(gnuplot) \space\space\space\@spaces}{%
      Package color not loaded in conjunction with
      terminal option `colourtext'%
    }{See the gnuplot documentation for explanation.%
    }{Either use 'blacktext' in gnuplot or load the package
      color.sty in LaTeX.}%
    \renewcommand\color[2][]{}%
  }%
  \providecommand\includegraphics[2][]{%
    \GenericError{(gnuplot) \space\space\space\@spaces}{%
      Package graphicx or graphics not loaded%
    }{See the gnuplot documentation for explanation.%
    }{The gnuplot epslatex terminal needs graphicx.sty or graphics.sty.}%
    \renewcommand\includegraphics[2][]{}%
  }%
  \providecommand\rotatebox[2]{#2}%
  \@ifundefined{ifGPcolor}{%
    \newif\ifGPcolor
    \GPcolortrue
  }{}%
  \@ifundefined{ifGPblacktext}{%
    \newif\ifGPblacktext
    \GPblacktexttrue
  }{}%
  \let\gplgaddtomacro\g@addto@macro
  \gdef\gplbacktext{}%
  \gdef\gplfronttext{}%
  \makeatother
  \ifGPblacktext
    \def\colorrgb#1{}%
    \def\colorgray#1{}%
  \else
    \ifGPcolor
      \def\colorrgb#1{\color[rgb]{#1}}%
      \def\colorgray#1{\color[gray]{#1}}%
      \expandafter\def\csname LTw\endcsname{\color{white}}%
      \expandafter\def\csname LTb\endcsname{\color{black}}%
      \expandafter\def\csname LTa\endcsname{\color{black}}%
      \expandafter\def\csname LT0\endcsname{\color[rgb]{1,0,0}}%
      \expandafter\def\csname LT1\endcsname{\color[rgb]{0,1,0}}%
      \expandafter\def\csname LT2\endcsname{\color[rgb]{0,0,1}}%
      \expandafter\def\csname LT3\endcsname{\color[rgb]{1,0,1}}%
      \expandafter\def\csname LT4\endcsname{\color[rgb]{0,1,1}}%
      \expandafter\def\csname LT5\endcsname{\color[rgb]{1,1,0}}%
      \expandafter\def\csname LT6\endcsname{\color[rgb]{0,0,0}}%
      \expandafter\def\csname LT7\endcsname{\color[rgb]{1,0.3,0}}%
      \expandafter\def\csname LT8\endcsname{\color[rgb]{0.5,0.5,0.5}}%
    \else
      \def\colorrgb#1{\color{black}}%
      \def\colorgray#1{\color[gray]{#1}}%
      \expandafter\def\csname LTw\endcsname{\color{white}}%
      \expandafter\def\csname LTb\endcsname{\color{black}}%
      \expandafter\def\csname LTa\endcsname{\color{black}}%
      \expandafter\def\csname LT0\endcsname{\color{black}}%
      \expandafter\def\csname LT1\endcsname{\color{black}}%
      \expandafter\def\csname LT2\endcsname{\color{black}}%
      \expandafter\def\csname LT3\endcsname{\color{black}}%
      \expandafter\def\csname LT4\endcsname{\color{black}}%
      \expandafter\def\csname LT5\endcsname{\color{black}}%
      \expandafter\def\csname LT6\endcsname{\color{black}}%
      \expandafter\def\csname LT7\endcsname{\color{black}}%
      \expandafter\def\csname LT8\endcsname{\color{black}}%
    \fi
  \fi
  \setlength{\unitlength}{0.0500bp}%
  \begin{picture}(11338.00,3968.00)%
    \gplgaddtomacro\gplbacktext{%
      \csname LTb\endcsname%
      \put(946,704){\makebox(0,0)[r]{\strut{} 1}}%
      \put(946,1304){\makebox(0,0)[r]{\strut{} 1.2}}%
      \put(946,1904){\makebox(0,0)[r]{\strut{} 1.4}}%
      \put(946,2503){\makebox(0,0)[r]{\strut{} 1.6}}%
      \put(946,3103){\makebox(0,0)[r]{\strut{} 1.8}}%
      \put(946,3703){\makebox(0,0)[r]{\strut{} 2}}%
      \put(1613,484){\makebox(0,0){\strut{} 20}}%
      \put(2176,484){\makebox(0,0){\strut{} 40}}%
      \put(2739,484){\makebox(0,0){\strut{} 60}}%
      \put(3302,484){\makebox(0,0){\strut{} 80}}%
      \put(3865,484){\makebox(0,0){\strut{} 100}}%
      \put(4428,484){\makebox(0,0){\strut{} 120}}%
      \put(4991,484){\makebox(0,0){\strut{} 140}}%
      \put(176,2203){\rotatebox{-270}{\makebox(0,0){\strut{}$\partial_{\rho} F$}}}%
      \put(3175,154){\makebox(0,0){\strut{}$\rho$}}%
    }%
    \gplgaddtomacro\gplfronttext{%
    }%
    \gplgaddtomacro\gplbacktext{%
      \csname LTb\endcsname%
      \put(6615,704){\makebox(0,0)[r]{\strut{} 0}}%
      \put(6615,2204){\makebox(0,0)[r]{\strut{} 1.5}}%
      \put(6615,3703){\makebox(0,0)[r]{\strut{} 3}}%
      \put(7282,484){\makebox(0,0){\strut{} 20}}%
      \put(7845,484){\makebox(0,0){\strut{} 40}}%
      \put(8408,484){\makebox(0,0){\strut{} 60}}%
      \put(8971,484){\makebox(0,0){\strut{} 80}}%
      \put(9534,484){\makebox(0,0){\strut{} 100}}%
      \put(10097,484){\makebox(0,0){\strut{} 120}}%
      \put(10660,484){\makebox(0,0){\strut{} 140}}%
      \put(5845,2203){\rotatebox{-270}{\makebox(0,0){\strut{}$\alpha$}}}%
      \put(8844,154){\makebox(0,0){\strut{}$\rho$}}%
    }%
    \gplgaddtomacro\gplfronttext{%
    }%
    \gplbacktext
    \put(0,0){\includegraphics{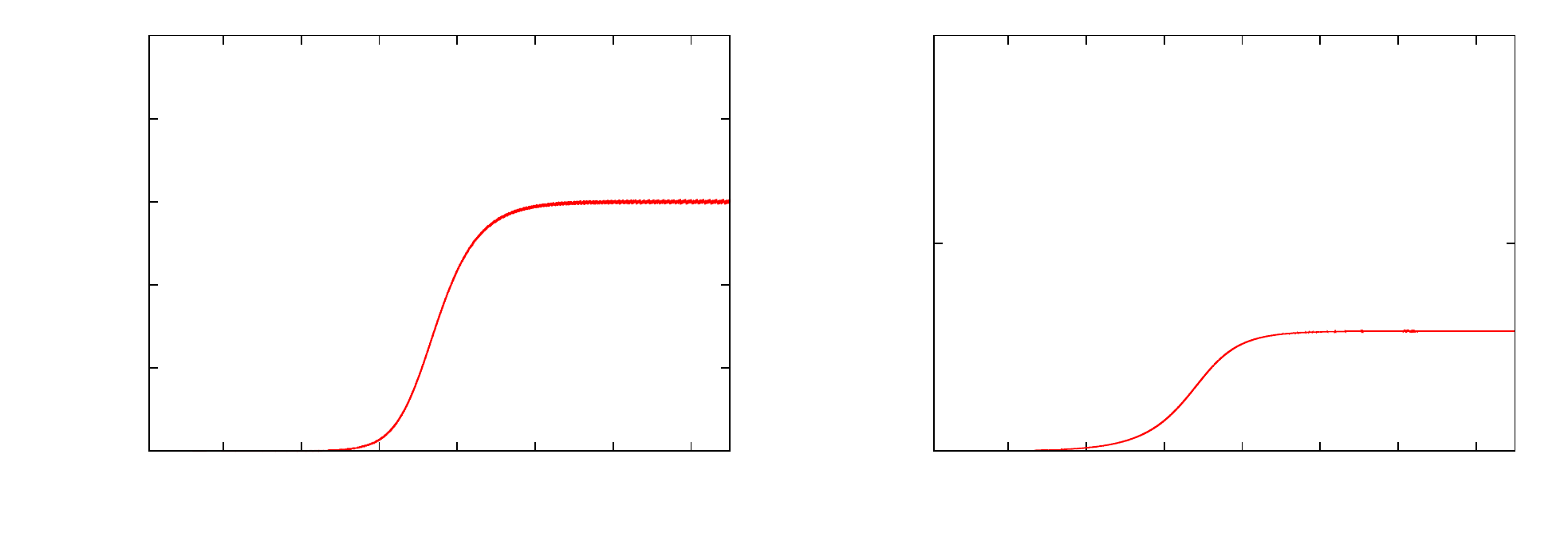}}%
    \gplfronttext
  \end{picture}%
\endgroup

%% file: MVM_Li_Li_1.tex
\begingroup
  \makeatletter
  \providecommand\color[2][]{%
    \GenericError{(gnuplot) \space\space\space\@spaces}{%
      Package color not loaded in conjunction with
      terminal option `colourtext'%
    }{See the gnuplot documentation for explanation.%
    }{Either use 'blacktext' in gnuplot or load the package
      color.sty in LaTeX.}%
    \renewcommand\color[2][]{}%
  }%
  \providecommand\includegraphics[2][]{%
    \GenericError{(gnuplot) \space\space\space\@spaces}{%
      Package graphicx or graphics not loaded%
    }{See the gnuplot documentation for explanation.%
    }{The gnuplot epslatex terminal needs graphicx.sty or graphics.sty.}%
    \renewcommand\includegraphics[2][]{}%
  }%
  \providecommand\rotatebox[2]{#2}%
  \@ifundefined{ifGPcolor}{%
    \newif\ifGPcolor
    \GPcolortrue
  }{}%
  \@ifundefined{ifGPblacktext}{%
    \newif\ifGPblacktext
    \GPblacktexttrue
  }{}%
  \let\gplgaddtomacro\g@addto@macro
  \gdef\gplbacktext{}%
  \gdef\gplfronttext{}%
  \makeatother
  \ifGPblacktext
    \def\colorrgb#1{}%
    \def\colorgray#1{}%
  \else
    \ifGPcolor
      \def\colorrgb#1{\color[rgb]{#1}}%
      \def\colorgray#1{\color[gray]{#1}}%
      \expandafter\def\csname LTw\endcsname{\color{white}}%
      \expandafter\def\csname LTb\endcsname{\color{black}}%
      \expandafter\def\csname LTa\endcsname{\color{black}}%
      \expandafter\def\csname LT0\endcsname{\color[rgb]{1,0,0}}%
      \expandafter\def\csname LT1\endcsname{\color[rgb]{0,1,0}}%
      \expandafter\def\csname LT2\endcsname{\color[rgb]{0,0,1}}%
      \expandafter\def\csname LT3\endcsname{\color[rgb]{1,0,1}}%
      \expandafter\def\csname LT4\endcsname{\color[rgb]{0,1,1}}%
      \expandafter\def\csname LT5\endcsname{\color[rgb]{1,1,0}}%
      \expandafter\def\csname LT6\endcsname{\color[rgb]{0,0,0}}%
      \expandafter\def\csname LT7\endcsname{\color[rgb]{1,0.3,0}}%
      \expandafter\def\csname LT8\endcsname{\color[rgb]{0.5,0.5,0.5}}%
    \else
      \def\colorrgb#1{\color{black}}%
      \def\colorgray#1{\color[gray]{#1}}%
      \expandafter\def\csname LTw\endcsname{\color{white}}%
      \expandafter\def\csname LTb\endcsname{\color{black}}%
      \expandafter\def\csname LTa\endcsname{\color{black}}%
      \expandafter\def\csname LT0\endcsname{\color{black}}%
      \expandafter\def\csname LT1\endcsname{\color{black}}%
      \expandafter\def\csname LT2\endcsname{\color{black}}%
      \expandafter\def\csname LT3\endcsname{\color{black}}%
      \expandafter\def\csname LT4\endcsname{\color{black}}%
      \expandafter\def\csname LT5\endcsname{\color{black}}%
      \expandafter\def\csname LT6\endcsname{\color{black}}%
      \expandafter\def\csname LT7\endcsname{\color{black}}%
      \expandafter\def\csname LT8\endcsname{\color{black}}%
    \fi
  \fi
  \setlength{\unitlength}{0.0500bp}%
  \begin{picture}(11338.00,3968.00)%
    \gplgaddtomacro\gplbacktext{%
      \csname LTb\endcsname%
      \put(946,704){\makebox(0,0)[r]{\strut{} 1}}%
      \put(946,1204){\makebox(0,0)[r]{\strut{} 1.5}}%
      \put(946,1704){\makebox(0,0)[r]{\strut{} 2}}%
      \put(946,2204){\makebox(0,0)[r]{\strut{} 2.5}}%
      \put(946,2703){\makebox(0,0)[r]{\strut{} 3}}%
      \put(946,3203){\makebox(0,0)[r]{\strut{} 3.5}}%
      \put(946,3703){\makebox(0,0)[r]{\strut{} 4}}%
      \put(1508,484){\makebox(0,0){\strut{} 5}}%
      \put(2046,484){\makebox(0,0){\strut{} 10}}%
      \put(2584,484){\makebox(0,0){\strut{} 15}}%
      \put(3121,484){\makebox(0,0){\strut{} 20}}%
      \put(3659,484){\makebox(0,0){\strut{} 25}}%
      \put(4197,484){\makebox(0,0){\strut{} 30}}%
      \put(4734,484){\makebox(0,0){\strut{} 35}}%
      \put(5272,484){\makebox(0,0){\strut{} 40}}%
      \put(176,2203){\rotatebox{-270}{\makebox(0,0){\strut{}$\partial_{\rho} F$}}}%
      \put(3175,154){\makebox(0,0){\strut{}$\rho$}}%
    }%
    \gplgaddtomacro\gplfronttext{%
    }%
    \gplgaddtomacro\gplbacktext{%
      \csname LTb\endcsname%
      \put(6615,704){\makebox(0,0)[r]{\strut{} 0}}%
      \put(6615,2204){\makebox(0,0)[r]{\strut{} 1.5}}%
      \put(6615,3703){\makebox(0,0)[r]{\strut{} 3}}%
      \put(7177,484){\makebox(0,0){\strut{} 5}}%
      \put(7715,484){\makebox(0,0){\strut{} 10}}%
      \put(8253,484){\makebox(0,0){\strut{} 15}}%
      \put(8790,484){\makebox(0,0){\strut{} 20}}%
      \put(9328,484){\makebox(0,0){\strut{} 25}}%
      \put(9866,484){\makebox(0,0){\strut{} 30}}%
      \put(10403,484){\makebox(0,0){\strut{} 35}}%
      \put(10941,484){\makebox(0,0){\strut{} 40}}%
      \put(5845,2203){\rotatebox{-270}{\makebox(0,0){\strut{}$\alpha$}}}%
      \put(8844,154){\makebox(0,0){\strut{}$\rho$}}%
    }%
    \gplgaddtomacro\gplfronttext{%
    }%
    \gplbacktext
    \put(0,0){\includegraphics{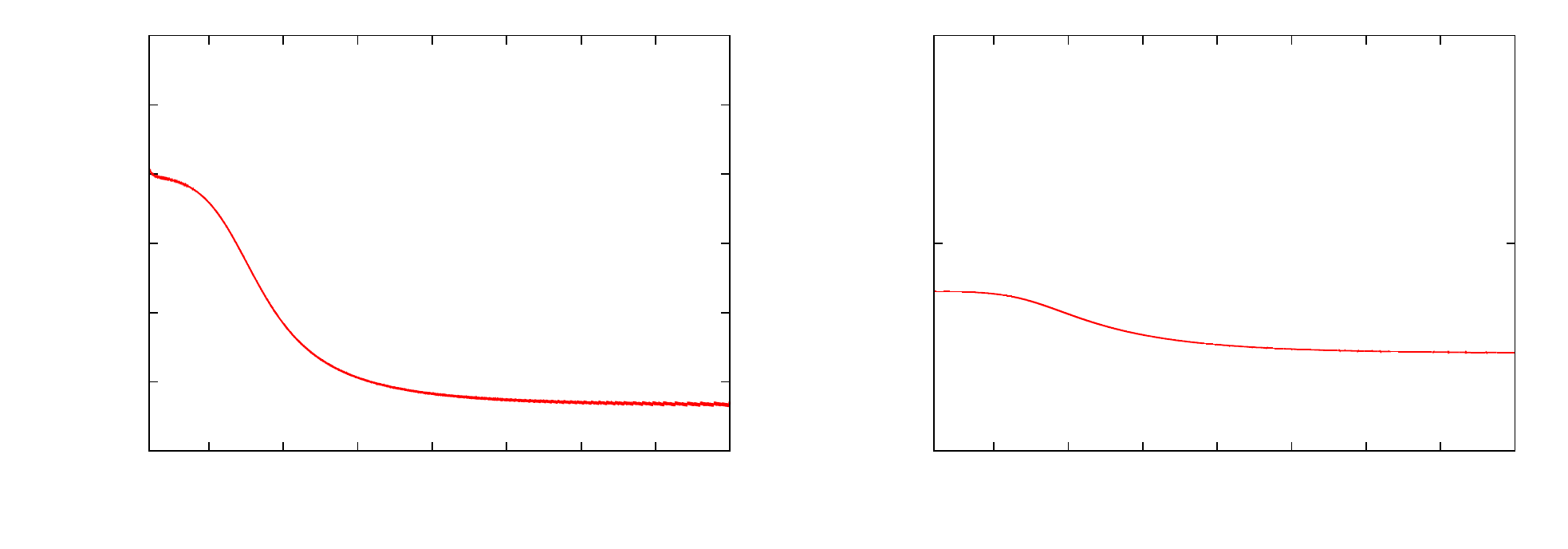}}%
    \gplfronttext
  \end{picture}%
\endgroup

%% file: MVM_AdS_Li_1.tex
\begingroup
  \makeatletter
  \providecommand\color[2][]{%
    \GenericError{(gnuplot) \space\space\space\@spaces}{%
      Package color not loaded in conjunction with
      terminal option `colourtext'%
    }{See the gnuplot documentation for explanation.%
    }{Either use 'blacktext' in gnuplot or load the package
      color.sty in LaTeX.}%
    \renewcommand\color[2][]{}%
  }%
  \providecommand\includegraphics[2][]{%
    \GenericError{(gnuplot) \space\space\space\@spaces}{%
      Package graphicx or graphics not loaded%
    }{See the gnuplot documentation for explanation.%
    }{The gnuplot epslatex terminal needs graphicx.sty or graphics.sty.}%
    \renewcommand\includegraphics[2][]{}%
  }%
  \providecommand\rotatebox[2]{#2}%
  \@ifundefined{ifGPcolor}{%
    \newif\ifGPcolor
    \GPcolortrue
  }{}%
  \@ifundefined{ifGPblacktext}{%
    \newif\ifGPblacktext
    \GPblacktexttrue
  }{}%
  \let\gplgaddtomacro\g@addto@macro
  \gdef\gplbacktext{}%
  \gdef\gplfronttext{}%
  \makeatother
  \ifGPblacktext
    \def\colorrgb#1{}%
    \def\colorgray#1{}%
  \else
    \ifGPcolor
      \def\colorrgb#1{\color[rgb]{#1}}%
      \def\colorgray#1{\color[gray]{#1}}%
      \expandafter\def\csname LTw\endcsname{\color{white}}%
      \expandafter\def\csname LTb\endcsname{\color{black}}%
      \expandafter\def\csname LTa\endcsname{\color{black}}%
      \expandafter\def\csname LT0\endcsname{\color[rgb]{1,0,0}}%
      \expandafter\def\csname LT1\endcsname{\color[rgb]{0,1,0}}%
      \expandafter\def\csname LT2\endcsname{\color[rgb]{0,0,1}}%
      \expandafter\def\csname LT3\endcsname{\color[rgb]{1,0,1}}%
      \expandafter\def\csname LT4\endcsname{\color[rgb]{0,1,1}}%
      \expandafter\def\csname LT5\endcsname{\color[rgb]{1,1,0}}%
      \expandafter\def\csname LT6\endcsname{\color[rgb]{0,0,0}}%
      \expandafter\def\csname LT7\endcsname{\color[rgb]{1,0.3,0}}%
      \expandafter\def\csname LT8\endcsname{\color[rgb]{0.5,0.5,0.5}}%
    \else
      \def\colorrgb#1{\color{black}}%
      \def\colorgray#1{\color[gray]{#1}}%
      \expandafter\def\csname LTw\endcsname{\color{white}}%
      \expandafter\def\csname LTb\endcsname{\color{black}}%
      \expandafter\def\csname LTa\endcsname{\color{black}}%
      \expandafter\def\csname LT0\endcsname{\color{black}}%
      \expandafter\def\csname LT1\endcsname{\color{black}}%
      \expandafter\def\csname LT2\endcsname{\color{black}}%
      \expandafter\def\csname LT3\endcsname{\color{black}}%
      \expandafter\def\csname LT4\endcsname{\color{black}}%
      \expandafter\def\csname LT5\endcsname{\color{black}}%
      \expandafter\def\csname LT6\endcsname{\color{black}}%
      \expandafter\def\csname LT7\endcsname{\color{black}}%
      \expandafter\def\csname LT8\endcsname{\color{black}}%
    \fi
  \fi
  \setlength{\unitlength}{0.0500bp}%
  \begin{picture}(11338.00,3968.00)%
    \gplgaddtomacro\gplbacktext{%
      \csname LTb\endcsname%
      \put(682,704){\makebox(0,0)[r]{\strut{} 1}}%
      \put(682,1132){\makebox(0,0)[r]{\strut{} 2}}%
      \put(682,1561){\makebox(0,0)[r]{\strut{} 3}}%
      \put(682,1989){\makebox(0,0)[r]{\strut{} 4}}%
      \put(682,2418){\makebox(0,0)[r]{\strut{} 5}}%
      \put(682,2846){\makebox(0,0)[r]{\strut{} 6}}%
      \put(682,3275){\makebox(0,0)[r]{\strut{} 7}}%
      \put(682,3703){\makebox(0,0)[r]{\strut{} 8}}%
      \put(1429,484){\makebox(0,0){\strut{} 5}}%
      \put(2198,484){\makebox(0,0){\strut{} 10}}%
      \put(2966,484){\makebox(0,0){\strut{} 15}}%
      \put(3735,484){\makebox(0,0){\strut{} 20}}%
      \put(4503,484){\makebox(0,0){\strut{} 25}}%
      \put(5272,484){\makebox(0,0){\strut{} 30}}%
      \put(176,2203){\rotatebox{-270}{\makebox(0,0){\strut{}$\partial_{\rho} F$}}}%
      \put(3043,154){\makebox(0,0){\strut{}$\rho$}}%
    }%
    \gplgaddtomacro\gplfronttext{%
    }%
    \gplgaddtomacro\gplbacktext{%
      \csname LTb\endcsname%
      \put(6615,704){\makebox(0,0)[r]{\strut{} 0}}%
      \put(6615,2204){\makebox(0,0)[r]{\strut{} 1.5}}%
      \put(6615,3703){\makebox(0,0)[r]{\strut{} 3}}%
      \put(7325,484){\makebox(0,0){\strut{} 5}}%
      \put(8049,484){\makebox(0,0){\strut{} 10}}%
      \put(8772,484){\makebox(0,0){\strut{} 15}}%
      \put(9495,484){\makebox(0,0){\strut{} 20}}%
      \put(10218,484){\makebox(0,0){\strut{} 25}}%
      \put(10941,484){\makebox(0,0){\strut{} 30}}%
      \put(5845,2203){\rotatebox{-270}{\makebox(0,0){\strut{}$\alpha$}}}%
      \put(8844,154){\makebox(0,0){\strut{}$\rho$}}%
    }%
    \gplgaddtomacro\gplfronttext{%
    }%
    \gplbacktext
    \put(0,0){\includegraphics{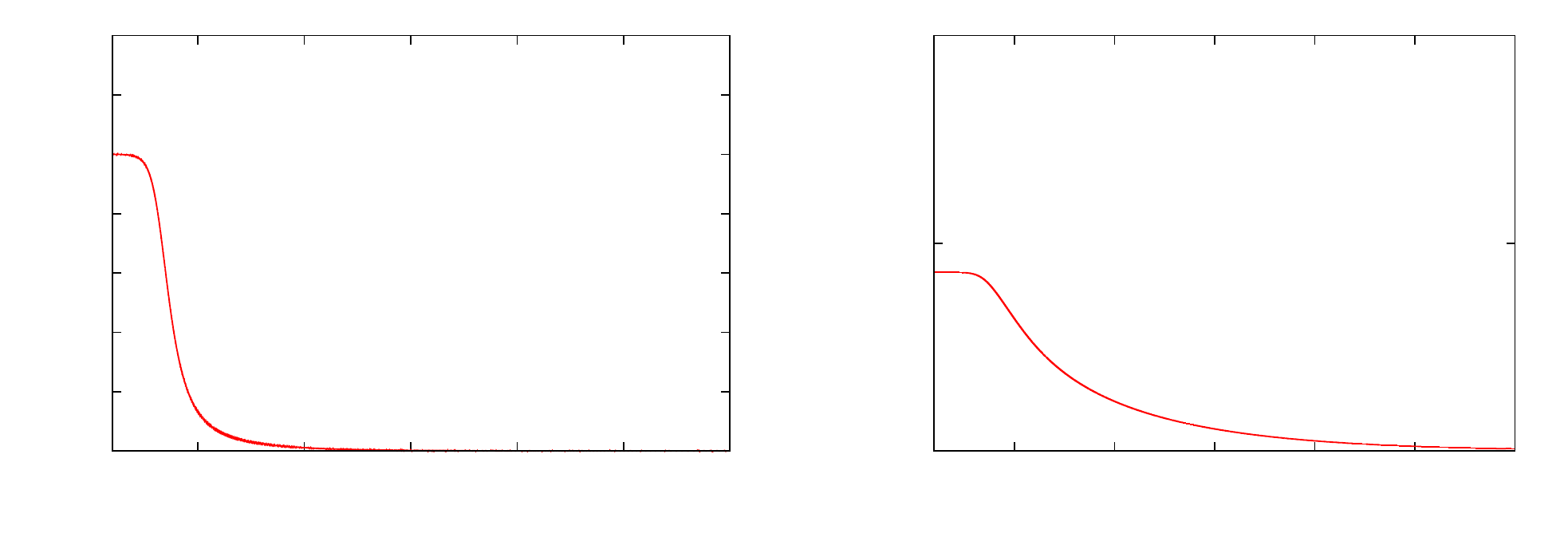}}%
    \gplfronttext
  \end{picture}%
\endgroup

%% file: AdS_evals.tex
\begingroup
  \makeatletter
  \providecommand\color[2][]{%
    \GenericError{(gnuplot) \space\space\space\@spaces}{%
      Package color not loaded in conjunction with
      terminal option `colourtext'%
    }{See the gnuplot documentation for explanation.%
    }{Either use 'blacktext' in gnuplot or load the package
      color.sty in LaTeX.}%
    \renewcommand\color[2][]{}%
  }%
  \providecommand\includegraphics[2][]{%
    \GenericError{(gnuplot) \space\space\space\@spaces}{%
      Package graphicx or graphics not loaded%
    }{See the gnuplot documentation for explanation.%
    }{The gnuplot epslatex terminal needs graphicx.sty or graphics.sty.}%
    \renewcommand\includegraphics[2][]{}%
  }%
  \providecommand\rotatebox[2]{#2}%
  \@ifundefined{ifGPcolor}{%
    \newif\ifGPcolor
    \GPcolortrue
  }{}%
  \@ifundefined{ifGPblacktext}{%
    \newif\ifGPblacktext
    \GPblacktexttrue
  }{}%
  \let\gplgaddtomacro\g@addto@macro
  \gdef\gplbacktext{}%
  \gdef\gplfronttext{}%
  \makeatother
  \ifGPblacktext
    \def\colorrgb#1{}%
    \def\colorgray#1{}%
  \else
    \ifGPcolor
      \def\colorrgb#1{\color[rgb]{#1}}%
      \def\colorgray#1{\color[gray]{#1}}%
      \expandafter\def\csname LTw\endcsname{\color{white}}%
      \expandafter\def\csname LTb\endcsname{\color{black}}%
      \expandafter\def\csname LTa\endcsname{\color{black}}%
      \expandafter\def\csname LT0\endcsname{\color[rgb]{1,0,0}}%
      \expandafter\def\csname LT1\endcsname{\color[rgb]{0,1,0}}%
      \expandafter\def\csname LT2\endcsname{\color[rgb]{0,0,1}}%
      \expandafter\def\csname LT3\endcsname{\color[rgb]{1,0,1}}%
      \expandafter\def\csname LT4\endcsname{\color[rgb]{0,1,1}}%
      \expandafter\def\csname LT5\endcsname{\color[rgb]{1,1,0}}%
      \expandafter\def\csname LT6\endcsname{\color[rgb]{0,0,0}}%
      \expandafter\def\csname LT7\endcsname{\color[rgb]{1,0.3,0}}%
      \expandafter\def\csname LT8\endcsname{\color[rgb]{0.5,0.5,0.5}}%
    \else
      \def\colorrgb#1{\color{black}}%
      \def\colorgray#1{\color[gray]{#1}}%
      \expandafter\def\csname LTw\endcsname{\color{white}}%
      \expandafter\def\csname LTb\endcsname{\color{black}}%
      \expandafter\def\csname LTa\endcsname{\color{black}}%
      \expandafter\def\csname LT0\endcsname{\color{black}}%
      \expandafter\def\csname LT1\endcsname{\color{black}}%
      \expandafter\def\csname LT2\endcsname{\color{black}}%
      \expandafter\def\csname LT3\endcsname{\color{black}}%
      \expandafter\def\csname LT4\endcsname{\color{black}}%
      \expandafter\def\csname LT5\endcsname{\color{black}}%
      \expandafter\def\csname LT6\endcsname{\color{black}}%
      \expandafter\def\csname LT7\endcsname{\color{black}}%
      \expandafter\def\csname LT8\endcsname{\color{black}}%
    \fi
  \fi
  \setlength{\unitlength}{0.0500bp}%
  \begin{picture}(7652.00,4250.00)%
    \gplgaddtomacro\gplbacktext{%
      \csname LTb\endcsname%
      \put(814,704){\makebox(0,0)[r]{\strut{} 0}}%
      \put(814,1688){\makebox(0,0)[r]{\strut{}3/2}}%
      \put(814,2673){\makebox(0,0)[r]{\strut{} 3}}%
      \put(814,3985){\makebox(0,0)[r]{\strut{} 5}}%
      \put(946,484){\makebox(0,0){\strut{}$\frac{1}{3}$}}%
      \put(1270,484){\makebox(0,0){\strut{}$\;\;\;\;\;0.368$}}%
      \put(3392,484){\makebox(0,0){\strut{}$0.592$}}%
      \put(7255,484){\makebox(0,0){\strut{}1}}%
      \put(176,2344){\rotatebox{-270}{\makebox(0,0){\strut{}Operator dimensions (real part)}}}%
      \put(4100,154){\makebox(0,0){\strut{}$\varphi_0^2$}}%
    }%
    \gplgaddtomacro\gplfronttext{%
      \csname LTb\endcsname%
      \put(6268,1537){\makebox(0,0)[r]{\strut{}$\Delta_1$}}%
      \csname LTb\endcsname%
      \put(6268,1317){\makebox(0,0)[r]{\strut{}$\Delta_2$}}%
      \csname LTb\endcsname%
      \put(6268,1097){\makebox(0,0)[r]{\strut{}$\Delta_3$}}%
      \csname LTb\endcsname%
      \put(6268,877){\makebox(0,0)[r]{\strut{}$\Delta_4$}}%
    }%
    \gplbacktext
    \put(0,0){\includegraphics{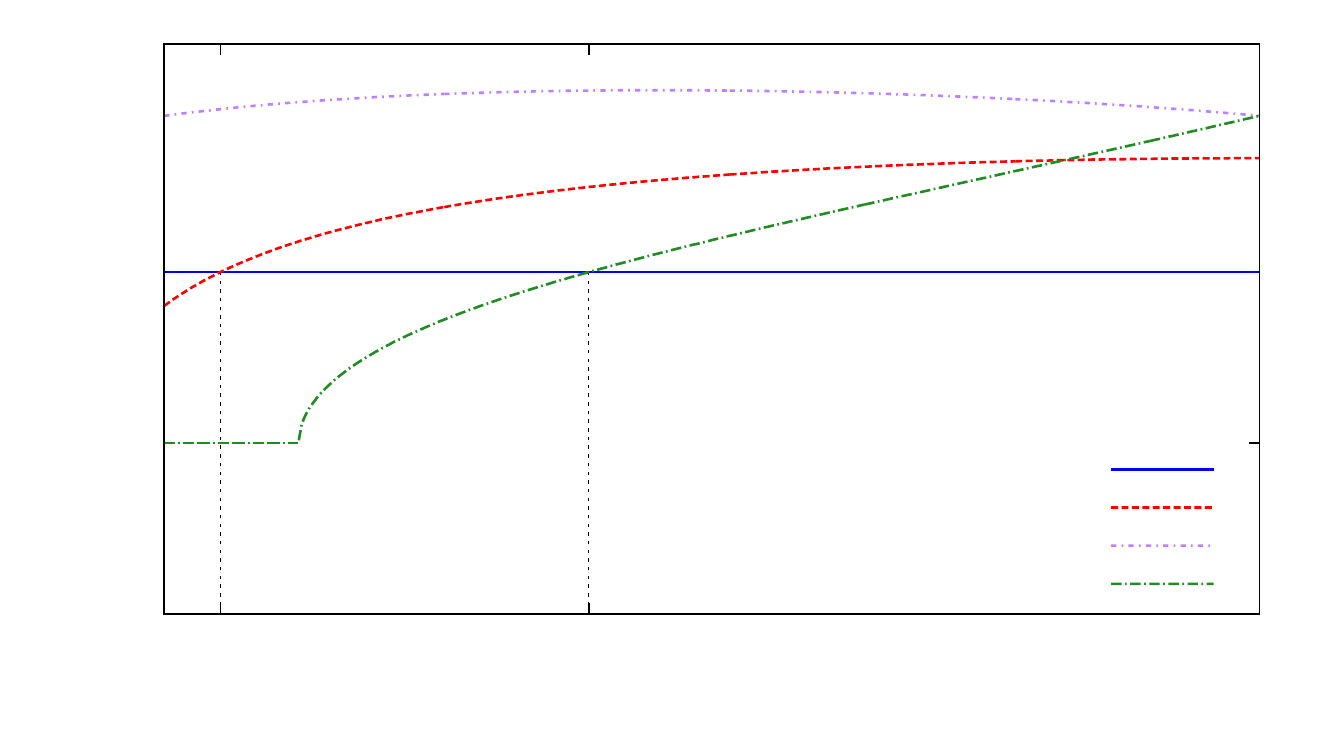}}%
    \gplfronttext
  \end{picture}%
\endgroup

%% file: lower_Li_evals.tex
\begingroup
  \makeatletter
  \providecommand\color[2][]{%
    \GenericError{(gnuplot) \space\space\space\@spaces}{%
      Package color not loaded in conjunction with
      terminal option `colourtext'%
    }{See the gnuplot documentation for explanation.%
    }{Either use 'blacktext' in gnuplot or load the package
      color.sty in LaTeX.}%
    \renewcommand\color[2][]{}%
  }%
  \providecommand\includegraphics[2][]{%
    \GenericError{(gnuplot) \space\space\space\@spaces}{%
      Package graphicx or graphics not loaded%
    }{See the gnuplot documentation for explanation.%
    }{The gnuplot epslatex terminal needs graphicx.sty or graphics.sty.}%
    \renewcommand\includegraphics[2][]{}%
  }%
  \providecommand\rotatebox[2]{#2}%
  \@ifundefined{ifGPcolor}{%
    \newif\ifGPcolor
    \GPcolortrue
  }{}%
  \@ifundefined{ifGPblacktext}{%
    \newif\ifGPblacktext
    \GPblacktexttrue
  }{}%
  \let\gplgaddtomacro\g@addto@macro
  \gdef\gplbacktext{}%
  \gdef\gplfronttext{}%
  \makeatother
  \ifGPblacktext
    \def\colorrgb#1{}%
    \def\colorgray#1{}%
  \else
    \ifGPcolor
      \def\colorrgb#1{\color[rgb]{#1}}%
      \def\colorgray#1{\color[gray]{#1}}%
      \expandafter\def\csname LTw\endcsname{\color{white}}%
      \expandafter\def\csname LTb\endcsname{\color{black}}%
      \expandafter\def\csname LTa\endcsname{\color{black}}%
      \expandafter\def\csname LT0\endcsname{\color[rgb]{1,0,0}}%
      \expandafter\def\csname LT1\endcsname{\color[rgb]{0,1,0}}%
      \expandafter\def\csname LT2\endcsname{\color[rgb]{0,0,1}}%
      \expandafter\def\csname LT3\endcsname{\color[rgb]{1,0,1}}%
      \expandafter\def\csname LT4\endcsname{\color[rgb]{0,1,1}}%
      \expandafter\def\csname LT5\endcsname{\color[rgb]{1,1,0}}%
      \expandafter\def\csname LT6\endcsname{\color[rgb]{0,0,0}}%
      \expandafter\def\csname LT7\endcsname{\color[rgb]{1,0.3,0}}%
      \expandafter\def\csname LT8\endcsname{\color[rgb]{0.5,0.5,0.5}}%
    \else
      \def\colorrgb#1{\color{black}}%
      \def\colorgray#1{\color[gray]{#1}}%
      \expandafter\def\csname LTw\endcsname{\color{white}}%
      \expandafter\def\csname LTb\endcsname{\color{black}}%
      \expandafter\def\csname LTa\endcsname{\color{black}}%
      \expandafter\def\csname LT0\endcsname{\color{black}}%
      \expandafter\def\csname LT1\endcsname{\color{black}}%
      \expandafter\def\csname LT2\endcsname{\color{black}}%
      \expandafter\def\csname LT3\endcsname{\color{black}}%
      \expandafter\def\csname LT4\endcsname{\color{black}}%
      \expandafter\def\csname LT5\endcsname{\color{black}}%
      \expandafter\def\csname LT6\endcsname{\color{black}}%
      \expandafter\def\csname LT7\endcsname{\color{black}}%
      \expandafter\def\csname LT8\endcsname{\color{black}}%
    \fi
  \fi
  \setlength{\unitlength}{0.0500bp}%
  \begin{picture}(7652.00,4250.00)%
    \gplgaddtomacro\gplbacktext{%
      \csname LTb\endcsname%
      \put(1078,704){\makebox(0,0)[r]{\strut{} 0}}%
      \put(1078,1524){\makebox(0,0)[r]{\strut{}$1/2$}}%
      \put(1078,2345){\makebox(0,0)[r]{\strut{} 1}}%
      \put(1078,3985){\makebox(0,0)[r]{\strut{} 2}}%
      \put(1210,484){\makebox(0,0){\strut{} 4.29}}%
      \put(3407,484){\makebox(0,0){\strut{} 10}}%
      \put(7255,484){\makebox(0,0){\strut{} 20}}%
      \put(176,2344){\rotatebox{-270}{\makebox(0,0){\strut{}Operator dimensions (real part) $/z+2$}}}%
      \put(4232,154){\makebox(0,0){\strut{}$z$}}%
    }%
    \gplgaddtomacro\gplfronttext{%
      \csname LTb\endcsname%
      \put(6268,3812){\makebox(0,0)[r]{\strut{}$\Delta_1 / z+2$}}%
      \csname LTb\endcsname%
      \put(6268,3592){\makebox(0,0)[r]{\strut{}$\Delta_2 / z+2$}}%
      \csname LTb\endcsname%
      \put(6268,3372){\makebox(0,0)[r]{\strut{}$\Delta_3 / z+2$}}%
      \csname LTb\endcsname%
      \put(6268,3152){\makebox(0,0)[r]{\strut{}$\Delta_4 / z+2$}}%
    }%
    \gplbacktext
    \put(0,0){\includegraphics{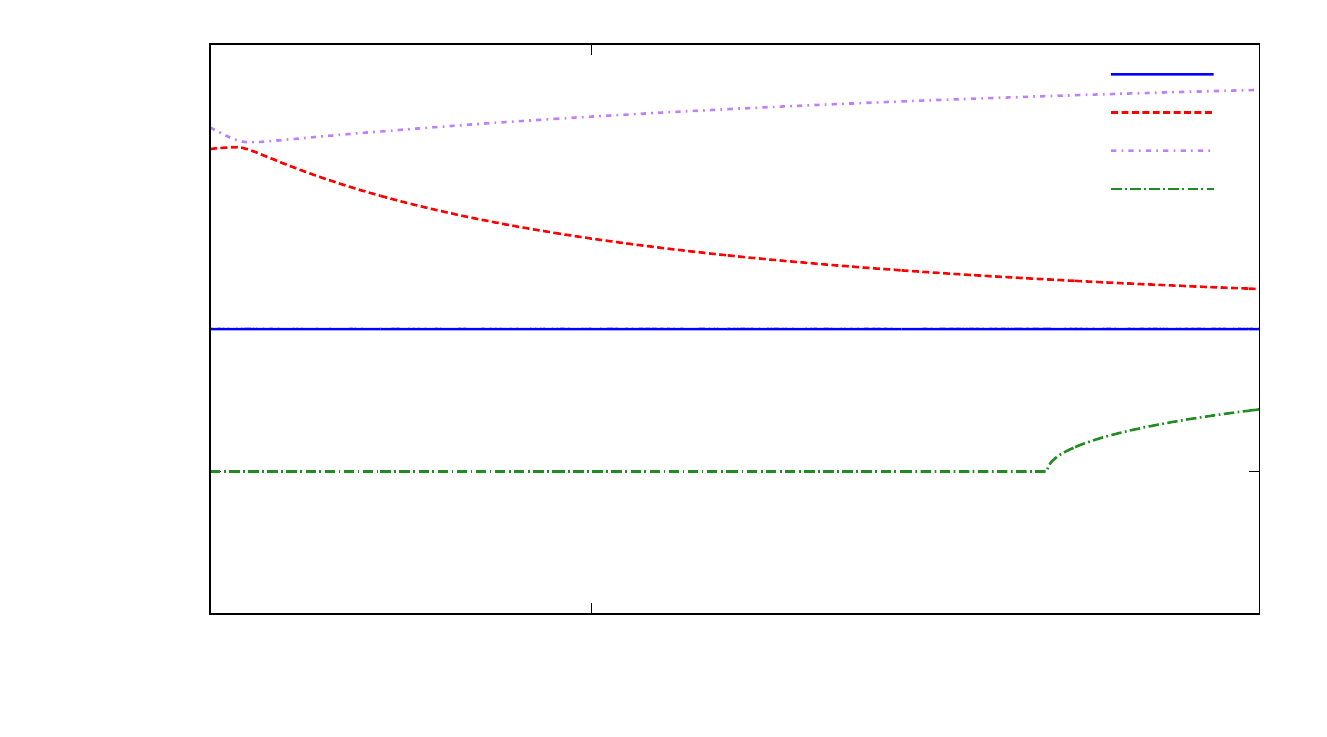}}%
    \gplfronttext
  \end{picture}%
\endgroup

%% file: upper_Li_evals.tex
\begingroup
  \makeatletter
  \providecommand\color[2][]{%
    \GenericError{(gnuplot) \space\space\space\@spaces}{%
      Package color not loaded in conjunction with
      terminal option `colourtext'%
    }{See the gnuplot documentation for explanation.%
    }{Either use 'blacktext' in gnuplot or load the package
      color.sty in LaTeX.}%
    \renewcommand\color[2][]{}%
  }%
  \providecommand\includegraphics[2][]{%
    \GenericError{(gnuplot) \space\space\space\@spaces}{%
      Package graphicx or graphics not loaded%
    }{See the gnuplot documentation for explanation.%
    }{The gnuplot epslatex terminal needs graphicx.sty or graphics.sty.}%
    \renewcommand\includegraphics[2][]{}%
  }%
  \providecommand\rotatebox[2]{#2}%
  \@ifundefined{ifGPcolor}{%
    \newif\ifGPcolor
    \GPcolortrue
  }{}%
  \@ifundefined{ifGPblacktext}{%
    \newif\ifGPblacktext
    \GPblacktexttrue
  }{}%
  \let\gplgaddtomacro\g@addto@macro
  \gdef\gplbacktext{}%
  \gdef\gplfronttext{}%
  \makeatother
  \ifGPblacktext
    \def\colorrgb#1{}%
    \def\colorgray#1{}%
  \else
    \ifGPcolor
      \def\colorrgb#1{\color[rgb]{#1}}%
      \def\colorgray#1{\color[gray]{#1}}%
      \expandafter\def\csname LTw\endcsname{\color{white}}%
      \expandafter\def\csname LTb\endcsname{\color{black}}%
      \expandafter\def\csname LTa\endcsname{\color{black}}%
      \expandafter\def\csname LT0\endcsname{\color[rgb]{1,0,0}}%
      \expandafter\def\csname LT1\endcsname{\color[rgb]{0,1,0}}%
      \expandafter\def\csname LT2\endcsname{\color[rgb]{0,0,1}}%
      \expandafter\def\csname LT3\endcsname{\color[rgb]{1,0,1}}%
      \expandafter\def\csname LT4\endcsname{\color[rgb]{0,1,1}}%
      \expandafter\def\csname LT5\endcsname{\color[rgb]{1,1,0}}%
      \expandafter\def\csname LT6\endcsname{\color[rgb]{0,0,0}}%
      \expandafter\def\csname LT7\endcsname{\color[rgb]{1,0.3,0}}%
      \expandafter\def\csname LT8\endcsname{\color[rgb]{0.5,0.5,0.5}}%
    \else
      \def\colorrgb#1{\color{black}}%
      \def\colorgray#1{\color[gray]{#1}}%
      \expandafter\def\csname LTw\endcsname{\color{white}}%
      \expandafter\def\csname LTb\endcsname{\color{black}}%
      \expandafter\def\csname LTa\endcsname{\color{black}}%
      \expandafter\def\csname LT0\endcsname{\color{black}}%
      \expandafter\def\csname LT1\endcsname{\color{black}}%
      \expandafter\def\csname LT2\endcsname{\color{black}}%
      \expandafter\def\csname LT3\endcsname{\color{black}}%
      \expandafter\def\csname LT4\endcsname{\color{black}}%
      \expandafter\def\csname LT5\endcsname{\color{black}}%
      \expandafter\def\csname LT6\endcsname{\color{black}}%
      \expandafter\def\csname LT7\endcsname{\color{black}}%
      \expandafter\def\csname LT8\endcsname{\color{black}}%
    \fi
  \fi
  \setlength{\unitlength}{0.0500bp}%
  \begin{picture}(7652.00,4250.00)%
    \gplgaddtomacro\gplbacktext{%
      \csname LTb\endcsname%
      \put(1078,704){\makebox(0,0)[r]{\strut{} 0}}%
      \put(1078,1524){\makebox(0,0)[r]{\strut{}$1/2$}}%
      \put(1078,2345){\makebox(0,0)[r]{\strut{} 1}}%
      \put(1078,3985){\makebox(0,0)[r]{\strut{} 2}}%
      \put(1210,484){\makebox(0,0){\strut{} 1}}%
      \put(4073,484){\makebox(0,0){\strut{} 10}}%
      \put(7255,484){\makebox(0,0){\strut{} 20}}%
      \put(176,2344){\rotatebox{-270}{\makebox(0,0){\strut{}Operator dimensions (real part) $/z+2$}}}%
      \put(4232,154){\makebox(0,0){\strut{}$z$}}%
    }%
    \gplgaddtomacro\gplfronttext{%
      \csname LTb\endcsname%
      \put(6268,3812){\makebox(0,0)[r]{\strut{}$\Delta_1 / z+2$}}%
      \csname LTb\endcsname%
      \put(6268,3592){\makebox(0,0)[r]{\strut{}$\Delta_2 / z+2$}}%
      \csname LTb\endcsname%
      \put(6268,3372){\makebox(0,0)[r]{\strut{}$\Delta_3 / z+2$}}%
      \csname LTb\endcsname%
      \put(6268,3152){\makebox(0,0)[r]{\strut{}$\Delta_4 / z+2$}}%
    }%
    \gplbacktext
    \put(0,0){\includegraphics{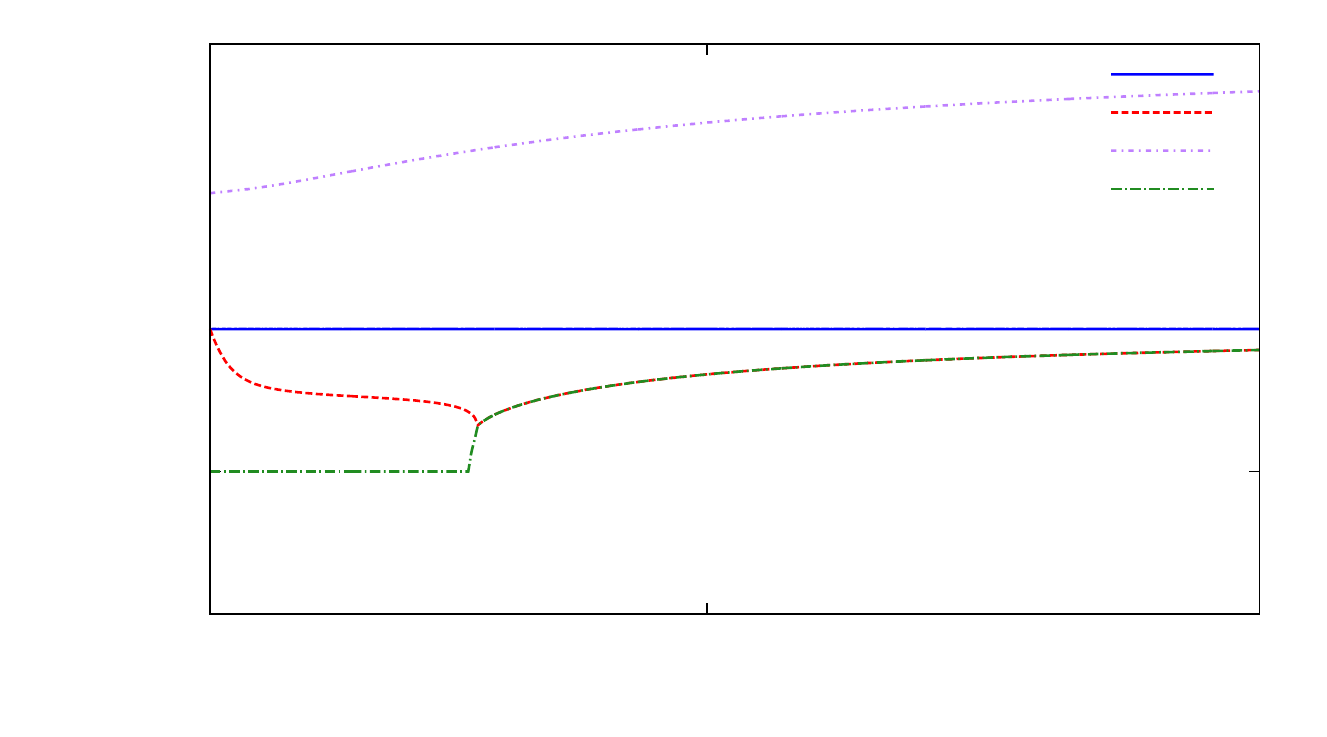}}%
    \gplfronttext
  \end{picture}%
\endgroup

%% file: 6D_AdS_2.tex
\begingroup
  \makeatletter
  \providecommand\color[2][]{%
    \GenericError{(gnuplot) \space\space\space\@spaces}{%
      Package color not loaded in conjunction with
      terminal option `colourtext'%
    }{See the gnuplot documentation for explanation.%
    }{Either use 'blacktext' in gnuplot or load the package
      color.sty in LaTeX.}%
    \renewcommand\color[2][]{}%
  }%
  \providecommand\includegraphics[2][]{%
    \GenericError{(gnuplot) \space\space\space\@spaces}{%
      Package graphicx or graphics not loaded%
    }{See the gnuplot documentation for explanation.%
    }{The gnuplot epslatex terminal needs graphicx.sty or graphics.sty.}%
    \renewcommand\includegraphics[2][]{}%
  }%
  \providecommand\rotatebox[2]{#2}%
  \@ifundefined{ifGPcolor}{%
    \newif\ifGPcolor
    \GPcolortrue
  }{}%
  \@ifundefined{ifGPblacktext}{%
    \newif\ifGPblacktext
    \GPblacktexttrue
  }{}%
  \let\gplgaddtomacro\g@addto@macro
  \gdef\gplbacktext{}%
  \gdef\gplfronttext{}%
  \makeatother
  \ifGPblacktext
    \def\colorrgb#1{}%
    \def\colorgray#1{}%
  \else
    \ifGPcolor
      \def\colorrgb#1{\color[rgb]{#1}}%
      \def\colorgray#1{\color[gray]{#1}}%
      \expandafter\def\csname LTw\endcsname{\color{white}}%
      \expandafter\def\csname LTb\endcsname{\color{black}}%
      \expandafter\def\csname LTa\endcsname{\color{black}}%
      \expandafter\def\csname LT0\endcsname{\color[rgb]{1,0,0}}%
      \expandafter\def\csname LT1\endcsname{\color[rgb]{0,1,0}}%
      \expandafter\def\csname LT2\endcsname{\color[rgb]{0,0,1}}%
      \expandafter\def\csname LT3\endcsname{\color[rgb]{1,0,1}}%
      \expandafter\def\csname LT4\endcsname{\color[rgb]{0,1,1}}%
      \expandafter\def\csname LT5\endcsname{\color[rgb]{1,1,0}}%
      \expandafter\def\csname LT6\endcsname{\color[rgb]{0,0,0}}%
      \expandafter\def\csname LT7\endcsname{\color[rgb]{1,0.3,0}}%
      \expandafter\def\csname LT8\endcsname{\color[rgb]{0.5,0.5,0.5}}%
    \else
      \def\colorrgb#1{\color{black}}%
      \def\colorgray#1{\color[gray]{#1}}%
      \expandafter\def\csname LTw\endcsname{\color{white}}%
      \expandafter\def\csname LTb\endcsname{\color{black}}%
      \expandafter\def\csname LTa\endcsname{\color{black}}%
      \expandafter\def\csname LT0\endcsname{\color{black}}%
      \expandafter\def\csname LT1\endcsname{\color{black}}%
      \expandafter\def\csname LT2\endcsname{\color{black}}%
      \expandafter\def\csname LT3\endcsname{\color{black}}%
      \expandafter\def\csname LT4\endcsname{\color{black}}%
      \expandafter\def\csname LT5\endcsname{\color{black}}%
      \expandafter\def\csname LT6\endcsname{\color{black}}%
      \expandafter\def\csname LT7\endcsname{\color{black}}%
      \expandafter\def\csname LT8\endcsname{\color{black}}%
    \fi
  \fi
  \setlength{\unitlength}{0.0500bp}%
  \begin{picture}(11338.00,3968.00)%
    \gplgaddtomacro\gplbacktext{%
      \csname LTb\endcsname%
      \put(1078,704){\makebox(0,0)[r]{\strut{} 0.5}}%
      \put(1078,1864){\makebox(0,0)[r]{\strut{}0.577}}%
      \put(1078,3703){\makebox(0,0)[r]{\strut{} 0.7}}%
      \put(1210,484){\makebox(0,0){\strut{} 1}}%
      \put(1790,484){\makebox(0,0){\strut{} 2}}%
      \put(2371,484){\makebox(0,0){\strut{} 3}}%
      \put(2951,484){\makebox(0,0){\strut{} 4}}%
      \put(3531,484){\makebox(0,0){\strut{} 5}}%
      \put(4111,484){\makebox(0,0){\strut{} 6}}%
      \put(4692,484){\makebox(0,0){\strut{} 7}}%
      \put(5272,484){\makebox(0,0){\strut{} 8}}%
      \put(176,2203){\rotatebox{-270}{\makebox(0,0){\strut{}$\varphi$}}}%
      \put(3241,154){\makebox(0,0){\strut{}$\rho$}}%
    }%
    \gplgaddtomacro\gplfronttext{%
    }%
    \gplgaddtomacro\gplbacktext{%
      \csname LTb\endcsname%
      \put(6351,704){\makebox(0,0)[r]{\strut{} 0}}%
      \put(6351,2703){\makebox(0,0)[r]{\strut{} 2}}%
      \put(6351,3703){\makebox(0,0)[r]{\strut{} 3}}%
      \put(6483,484){\makebox(0,0){\strut{} 1}}%
      \put(7120,484){\makebox(0,0){\strut{} 2}}%
      \put(7757,484){\makebox(0,0){\strut{} 3}}%
      \put(8394,484){\makebox(0,0){\strut{} 4}}%
      \put(9030,484){\makebox(0,0){\strut{} 5}}%
      \put(9667,484){\makebox(0,0){\strut{} 6}}%
      \put(10304,484){\makebox(0,0){\strut{} 7}}%
      \put(10941,484){\makebox(0,0){\strut{} 8}}%
      \put(5845,2203){\rotatebox{-270}{\makebox(0,0){\strut{}$- \partial_{\rho} e^{-2H} / e^{-2H}$}}}%
      \put(8712,154){\makebox(0,0){\strut{}$\rho$}}%
    }%
    \gplgaddtomacro\gplfronttext{%
    }%
    \gplbacktext
    \put(0,0){\includegraphics{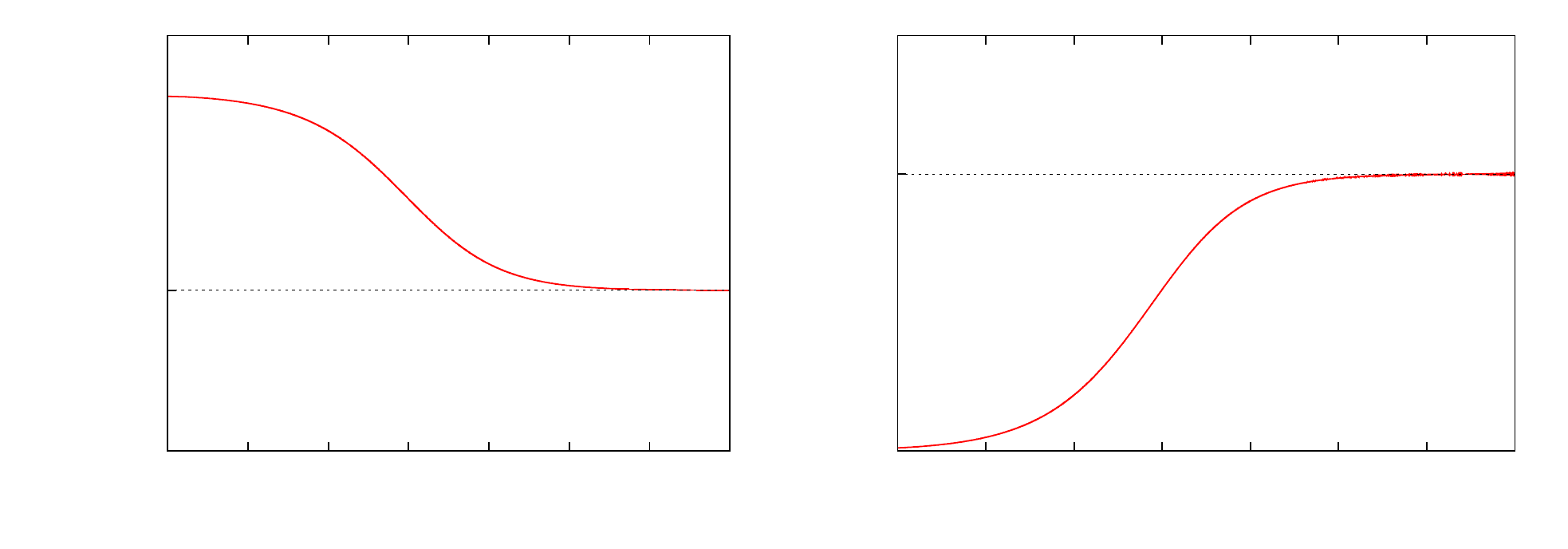}}%
    \gplfronttext
  \end{picture}%
\endgroup

%% file: 6D_us_Li_1.tex
\begingroup
  \makeatletter
  \providecommand\color[2][]{%
    \GenericError{(gnuplot) \space\space\space\@spaces}{%
      Package color not loaded in conjunction with
      terminal option `colourtext'%
    }{See the gnuplot documentation for explanation.%
    }{Either use 'blacktext' in gnuplot or load the package
      color.sty in LaTeX.}%
    \renewcommand\color[2][]{}%
  }%
  \providecommand\includegraphics[2][]{%
    \GenericError{(gnuplot) \space\space\space\@spaces}{%
      Package graphicx or graphics not loaded%
    }{See the gnuplot documentation for explanation.%
    }{The gnuplot epslatex terminal needs graphicx.sty or graphics.sty.}%
    \renewcommand\includegraphics[2][]{}%
  }%
  \providecommand\rotatebox[2]{#2}%
  \@ifundefined{ifGPcolor}{%
    \newif\ifGPcolor
    \GPcolortrue
  }{}%
  \@ifundefined{ifGPblacktext}{%
    \newif\ifGPblacktext
    \GPblacktexttrue
  }{}%
  \let\gplgaddtomacro\g@addto@macro
  \gdef\gplbacktext{}%
  \gdef\gplfronttext{}%
  \makeatother
  \ifGPblacktext
    \def\colorrgb#1{}%
    \def\colorgray#1{}%
  \else
    \ifGPcolor
      \def\colorrgb#1{\color[rgb]{#1}}%
      \def\colorgray#1{\color[gray]{#1}}%
      \expandafter\def\csname LTw\endcsname{\color{white}}%
      \expandafter\def\csname LTb\endcsname{\color{black}}%
      \expandafter\def\csname LTa\endcsname{\color{black}}%
      \expandafter\def\csname LT0\endcsname{\color[rgb]{1,0,0}}%
      \expandafter\def\csname LT1\endcsname{\color[rgb]{0,1,0}}%
      \expandafter\def\csname LT2\endcsname{\color[rgb]{0,0,1}}%
      \expandafter\def\csname LT3\endcsname{\color[rgb]{1,0,1}}%
      \expandafter\def\csname LT4\endcsname{\color[rgb]{0,1,1}}%
      \expandafter\def\csname LT5\endcsname{\color[rgb]{1,1,0}}%
      \expandafter\def\csname LT6\endcsname{\color[rgb]{0,0,0}}%
      \expandafter\def\csname LT7\endcsname{\color[rgb]{1,0.3,0}}%
      \expandafter\def\csname LT8\endcsname{\color[rgb]{0.5,0.5,0.5}}%
    \else
      \def\colorrgb#1{\color{black}}%
      \def\colorgray#1{\color[gray]{#1}}%
      \expandafter\def\csname LTw\endcsname{\color{white}}%
      \expandafter\def\csname LTb\endcsname{\color{black}}%
      \expandafter\def\csname LTa\endcsname{\color{black}}%
      \expandafter\def\csname LT0\endcsname{\color{black}}%
      \expandafter\def\csname LT1\endcsname{\color{black}}%
      \expandafter\def\csname LT2\endcsname{\color{black}}%
      \expandafter\def\csname LT3\endcsname{\color{black}}%
      \expandafter\def\csname LT4\endcsname{\color{black}}%
      \expandafter\def\csname LT5\endcsname{\color{black}}%
      \expandafter\def\csname LT6\endcsname{\color{black}}%
      \expandafter\def\csname LT7\endcsname{\color{black}}%
      \expandafter\def\csname LT8\endcsname{\color{black}}%
    \fi
  \fi
  \setlength{\unitlength}{0.0500bp}%
  \begin{picture}(11338.00,7936.00)%
    \gplgaddtomacro\gplbacktext{%
      \csname LTb\endcsname%
      \put(1078,4672){\makebox(0,0)[r]{\strut{} 0.4}}%
      \put(1078,7331){\makebox(0,0)[r]{\strut{}0.557}}%
      \put(1078,7671){\makebox(0,0)[r]{\strut{} 0.6}}%
      \put(1210,4452){\makebox(0,0){\strut{} 1}}%
      \put(1887,4452){\makebox(0,0){\strut{} 2}}%
      \put(2564,4452){\makebox(0,0){\strut{} 3}}%
      \put(3241,4452){\makebox(0,0){\strut{} 4}}%
      \put(3918,4452){\makebox(0,0){\strut{} 5}}%
      \put(4595,4452){\makebox(0,0){\strut{} 6}}%
      \put(5272,4452){\makebox(0,0){\strut{} 7}}%
      \put(176,6171){\rotatebox{-270}{\makebox(0,0){\strut{}$\varphi$}}}%
      \put(3241,4122){\makebox(0,0){\strut{}$\rho$}}%
    }%
    \gplgaddtomacro\gplfronttext{%
    }%
    \gplgaddtomacro\gplbacktext{%
      \csname LTb\endcsname%
      \put(6351,4672){\makebox(0,0)[r]{\strut{} 0}}%
      \put(6351,6671){\makebox(0,0)[r]{\strut{} 2}}%
      \put(6351,7671){\makebox(0,0)[r]{\strut{} 3}}%
      \put(6483,4452){\makebox(0,0){\strut{} 1}}%
      \put(7226,4452){\makebox(0,0){\strut{} 2}}%
      \put(7969,4452){\makebox(0,0){\strut{} 3}}%
      \put(8712,4452){\makebox(0,0){\strut{} 4}}%
      \put(9455,4452){\makebox(0,0){\strut{} 5}}%
      \put(10198,4452){\makebox(0,0){\strut{} 6}}%
      \put(10941,4452){\makebox(0,0){\strut{} 7}}%
      \put(5845,6171){\rotatebox{-270}{\makebox(0,0){\strut{}$- \partial_{\rho} e^{-2H} / e^{-2H}$}}}%
      \put(8712,4122){\makebox(0,0){\strut{}$\rho$}}%
    }%
    \gplgaddtomacro\gplfronttext{%
    }%
    \gplgaddtomacro\gplbacktext{%
      \csname LTb\endcsname%
      \put(682,704){\makebox(0,0)[r]{\strut{} 0}}%
      \put(682,1704){\makebox(0,0)[r]{\strut{} 1}}%
      \put(682,2704){\makebox(0,0)[r]{\strut{} 2}}%
      \put(814,484){\makebox(0,0){\strut{} 1}}%
      \put(1557,484){\makebox(0,0){\strut{} 2}}%
      \put(2300,484){\makebox(0,0){\strut{} 3}}%
      \put(3043,484){\makebox(0,0){\strut{} 4}}%
      \put(3786,484){\makebox(0,0){\strut{} 5}}%
      \put(4529,484){\makebox(0,0){\strut{} 6}}%
      \put(5272,484){\makebox(0,0){\strut{} 7}}%
      \put(176,2204){\rotatebox{-270}{\makebox(0,0){\strut{}$\partial_{\rho} F$}}}%
      \put(3043,154){\makebox(0,0){\strut{}$\rho$}}%
    }%
    \gplgaddtomacro\gplfronttext{%
    }%
    \gplgaddtomacro\gplbacktext{%
      \csname LTb\endcsname%
      \put(6615,704){\makebox(0,0)[r]{\strut{} 0}}%
      \put(6615,2204){\makebox(0,0)[r]{\strut{} 0.5}}%
      \put(6615,3704){\makebox(0,0)[r]{\strut{} 1}}%
      \put(6747,484){\makebox(0,0){\strut{} 1}}%
      \put(7446,484){\makebox(0,0){\strut{} 2}}%
      \put(8145,484){\makebox(0,0){\strut{} 3}}%
      \put(8844,484){\makebox(0,0){\strut{} 4}}%
      \put(9543,484){\makebox(0,0){\strut{} 5}}%
      \put(10242,484){\makebox(0,0){\strut{} 6}}%
      \put(10941,484){\makebox(0,0){\strut{} 7}}%
      \put(5845,2204){\rotatebox{-270}{\makebox(0,0){\strut{}$\beta$}}}%
      \put(8844,154){\makebox(0,0){\strut{}$\rho$}}%
    }%
    \gplgaddtomacro\gplfronttext{%
    }%
    \gplbacktext
    \put(0,0){\includegraphics{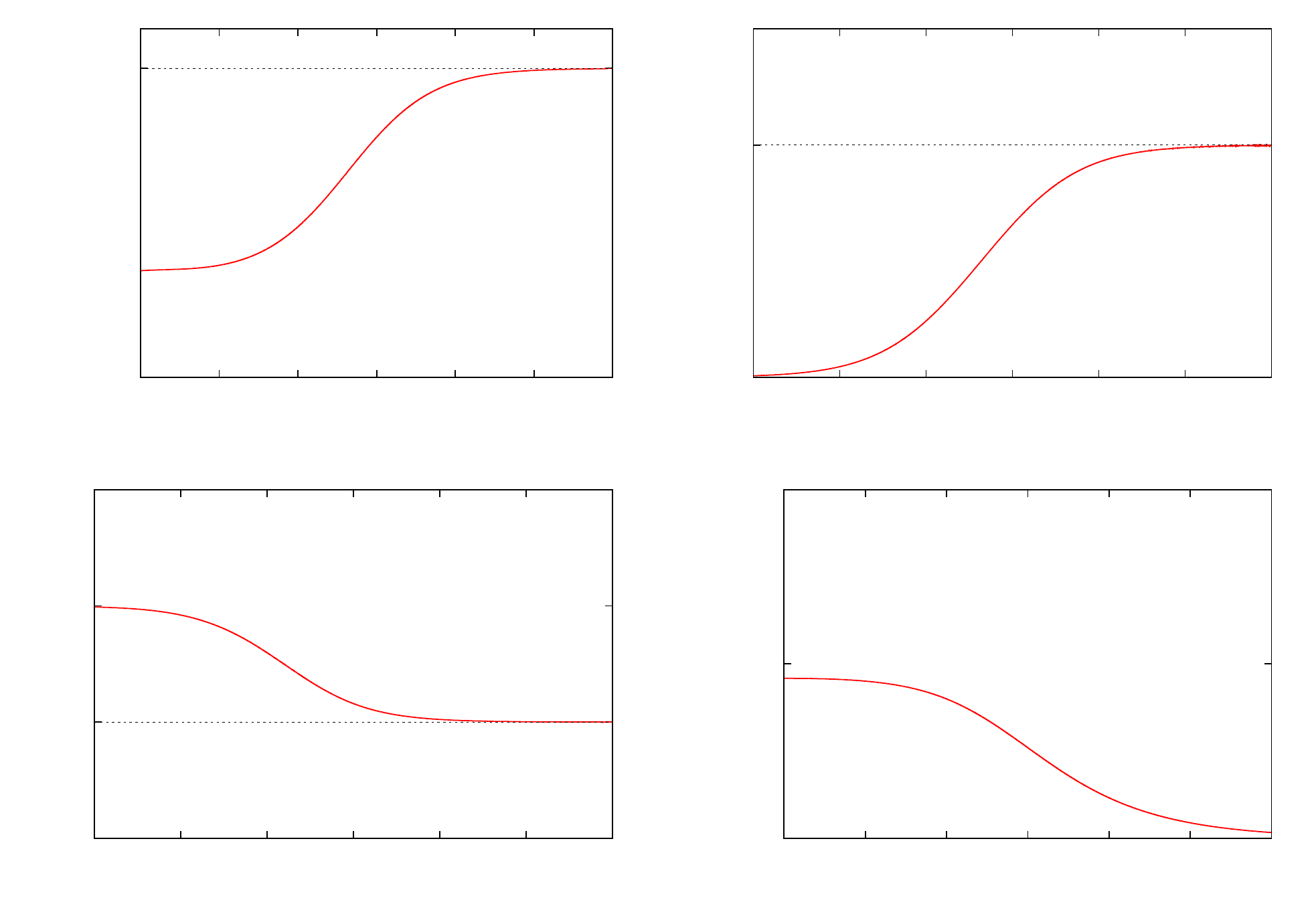}}%
    \gplfronttext
  \end{picture}%
\endgroup

%% file: AdS_AdS_1.tex
\begingroup
  \makeatletter
  \providecommand\color[2][]{%
    \GenericError{(gnuplot) \space\space\space\@spaces}{%
      Package color not loaded in conjunction with
      terminal option `colourtext'%
    }{See the gnuplot documentation for explanation.%
    }{Either use 'blacktext' in gnuplot or load the package
      color.sty in LaTeX.}%
    \renewcommand\color[2][]{}%
  }%
  \providecommand\includegraphics[2][]{%
    \GenericError{(gnuplot) \space\space\space\@spaces}{%
      Package graphicx or graphics not loaded%
    }{See the gnuplot documentation for explanation.%
    }{The gnuplot epslatex terminal needs graphicx.sty or graphics.sty.}%
    \renewcommand\includegraphics[2][]{}%
  }%
  \providecommand\rotatebox[2]{#2}%
  \@ifundefined{ifGPcolor}{%
    \newif\ifGPcolor
    \GPcolortrue
  }{}%
  \@ifundefined{ifGPblacktext}{%
    \newif\ifGPblacktext
    \GPblacktexttrue
  }{}%
  \let\gplgaddtomacro\g@addto@macro
  \gdef\gplbacktext{}%
  \gdef\gplfronttext{}%
  \makeatother
  \ifGPblacktext
    \def\colorrgb#1{}%
    \def\colorgray#1{}%
  \else
    \ifGPcolor
      \def\colorrgb#1{\color[rgb]{#1}}%
      \def\colorgray#1{\color[gray]{#1}}%
      \expandafter\def\csname LTw\endcsname{\color{white}}%
      \expandafter\def\csname LTb\endcsname{\color{black}}%
      \expandafter\def\csname LTa\endcsname{\color{black}}%
      \expandafter\def\csname LT0\endcsname{\color[rgb]{1,0,0}}%
      \expandafter\def\csname LT1\endcsname{\color[rgb]{0,1,0}}%
      \expandafter\def\csname LT2\endcsname{\color[rgb]{0,0,1}}%
      \expandafter\def\csname LT3\endcsname{\color[rgb]{1,0,1}}%
      \expandafter\def\csname LT4\endcsname{\color[rgb]{0,1,1}}%
      \expandafter\def\csname LT5\endcsname{\color[rgb]{1,1,0}}%
      \expandafter\def\csname LT6\endcsname{\color[rgb]{0,0,0}}%
      \expandafter\def\csname LT7\endcsname{\color[rgb]{1,0.3,0}}%
      \expandafter\def\csname LT8\endcsname{\color[rgb]{0.5,0.5,0.5}}%
    \else
      \def\colorrgb#1{\color{black}}%
      \def\colorgray#1{\color[gray]{#1}}%
      \expandafter\def\csname LTw\endcsname{\color{white}}%
      \expandafter\def\csname LTb\endcsname{\color{black}}%
      \expandafter\def\csname LTa\endcsname{\color{black}}%
      \expandafter\def\csname LT0\endcsname{\color{black}}%
      \expandafter\def\csname LT1\endcsname{\color{black}}%
      \expandafter\def\csname LT2\endcsname{\color{black}}%
      \expandafter\def\csname LT3\endcsname{\color{black}}%
      \expandafter\def\csname LT4\endcsname{\color{black}}%
      \expandafter\def\csname LT5\endcsname{\color{black}}%
      \expandafter\def\csname LT6\endcsname{\color{black}}%
      \expandafter\def\csname LT7\endcsname{\color{black}}%
      \expandafter\def\csname LT8\endcsname{\color{black}}%
    \fi
  \fi
  \setlength{\unitlength}{0.0500bp}%
  \begin{picture}(11338.00,4250.00)%
    \gplgaddtomacro\gplbacktext{%
      \csname LTb\endcsname%
      \put(1078,704){\makebox(0,0)[r]{\strut{} 0.6}}%
      \put(1078,1251){\makebox(0,0)[r]{\strut{} 0.65}}%
      \put(1078,1798){\makebox(0,0)[r]{\strut{} 0.7}}%
      \put(1078,2345){\makebox(0,0)[r]{\strut{} 0.75}}%
      \put(1078,2891){\makebox(0,0)[r]{\strut{} 0.8}}%
      \put(1078,3438){\makebox(0,0)[r]{\strut{} 0.85}}%
      \put(1078,3985){\makebox(0,0)[r]{\strut{} 0.9}}%
      \put(1522,484){\makebox(0,0){\strut{} 2}}%
      \put(2147,484){\makebox(0,0){\strut{} 4}}%
      \put(2772,484){\makebox(0,0){\strut{} 6}}%
      \put(3397,484){\makebox(0,0){\strut{} 8}}%
      \put(4022,484){\makebox(0,0){\strut{} 10}}%
      \put(4647,484){\makebox(0,0){\strut{} 12}}%
      \put(5272,484){\makebox(0,0){\strut{} 14}}%
      \put(176,2344){\rotatebox{-270}{\makebox(0,0){\strut{}$\varphi$}}}%
      \put(3241,154){\makebox(0,0){\strut{}$\rho$}}%
    }%
    \gplgaddtomacro\gplfronttext{%
    }%
    \gplgaddtomacro\gplbacktext{%
      \csname LTb\endcsname%
      \put(6747,704){\makebox(0,0)[r]{\strut{} 0.25}}%
      \put(6747,2345){\makebox(0,0)[r]{\strut{} 0.3}}%
      \put(6747,3985){\makebox(0,0)[r]{\strut{} 0.35}}%
      \put(7191,484){\makebox(0,0){\strut{} 2}}%
      \put(7816,484){\makebox(0,0){\strut{} 4}}%
      \put(8441,484){\makebox(0,0){\strut{} 6}}%
      \put(9066,484){\makebox(0,0){\strut{} 8}}%
      \put(9691,484){\makebox(0,0){\strut{} 10}}%
      \put(10316,484){\makebox(0,0){\strut{} 12}}%
      \put(10941,484){\makebox(0,0){\strut{} 14}}%
      \put(5845,2344){\rotatebox{-270}{\makebox(0,0){\strut{}$e^{-2H}$}}}%
      \put(8910,154){\makebox(0,0){\strut{}$\rho$}}%
    }%
    \gplgaddtomacro\gplfronttext{%
    }%
    \gplbacktext
    \put(0,0){\includegraphics{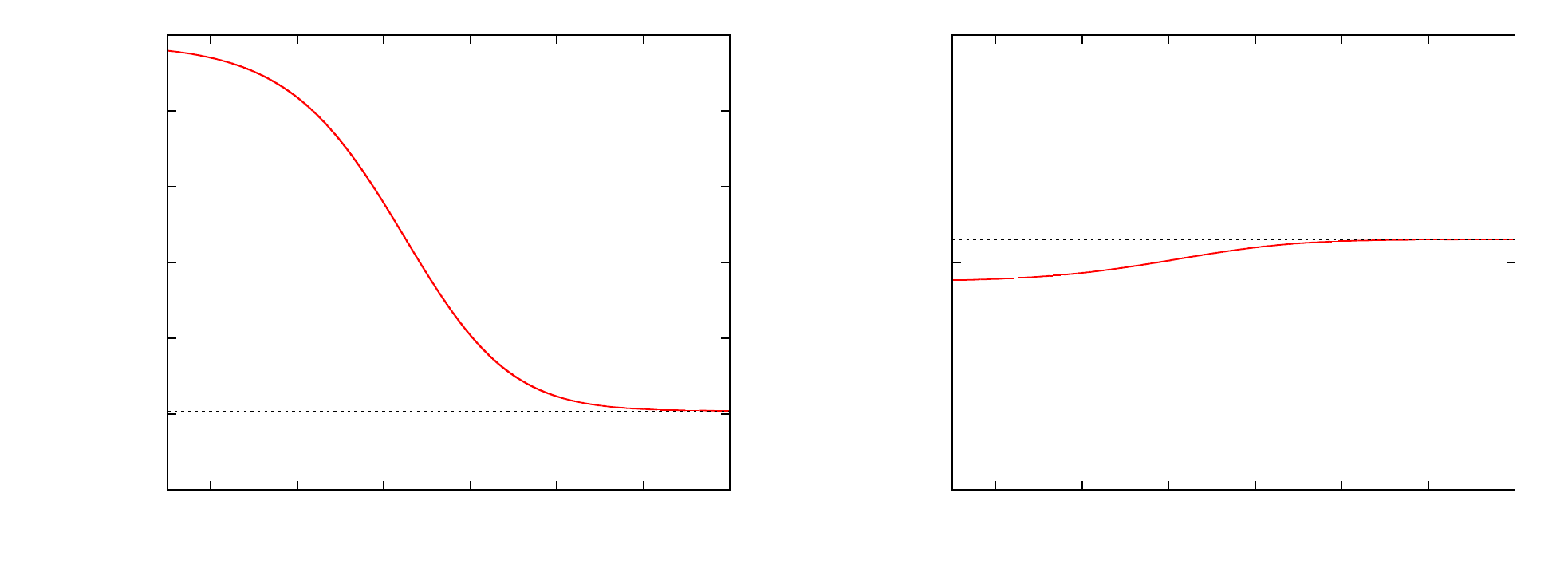}}%
    \gplfronttext
  \end{picture}%
\endgroup

%% file: AdS_Li_1.tex
\begingroup
  \makeatletter
  \providecommand\color[2][]{%
    \GenericError{(gnuplot) \space\space\space\@spaces}{%
      Package color not loaded in conjunction with
      terminal option `colourtext'%
    }{See the gnuplot documentation for explanation.%
    }{Either use 'blacktext' in gnuplot or load the package
      color.sty in LaTeX.}%
    \renewcommand\color[2][]{}%
  }%
  \providecommand\includegraphics[2][]{%
    \GenericError{(gnuplot) \space\space\space\@spaces}{%
      Package graphicx or graphics not loaded%
    }{See the gnuplot documentation for explanation.%
    }{The gnuplot epslatex terminal needs graphicx.sty or graphics.sty.}%
    \renewcommand\includegraphics[2][]{}%
  }%
  \providecommand\rotatebox[2]{#2}%
  \@ifundefined{ifGPcolor}{%
    \newif\ifGPcolor
    \GPcolortrue
  }{}%
  \@ifundefined{ifGPblacktext}{%
    \newif\ifGPblacktext
    \GPblacktexttrue
  }{}%
  \let\gplgaddtomacro\g@addto@macro
  \gdef\gplbacktext{}%
  \gdef\gplfronttext{}%
  \makeatother
  \ifGPblacktext
    \def\colorrgb#1{}%
    \def\colorgray#1{}%
  \else
    \ifGPcolor
      \def\colorrgb#1{\color[rgb]{#1}}%
      \def\colorgray#1{\color[gray]{#1}}%
      \expandafter\def\csname LTw\endcsname{\color{white}}%
      \expandafter\def\csname LTb\endcsname{\color{black}}%
      \expandafter\def\csname LTa\endcsname{\color{black}}%
      \expandafter\def\csname LT0\endcsname{\color[rgb]{1,0,0}}%
      \expandafter\def\csname LT1\endcsname{\color[rgb]{0,1,0}}%
      \expandafter\def\csname LT2\endcsname{\color[rgb]{0,0,1}}%
      \expandafter\def\csname LT3\endcsname{\color[rgb]{1,0,1}}%
      \expandafter\def\csname LT4\endcsname{\color[rgb]{0,1,1}}%
      \expandafter\def\csname LT5\endcsname{\color[rgb]{1,1,0}}%
      \expandafter\def\csname LT6\endcsname{\color[rgb]{0,0,0}}%
      \expandafter\def\csname LT7\endcsname{\color[rgb]{1,0.3,0}}%
      \expandafter\def\csname LT8\endcsname{\color[rgb]{0.5,0.5,0.5}}%
    \else
      \def\colorrgb#1{\color{black}}%
      \def\colorgray#1{\color[gray]{#1}}%
      \expandafter\def\csname LTw\endcsname{\color{white}}%
      \expandafter\def\csname LTb\endcsname{\color{black}}%
      \expandafter\def\csname LTa\endcsname{\color{black}}%
      \expandafter\def\csname LT0\endcsname{\color{black}}%
      \expandafter\def\csname LT1\endcsname{\color{black}}%
      \expandafter\def\csname LT2\endcsname{\color{black}}%
      \expandafter\def\csname LT3\endcsname{\color{black}}%
      \expandafter\def\csname LT4\endcsname{\color{black}}%
      \expandafter\def\csname LT5\endcsname{\color{black}}%
      \expandafter\def\csname LT6\endcsname{\color{black}}%
      \expandafter\def\csname LT7\endcsname{\color{black}}%
      \expandafter\def\csname LT8\endcsname{\color{black}}%
    \fi
  \fi
  \setlength{\unitlength}{0.0500bp}%
  \begin{picture}(11338.00,8502.00)%
    \gplgaddtomacro\gplbacktext{%
      \csname LTb\endcsname%
      \put(1078,4955){\makebox(0,0)[r]{\strut{} 0.4}}%
      \put(1078,5365){\makebox(0,0)[r]{\strut{} 0.45}}%
      \put(1078,5776){\makebox(0,0)[r]{\strut{} 0.5}}%
      \put(1078,6186){\makebox(0,0)[r]{\strut{} 0.55}}%
      \put(1078,6596){\makebox(0,0)[r]{\strut{} 0.6}}%
      \put(1078,7006){\makebox(0,0)[r]{\strut{} 0.65}}%
      \put(1078,7417){\makebox(0,0)[r]{\strut{} 0.7}}%
      \put(1078,7827){\makebox(0,0)[r]{\strut{} 0.75}}%
      \put(1078,8237){\makebox(0,0)[r]{\strut{} 0.8}}%
      \put(1481,4735){\makebox(0,0){\strut{} 2}}%
      \put(2022,4735){\makebox(0,0){\strut{} 4}}%
      \put(2564,4735){\makebox(0,0){\strut{} 6}}%
      \put(3106,4735){\makebox(0,0){\strut{} 8}}%
      \put(3647,4735){\makebox(0,0){\strut{} 10}}%
      \put(4189,4735){\makebox(0,0){\strut{} 12}}%
      \put(4730,4735){\makebox(0,0){\strut{} 14}}%
      \put(5272,4735){\makebox(0,0){\strut{} 16}}%
      \put(176,6596){\rotatebox{-270}{\makebox(0,0){\strut{}$\varphi$}}}%
      \put(3241,4405){\makebox(0,0){\strut{}$\rho$}}%
    }%
    \gplgaddtomacro\gplfronttext{%
    }%
    \gplgaddtomacro\gplbacktext{%
      \csname LTb\endcsname%
      \put(6747,4955){\makebox(0,0)[r]{\strut{} 0.1}}%
      \put(6747,5611){\makebox(0,0)[r]{\strut{} 0.12}}%
      \put(6747,6268){\makebox(0,0)[r]{\strut{} 0.14}}%
      \put(6747,6924){\makebox(0,0)[r]{\strut{} 0.16}}%
      \put(6747,7581){\makebox(0,0)[r]{\strut{} 0.18}}%
      \put(6747,8237){\makebox(0,0)[r]{\strut{} 0.2}}%
      \put(7150,4735){\makebox(0,0){\strut{} 2}}%
      \put(7691,4735){\makebox(0,0){\strut{} 4}}%
      \put(8233,4735){\makebox(0,0){\strut{} 6}}%
      \put(8775,4735){\makebox(0,0){\strut{} 8}}%
      \put(9316,4735){\makebox(0,0){\strut{} 10}}%
      \put(9858,4735){\makebox(0,0){\strut{} 12}}%
      \put(10399,4735){\makebox(0,0){\strut{} 14}}%
      \put(10941,4735){\makebox(0,0){\strut{} 16}}%
      \put(5845,6596){\rotatebox{-270}{\makebox(0,0){\strut{}$e^{-2H}$}}}%
      \put(8910,4405){\makebox(0,0){\strut{}$\rho$}}%
    }%
    \gplgaddtomacro\gplfronttext{%
    }%
    \gplgaddtomacro\gplbacktext{%
      \csname LTb\endcsname%
      \put(682,704){\makebox(0,0)[r]{\strut{} 0}}%
      \put(682,1361){\makebox(0,0)[r]{\strut{} 1}}%
      \put(682,2017){\makebox(0,0)[r]{\strut{} 2}}%
      \put(682,2674){\makebox(0,0)[r]{\strut{} 3}}%
      \put(682,3330){\makebox(0,0)[r]{\strut{} 4}}%
      \put(682,3987){\makebox(0,0)[r]{\strut{} 5}}%
      \put(1111,484){\makebox(0,0){\strut{} 2}}%
      \put(1706,484){\makebox(0,0){\strut{} 4}}%
      \put(2300,484){\makebox(0,0){\strut{} 6}}%
      \put(2894,484){\makebox(0,0){\strut{} 8}}%
      \put(3489,484){\makebox(0,0){\strut{} 10}}%
      \put(4083,484){\makebox(0,0){\strut{} 12}}%
      \put(4678,484){\makebox(0,0){\strut{} 14}}%
      \put(5272,484){\makebox(0,0){\strut{} 16}}%
      \put(176,2345){\rotatebox{-270}{\makebox(0,0){\strut{}$\partial_{\rho} F$}}}%
      \put(3043,154){\makebox(0,0){\strut{}$\rho$}}%
    }%
    \gplgaddtomacro\gplfronttext{%
    }%
    \gplgaddtomacro\gplbacktext{%
      \csname LTb\endcsname%
      \put(6615,704){\makebox(0,0)[r]{\strut{} 0}}%
      \put(6615,1142){\makebox(0,0)[r]{\strut{} 0.2}}%
      \put(6615,1579){\makebox(0,0)[r]{\strut{} 0.4}}%
      \put(6615,2017){\makebox(0,0)[r]{\strut{} 0.6}}%
      \put(6615,2455){\makebox(0,0)[r]{\strut{} 0.8}}%
      \put(6615,2893){\makebox(0,0)[r]{\strut{} 1}}%
      \put(6615,3330){\makebox(0,0)[r]{\strut{} 1.2}}%
      \put(6615,3768){\makebox(0,0)[r]{\strut{} 1.4}}%
      \put(7027,484){\makebox(0,0){\strut{} 2}}%
      \put(7586,484){\makebox(0,0){\strut{} 4}}%
      \put(8145,484){\makebox(0,0){\strut{} 6}}%
      \put(8704,484){\makebox(0,0){\strut{} 8}}%
      \put(9263,484){\makebox(0,0){\strut{} 10}}%
      \put(9823,484){\makebox(0,0){\strut{} 12}}%
      \put(10382,484){\makebox(0,0){\strut{} 14}}%
      \put(10941,484){\makebox(0,0){\strut{} 16}}%
      \put(5845,2345){\rotatebox{-270}{\makebox(0,0){\strut{}$\beta$}}}%
      \put(8844,154){\makebox(0,0){\strut{}$\rho$}}%
    }%
    \gplgaddtomacro\gplfronttext{%
    }%
    \gplbacktext
    \put(0,0){\includegraphics{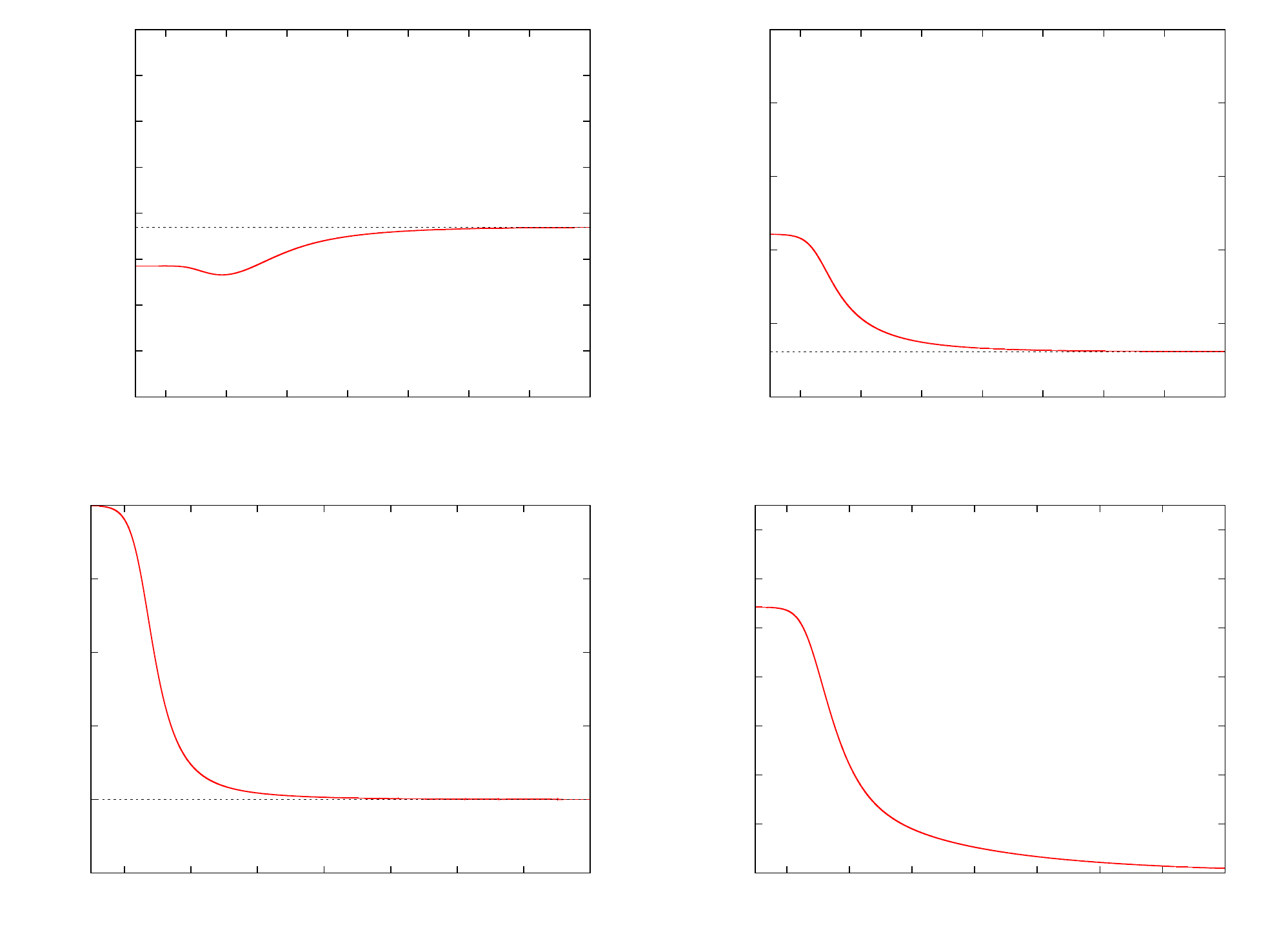}}%
    \gplfronttext
  \end{picture}%
\endgroup

%% file: Li_Li_2.tex
\begingroup
  \makeatletter
  \providecommand\color[2][]{%
    \GenericError{(gnuplot) \space\space\space\@spaces}{%
      Package color not loaded in conjunction with
      terminal option `colourtext'%
    }{See the gnuplot documentation for explanation.%
    }{Either use 'blacktext' in gnuplot or load the package
      color.sty in LaTeX.}%
    \renewcommand\color[2][]{}%
  }%
  \providecommand\includegraphics[2][]{%
    \GenericError{(gnuplot) \space\space\space\@spaces}{%
      Package graphicx or graphics not loaded%
    }{See the gnuplot documentation for explanation.%
    }{The gnuplot epslatex terminal needs graphicx.sty or graphics.sty.}%
    \renewcommand\includegraphics[2][]{}%
  }%
  \providecommand\rotatebox[2]{#2}%
  \@ifundefined{ifGPcolor}{%
    \newif\ifGPcolor
    \GPcolortrue
  }{}%
  \@ifundefined{ifGPblacktext}{%
    \newif\ifGPblacktext
    \GPblacktexttrue
  }{}%
  \let\gplgaddtomacro\g@addto@macro
  \gdef\gplbacktext{}%
  \gdef\gplfronttext{}%
  \makeatother
  \ifGPblacktext
    \def\colorrgb#1{}%
    \def\colorgray#1{}%
  \else
    \ifGPcolor
      \def\colorrgb#1{\color[rgb]{#1}}%
      \def\colorgray#1{\color[gray]{#1}}%
      \expandafter\def\csname LTw\endcsname{\color{white}}%
      \expandafter\def\csname LTb\endcsname{\color{black}}%
      \expandafter\def\csname LTa\endcsname{\color{black}}%
      \expandafter\def\csname LT0\endcsname{\color[rgb]{1,0,0}}%
      \expandafter\def\csname LT1\endcsname{\color[rgb]{0,1,0}}%
      \expandafter\def\csname LT2\endcsname{\color[rgb]{0,0,1}}%
      \expandafter\def\csname LT3\endcsname{\color[rgb]{1,0,1}}%
      \expandafter\def\csname LT4\endcsname{\color[rgb]{0,1,1}}%
      \expandafter\def\csname LT5\endcsname{\color[rgb]{1,1,0}}%
      \expandafter\def\csname LT6\endcsname{\color[rgb]{0,0,0}}%
      \expandafter\def\csname LT7\endcsname{\color[rgb]{1,0.3,0}}%
      \expandafter\def\csname LT8\endcsname{\color[rgb]{0.5,0.5,0.5}}%
    \else
      \def\colorrgb#1{\color{black}}%
      \def\colorgray#1{\color[gray]{#1}}%
      \expandafter\def\csname LTw\endcsname{\color{white}}%
      \expandafter\def\csname LTb\endcsname{\color{black}}%
      \expandafter\def\csname LTa\endcsname{\color{black}}%
      \expandafter\def\csname LT0\endcsname{\color{black}}%
      \expandafter\def\csname LT1\endcsname{\color{black}}%
      \expandafter\def\csname LT2\endcsname{\color{black}}%
      \expandafter\def\csname LT3\endcsname{\color{black}}%
      \expandafter\def\csname LT4\endcsname{\color{black}}%
      \expandafter\def\csname LT5\endcsname{\color{black}}%
      \expandafter\def\csname LT6\endcsname{\color{black}}%
      \expandafter\def\csname LT7\endcsname{\color{black}}%
      \expandafter\def\csname LT8\endcsname{\color{black}}%
    \fi
  \fi
  \setlength{\unitlength}{0.0500bp}%
  \begin{picture}(11338.00,8502.00)%
    \gplgaddtomacro\gplbacktext{%
      \csname LTb\endcsname%
      \put(946,4955){\makebox(0,0)[r]{\strut{} 0}}%
      \put(946,5611){\makebox(0,0)[r]{\strut{} 0.2}}%
      \put(946,6268){\makebox(0,0)[r]{\strut{} 0.4}}%
      \put(946,6924){\makebox(0,0)[r]{\strut{} 0.6}}%
      \put(946,7581){\makebox(0,0)[r]{\strut{} 0.8}}%
      \put(946,8237){\makebox(0,0)[r]{\strut{} 1}}%
      \put(1459,4735){\makebox(0,0){\strut{} 2}}%
      \put(2222,4735){\makebox(0,0){\strut{} 4}}%
      \put(2984,4735){\makebox(0,0){\strut{} 6}}%
      \put(3747,4735){\makebox(0,0){\strut{} 8}}%
      \put(4509,4735){\makebox(0,0){\strut{} 10}}%
      \put(5272,4735){\makebox(0,0){\strut{} 12}}%
      \put(176,6596){\rotatebox{-270}{\makebox(0,0){\strut{}$\varphi$}}}%
      \put(3175,4405){\makebox(0,0){\strut{}$\rho$}}%
    }%
    \gplgaddtomacro\gplfronttext{%
    }%
    \gplgaddtomacro\gplbacktext{%
      \csname LTb\endcsname%
      \put(6747,4955){\makebox(0,0)[r]{\strut{} 0.25}}%
      \put(6747,5283){\makebox(0,0)[r]{\strut{} 0.26}}%
      \put(6747,5611){\makebox(0,0)[r]{\strut{} 0.27}}%
      \put(6747,5940){\makebox(0,0)[r]{\strut{} 0.28}}%
      \put(6747,6268){\makebox(0,0)[r]{\strut{} 0.29}}%
      \put(6747,6596){\makebox(0,0)[r]{\strut{} 0.3}}%
      \put(6747,6924){\makebox(0,0)[r]{\strut{} 0.31}}%
      \put(6747,7252){\makebox(0,0)[r]{\strut{} 0.32}}%
      \put(6747,7581){\makebox(0,0)[r]{\strut{} 0.33}}%
      \put(6747,7909){\makebox(0,0)[r]{\strut{} 0.34}}%
      \put(6747,8237){\makebox(0,0)[r]{\strut{} 0.35}}%
      \put(7248,4735){\makebox(0,0){\strut{} 2}}%
      \put(7987,4735){\makebox(0,0){\strut{} 4}}%
      \put(8725,4735){\makebox(0,0){\strut{} 6}}%
      \put(9464,4735){\makebox(0,0){\strut{} 8}}%
      \put(10202,4735){\makebox(0,0){\strut{} 10}}%
      \put(10941,4735){\makebox(0,0){\strut{} 12}}%
      \put(5845,6596){\rotatebox{-270}{\makebox(0,0){\strut{}$e^{-2H}$}}}%
      \put(8910,4405){\makebox(0,0){\strut{}$\rho$}}%
    }%
    \gplgaddtomacro\gplfronttext{%
    }%
    \gplgaddtomacro\gplbacktext{%
      \csname LTb\endcsname%
      \put(814,704){\makebox(0,0)[r]{\strut{} 0}}%
      \put(814,1361){\makebox(0,0)[r]{\strut{} 2}}%
      \put(814,2017){\makebox(0,0)[r]{\strut{} 4}}%
      \put(814,2674){\makebox(0,0)[r]{\strut{} 6}}%
      \put(814,3330){\makebox(0,0)[r]{\strut{} 8}}%
      \put(814,3987){\makebox(0,0)[r]{\strut{} 10}}%
      \put(1339,484){\makebox(0,0){\strut{} 2}}%
      \put(2126,484){\makebox(0,0){\strut{} 4}}%
      \put(2912,484){\makebox(0,0){\strut{} 6}}%
      \put(3699,484){\makebox(0,0){\strut{} 8}}%
      \put(4485,484){\makebox(0,0){\strut{} 10}}%
      \put(5272,484){\makebox(0,0){\strut{} 12}}%
      \put(176,2345){\rotatebox{-270}{\makebox(0,0){\strut{}$\partial_{\rho} F$}}}%
      \put(3109,154){\makebox(0,0){\strut{}$\rho$}}%
    }%
    \gplgaddtomacro\gplfronttext{%
    }%
    \gplgaddtomacro\gplbacktext{%
      \csname LTb\endcsname%
      \put(6615,704){\makebox(0,0)[r]{\strut{} 0}}%
      \put(6615,1142){\makebox(0,0)[r]{\strut{} 0.2}}%
      \put(6615,1579){\makebox(0,0)[r]{\strut{} 0.4}}%
      \put(6615,2017){\makebox(0,0)[r]{\strut{} 0.6}}%
      \put(6615,2455){\makebox(0,0)[r]{\strut{} 0.8}}%
      \put(6615,2893){\makebox(0,0)[r]{\strut{} 1}}%
      \put(6615,3330){\makebox(0,0)[r]{\strut{} 1.2}}%
      \put(6615,3768){\makebox(0,0)[r]{\strut{} 1.4}}%
      \put(7128,484){\makebox(0,0){\strut{} 2}}%
      \put(7891,484){\makebox(0,0){\strut{} 4}}%
      \put(8653,484){\makebox(0,0){\strut{} 6}}%
      \put(9416,484){\makebox(0,0){\strut{} 8}}%
      \put(10178,484){\makebox(0,0){\strut{} 10}}%
      \put(10941,484){\makebox(0,0){\strut{} 12}}%
      \put(5845,2345){\rotatebox{-270}{\makebox(0,0){\strut{}$\beta$}}}%
      \put(8844,154){\makebox(0,0){\strut{}$\rho$}}%
    }%
    \gplgaddtomacro\gplfronttext{%
    }%
    \gplbacktext
    \put(0,0){\includegraphics{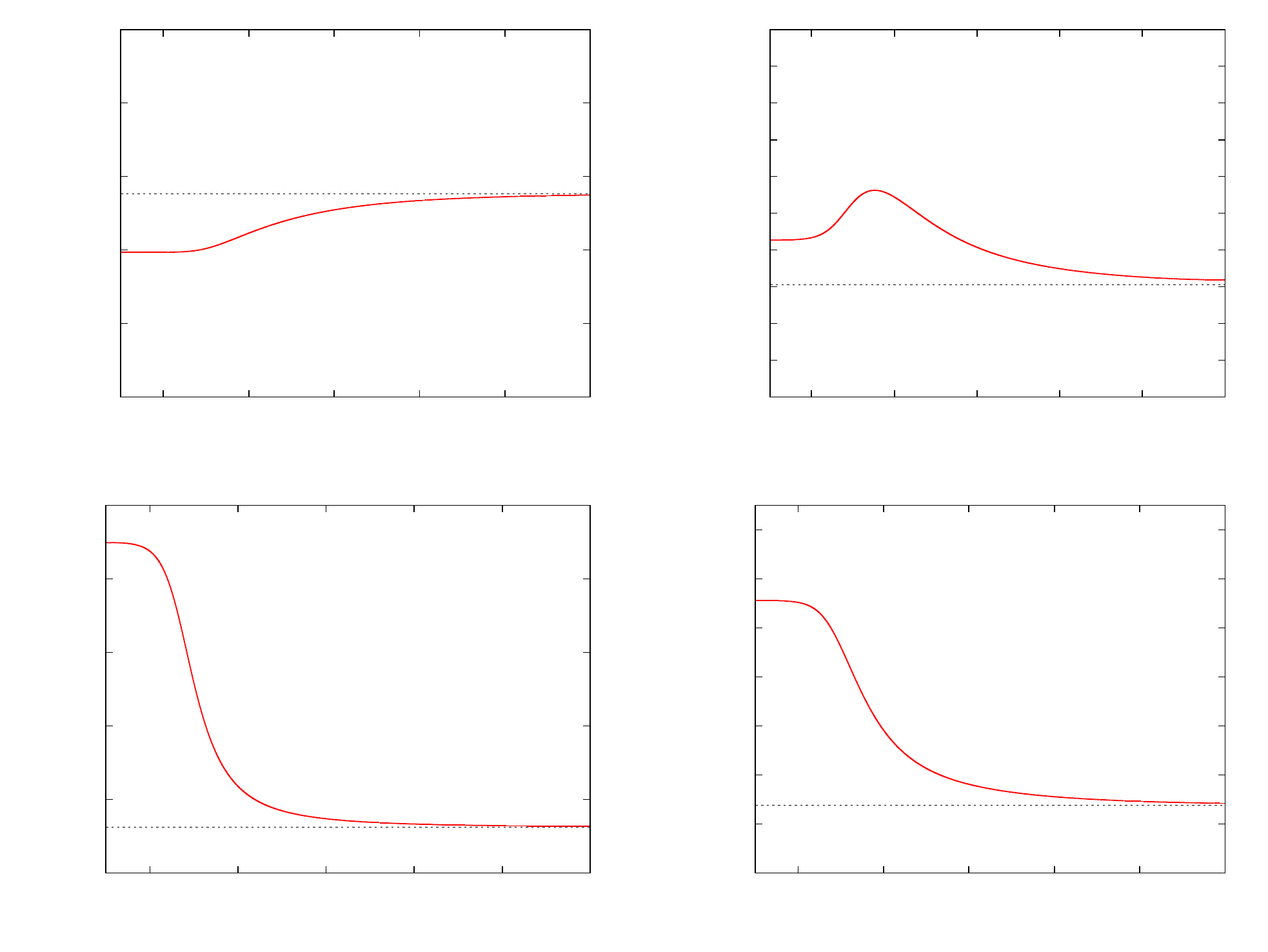}}%
    \gplfronttext
  \end{picture}%
\endgroup

%% file: Li_AdS_2.tex
\begingroup
  \makeatletter
  \providecommand\color[2][]{%
    \GenericError{(gnuplot) \space\space\space\@spaces}{%
      Package color not loaded in conjunction with
      terminal option `colourtext'%
    }{See the gnuplot documentation for explanation.%
    }{Either use 'blacktext' in gnuplot or load the package
      color.sty in LaTeX.}%
    \renewcommand\color[2][]{}%
  }%
  \providecommand\includegraphics[2][]{%
    \GenericError{(gnuplot) \space\space\space\@spaces}{%
      Package graphicx or graphics not loaded%
    }{See the gnuplot documentation for explanation.%
    }{The gnuplot epslatex terminal needs graphicx.sty or graphics.sty.}%
    \renewcommand\includegraphics[2][]{}%
  }%
  \providecommand\rotatebox[2]{#2}%
  \@ifundefined{ifGPcolor}{%
    \newif\ifGPcolor
    \GPcolortrue
  }{}%
  \@ifundefined{ifGPblacktext}{%
    \newif\ifGPblacktext
    \GPblacktexttrue
  }{}%
  \let\gplgaddtomacro\g@addto@macro
  \gdef\gplbacktext{}%
  \gdef\gplfronttext{}%
  \makeatother
  \ifGPblacktext
    \def\colorrgb#1{}%
    \def\colorgray#1{}%
  \else
    \ifGPcolor
      \def\colorrgb#1{\color[rgb]{#1}}%
      \def\colorgray#1{\color[gray]{#1}}%
      \expandafter\def\csname LTw\endcsname{\color{white}}%
      \expandafter\def\csname LTb\endcsname{\color{black}}%
      \expandafter\def\csname LTa\endcsname{\color{black}}%
      \expandafter\def\csname LT0\endcsname{\color[rgb]{1,0,0}}%
      \expandafter\def\csname LT1\endcsname{\color[rgb]{0,1,0}}%
      \expandafter\def\csname LT2\endcsname{\color[rgb]{0,0,1}}%
      \expandafter\def\csname LT3\endcsname{\color[rgb]{1,0,1}}%
      \expandafter\def\csname LT4\endcsname{\color[rgb]{0,1,1}}%
      \expandafter\def\csname LT5\endcsname{\color[rgb]{1,1,0}}%
      \expandafter\def\csname LT6\endcsname{\color[rgb]{0,0,0}}%
      \expandafter\def\csname LT7\endcsname{\color[rgb]{1,0.3,0}}%
      \expandafter\def\csname LT8\endcsname{\color[rgb]{0.5,0.5,0.5}}%
    \else
      \def\colorrgb#1{\color{black}}%
      \def\colorgray#1{\color[gray]{#1}}%
      \expandafter\def\csname LTw\endcsname{\color{white}}%
      \expandafter\def\csname LTb\endcsname{\color{black}}%
      \expandafter\def\csname LTa\endcsname{\color{black}}%
      \expandafter\def\csname LT0\endcsname{\color{black}}%
      \expandafter\def\csname LT1\endcsname{\color{black}}%
      \expandafter\def\csname LT2\endcsname{\color{black}}%
      \expandafter\def\csname LT3\endcsname{\color{black}}%
      \expandafter\def\csname LT4\endcsname{\color{black}}%
      \expandafter\def\csname LT5\endcsname{\color{black}}%
      \expandafter\def\csname LT6\endcsname{\color{black}}%
      \expandafter\def\csname LT7\endcsname{\color{black}}%
      \expandafter\def\csname LT8\endcsname{\color{black}}%
    \fi
  \fi
  \setlength{\unitlength}{0.0500bp}%
  \begin{picture}(11338.00,8502.00)%
    \gplgaddtomacro\gplbacktext{%
      \csname LTb\endcsname%
      \put(946,4955){\makebox(0,0)[r]{\strut{} 0}}%
      \put(946,5611){\makebox(0,0)[r]{\strut{} 0.2}}%
      \put(946,6268){\makebox(0,0)[r]{\strut{} 0.4}}%
      \put(946,6924){\makebox(0,0)[r]{\strut{} 0.6}}%
      \put(946,7581){\makebox(0,0)[r]{\strut{} 0.8}}%
      \put(946,8237){\makebox(0,0)[r]{\strut{} 1}}%
      \put(1340,4735){\makebox(0,0){\strut{} 2}}%
      \put(1864,4735){\makebox(0,0){\strut{} 4}}%
      \put(2389,4735){\makebox(0,0){\strut{} 6}}%
      \put(2913,4735){\makebox(0,0){\strut{} 8}}%
      \put(3437,4735){\makebox(0,0){\strut{} 10}}%
      \put(3961,4735){\makebox(0,0){\strut{} 12}}%
      \put(4486,4735){\makebox(0,0){\strut{} 14}}%
      \put(5010,4735){\makebox(0,0){\strut{} 16}}%
      \put(176,6596){\rotatebox{-270}{\makebox(0,0){\strut{}$\varphi$}}}%
      \put(3175,4405){\makebox(0,0){\strut{}$\rho$}}%
    }%
    \gplgaddtomacro\gplfronttext{%
    }%
    \gplgaddtomacro\gplbacktext{%
      \csname LTb\endcsname%
      \put(6747,4955){\makebox(0,0)[r]{\strut{} 0.3}}%
      \put(6747,5776){\makebox(0,0)[r]{\strut{} 0.35}}%
      \put(6747,6596){\makebox(0,0)[r]{\strut{} 0.4}}%
      \put(6747,7417){\makebox(0,0)[r]{\strut{} 0.45}}%
      \put(6747,8237){\makebox(0,0)[r]{\strut{} 0.5}}%
      \put(7133,4735){\makebox(0,0){\strut{} 2}}%
      \put(7641,4735){\makebox(0,0){\strut{} 4}}%
      \put(8148,4735){\makebox(0,0){\strut{} 6}}%
      \put(8656,4735){\makebox(0,0){\strut{} 8}}%
      \put(9164,4735){\makebox(0,0){\strut{} 10}}%
      \put(9672,4735){\makebox(0,0){\strut{} 12}}%
      \put(10179,4735){\makebox(0,0){\strut{} 14}}%
      \put(10687,4735){\makebox(0,0){\strut{} 16}}%
      \put(5845,6596){\rotatebox{-270}{\makebox(0,0){\strut{}$e^{-2H}$}}}%
      \put(8910,4405){\makebox(0,0){\strut{}$\rho$}}%
    }%
    \gplgaddtomacro\gplfronttext{%
    }%
    \gplgaddtomacro\gplbacktext{%
      \csname LTb\endcsname%
      \put(682,704){\makebox(0,0)[r]{\strut{} 0}}%
      \put(682,1361){\makebox(0,0)[r]{\strut{} 1}}%
      \put(682,2017){\makebox(0,0)[r]{\strut{} 2}}%
      \put(682,2674){\makebox(0,0)[r]{\strut{} 3}}%
      \put(682,3330){\makebox(0,0)[r]{\strut{} 4}}%
      \put(682,3987){\makebox(0,0)[r]{\strut{} 5}}%
      \put(1093,484){\makebox(0,0){\strut{} 2}}%
      \put(1650,484){\makebox(0,0){\strut{} 4}}%
      \put(2207,484){\makebox(0,0){\strut{} 6}}%
      \put(2764,484){\makebox(0,0){\strut{} 8}}%
      \put(3322,484){\makebox(0,0){\strut{} 10}}%
      \put(3879,484){\makebox(0,0){\strut{} 12}}%
      \put(4436,484){\makebox(0,0){\strut{} 14}}%
      \put(4993,484){\makebox(0,0){\strut{} 16}}%
      \put(176,2345){\rotatebox{-270}{\makebox(0,0){\strut{}$\partial_{\rho} F$}}}%
      \put(3043,154){\makebox(0,0){\strut{}$\rho$}}%
    }%
    \gplgaddtomacro\gplfronttext{%
    }%
    \gplgaddtomacro\gplbacktext{%
      \csname LTb\endcsname%
      \put(6615,704){\makebox(0,0)[r]{\strut{} 0}}%
      \put(6615,1142){\makebox(0,0)[r]{\strut{} 0.2}}%
      \put(6615,1579){\makebox(0,0)[r]{\strut{} 0.4}}%
      \put(6615,2017){\makebox(0,0)[r]{\strut{} 0.6}}%
      \put(6615,2455){\makebox(0,0)[r]{\strut{} 0.8}}%
      \put(6615,2893){\makebox(0,0)[r]{\strut{} 1}}%
      \put(6615,3330){\makebox(0,0)[r]{\strut{} 1.2}}%
      \put(6615,3768){\makebox(0,0)[r]{\strut{} 1.4}}%
      \put(7009,484){\makebox(0,0){\strut{} 2}}%
      \put(7533,484){\makebox(0,0){\strut{} 4}}%
      \put(8058,484){\makebox(0,0){\strut{} 6}}%
      \put(8582,484){\makebox(0,0){\strut{} 8}}%
      \put(9106,484){\makebox(0,0){\strut{} 10}}%
      \put(9630,484){\makebox(0,0){\strut{} 12}}%
      \put(10155,484){\makebox(0,0){\strut{} 14}}%
      \put(10679,484){\makebox(0,0){\strut{} 16}}%
      \put(5845,2345){\rotatebox{-270}{\makebox(0,0){\strut{}$\beta$}}}%
      \put(8844,154){\makebox(0,0){\strut{}$\rho$}}%
    }%
    \gplgaddtomacro\gplfronttext{%
    }%
    \gplbacktext
    \put(0,0){\includegraphics{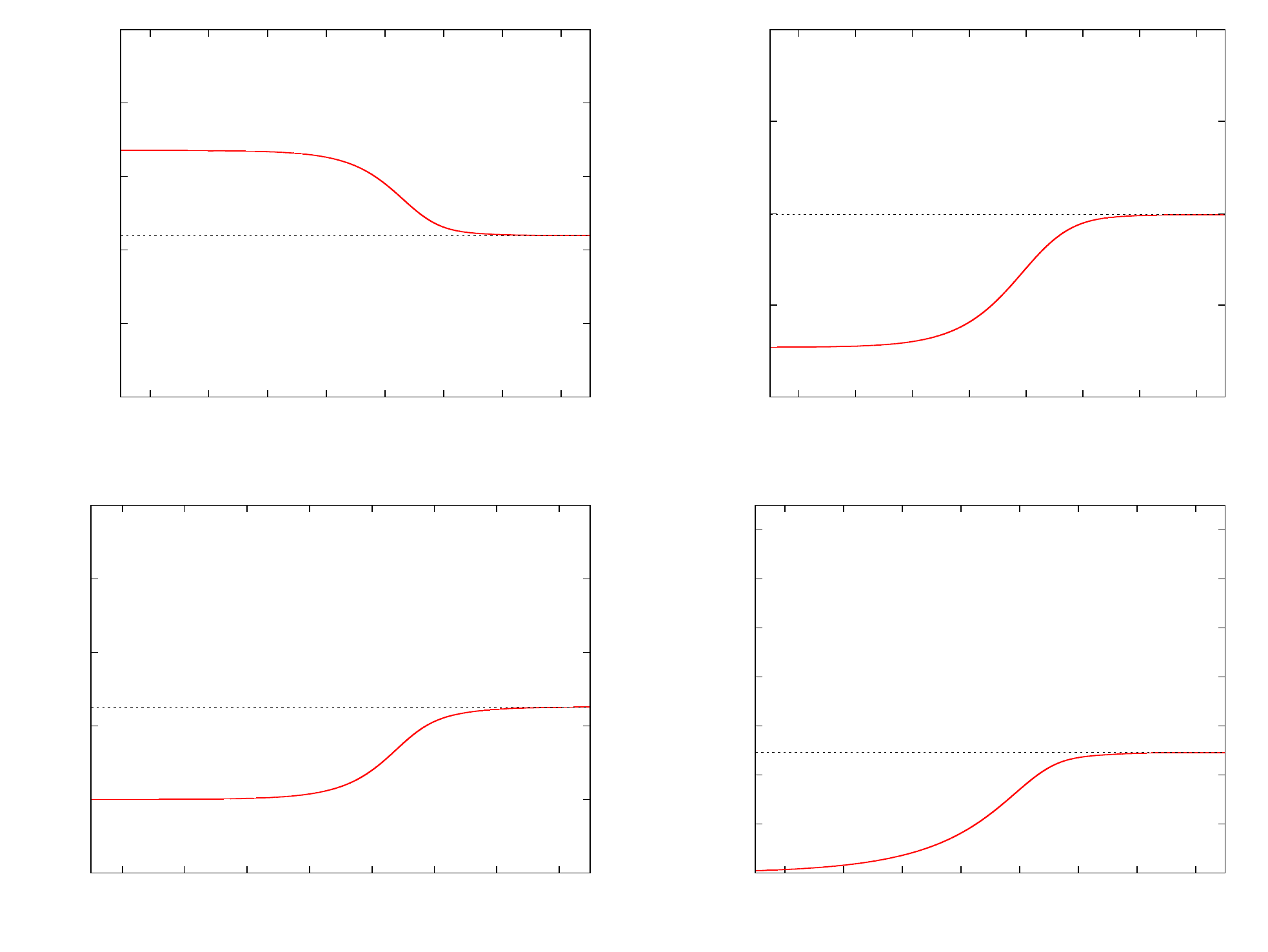}}%
    \gplfronttext
  \end{picture}%
\endgroup

%% file: Li_AdS_8.tex
\begingroup
  \makeatletter
  \providecommand\color[2][]{%
    \GenericError{(gnuplot) \space\space\space\@spaces}{%
      Package color not loaded in conjunction with
      terminal option `colourtext'%
    }{See the gnuplot documentation for explanation.%
    }{Either use 'blacktext' in gnuplot or load the package
      color.sty in LaTeX.}%
    \renewcommand\color[2][]{}%
  }%
  \providecommand\includegraphics[2][]{%
    \GenericError{(gnuplot) \space\space\space\@spaces}{%
      Package graphicx or graphics not loaded%
    }{See the gnuplot documentation for explanation.%
    }{The gnuplot epslatex terminal needs graphicx.sty or graphics.sty.}%
    \renewcommand\includegraphics[2][]{}%
  }%
  \providecommand\rotatebox[2]{#2}%
  \@ifundefined{ifGPcolor}{%
    \newif\ifGPcolor
    \GPcolortrue
  }{}%
  \@ifundefined{ifGPblacktext}{%
    \newif\ifGPblacktext
    \GPblacktexttrue
  }{}%
  \let\gplgaddtomacro\g@addto@macro
  \gdef\gplbacktext{}%
  \gdef\gplfronttext{}%
  \makeatother
  \ifGPblacktext
    \def\colorrgb#1{}%
    \def\colorgray#1{}%
  \else
    \ifGPcolor
      \def\colorrgb#1{\color[rgb]{#1}}%
      \def\colorgray#1{\color[gray]{#1}}%
      \expandafter\def\csname LTw\endcsname{\color{white}}%
      \expandafter\def\csname LTb\endcsname{\color{black}}%
      \expandafter\def\csname LTa\endcsname{\color{black}}%
      \expandafter\def\csname LT0\endcsname{\color[rgb]{1,0,0}}%
      \expandafter\def\csname LT1\endcsname{\color[rgb]{0,1,0}}%
      \expandafter\def\csname LT2\endcsname{\color[rgb]{0,0,1}}%
      \expandafter\def\csname LT3\endcsname{\color[rgb]{1,0,1}}%
      \expandafter\def\csname LT4\endcsname{\color[rgb]{0,1,1}}%
      \expandafter\def\csname LT5\endcsname{\color[rgb]{1,1,0}}%
      \expandafter\def\csname LT6\endcsname{\color[rgb]{0,0,0}}%
      \expandafter\def\csname LT7\endcsname{\color[rgb]{1,0.3,0}}%
      \expandafter\def\csname LT8\endcsname{\color[rgb]{0.5,0.5,0.5}}%
    \else
      \def\colorrgb#1{\color{black}}%
      \def\colorgray#1{\color[gray]{#1}}%
      \expandafter\def\csname LTw\endcsname{\color{white}}%
      \expandafter\def\csname LTb\endcsname{\color{black}}%
      \expandafter\def\csname LTa\endcsname{\color{black}}%
      \expandafter\def\csname LT0\endcsname{\color{black}}%
      \expandafter\def\csname LT1\endcsname{\color{black}}%
      \expandafter\def\csname LT2\endcsname{\color{black}}%
      \expandafter\def\csname LT3\endcsname{\color{black}}%
      \expandafter\def\csname LT4\endcsname{\color{black}}%
      \expandafter\def\csname LT5\endcsname{\color{black}}%
      \expandafter\def\csname LT6\endcsname{\color{black}}%
      \expandafter\def\csname LT7\endcsname{\color{black}}%
      \expandafter\def\csname LT8\endcsname{\color{black}}%
    \fi
  \fi
  \setlength{\unitlength}{0.0500bp}%
  \begin{picture}(11338.00,8502.00)%
    \gplgaddtomacro\gplbacktext{%
      \csname LTb\endcsname%
      \put(946,4955){\makebox(0,0)[r]{\strut{} 0}}%
      \put(946,5611){\makebox(0,0)[r]{\strut{} 0.2}}%
      \put(946,6268){\makebox(0,0)[r]{\strut{} 0.4}}%
      \put(946,6924){\makebox(0,0)[r]{\strut{} 0.6}}%
      \put(946,7581){\makebox(0,0)[r]{\strut{} 0.8}}%
      \put(946,8237){\makebox(0,0)[r]{\strut{} 1}}%
      \put(1497,4735){\makebox(0,0){\strut{} 2}}%
      \put(2336,4735){\makebox(0,0){\strut{} 4}}%
      \put(3175,4735){\makebox(0,0){\strut{} 6}}%
      \put(4014,4735){\makebox(0,0){\strut{} 8}}%
      \put(4853,4735){\makebox(0,0){\strut{} 10}}%
      \put(176,6596){\rotatebox{-270}{\makebox(0,0){\strut{}$\varphi$}}}%
      \put(3175,4405){\makebox(0,0){\strut{}$\rho$}}%
    }%
    \gplgaddtomacro\gplfronttext{%
    }%
    \gplgaddtomacro\gplbacktext{%
      \csname LTb\endcsname%
      \put(6747,4955){\makebox(0,0)[r]{\strut{} 0.2}}%
      \put(6747,5502){\makebox(0,0)[r]{\strut{} 0.25}}%
      \put(6747,6049){\makebox(0,0)[r]{\strut{} 0.3}}%
      \put(6747,6596){\makebox(0,0)[r]{\strut{} 0.35}}%
      \put(6747,7143){\makebox(0,0)[r]{\strut{} 0.4}}%
      \put(6747,7690){\makebox(0,0)[r]{\strut{} 0.45}}%
      \put(6747,8237){\makebox(0,0)[r]{\strut{} 0.5}}%
      \put(7285,4735){\makebox(0,0){\strut{} 2}}%
      \put(8098,4735){\makebox(0,0){\strut{} 4}}%
      \put(8910,4735){\makebox(0,0){\strut{} 6}}%
      \put(9722,4735){\makebox(0,0){\strut{} 8}}%
      \put(10535,4735){\makebox(0,0){\strut{} 10}}%
      \put(5845,6596){\rotatebox{-270}{\makebox(0,0){\strut{}$e^{-2H}$}}}%
      \put(8910,4405){\makebox(0,0){\strut{}$\rho$}}%
    }%
    \gplgaddtomacro\gplfronttext{%
    }%
    \gplgaddtomacro\gplbacktext{%
      \csname LTb\endcsname%
      \put(814,704){\makebox(0,0)[r]{\strut{} 0}}%
      \put(814,1361){\makebox(0,0)[r]{\strut{} 2}}%
      \put(814,2017){\makebox(0,0)[r]{\strut{} 4}}%
      \put(814,2674){\makebox(0,0)[r]{\strut{} 6}}%
      \put(814,3330){\makebox(0,0)[r]{\strut{} 8}}%
      \put(814,3987){\makebox(0,0)[r]{\strut{} 10}}%
      \put(1379,484){\makebox(0,0){\strut{} 2}}%
      \put(2244,484){\makebox(0,0){\strut{} 4}}%
      \put(3109,484){\makebox(0,0){\strut{} 6}}%
      \put(3974,484){\makebox(0,0){\strut{} 8}}%
      \put(4839,484){\makebox(0,0){\strut{} 10}}%
      \put(176,2345){\rotatebox{-270}{\makebox(0,0){\strut{}$\partial_{\rho} F$}}}%
      \put(3109,154){\makebox(0,0){\strut{}$\rho$}}%
    }%
    \gplgaddtomacro\gplfronttext{%
    }%
    \gplgaddtomacro\gplbacktext{%
      \csname LTb\endcsname%
      \put(6615,704){\makebox(0,0)[r]{\strut{} 0}}%
      \put(6615,1142){\makebox(0,0)[r]{\strut{} 0.2}}%
      \put(6615,1579){\makebox(0,0)[r]{\strut{} 0.4}}%
      \put(6615,2017){\makebox(0,0)[r]{\strut{} 0.6}}%
      \put(6615,2455){\makebox(0,0)[r]{\strut{} 0.8}}%
      \put(6615,2893){\makebox(0,0)[r]{\strut{} 1}}%
      \put(6615,3330){\makebox(0,0)[r]{\strut{} 1.2}}%
      \put(6615,3768){\makebox(0,0)[r]{\strut{} 1.4}}%
      \put(7166,484){\makebox(0,0){\strut{} 2}}%
      \put(8005,484){\makebox(0,0){\strut{} 4}}%
      \put(8844,484){\makebox(0,0){\strut{} 6}}%
      \put(9683,484){\makebox(0,0){\strut{} 8}}%
      \put(10522,484){\makebox(0,0){\strut{} 10}}%
      \put(5845,2345){\rotatebox{-270}{\makebox(0,0){\strut{}$\beta$}}}%
      \put(8844,154){\makebox(0,0){\strut{}$\rho$}}%
    }%
    \gplgaddtomacro\gplfronttext{%
    }%
    \gplbacktext
    \put(0,0){\includegraphics{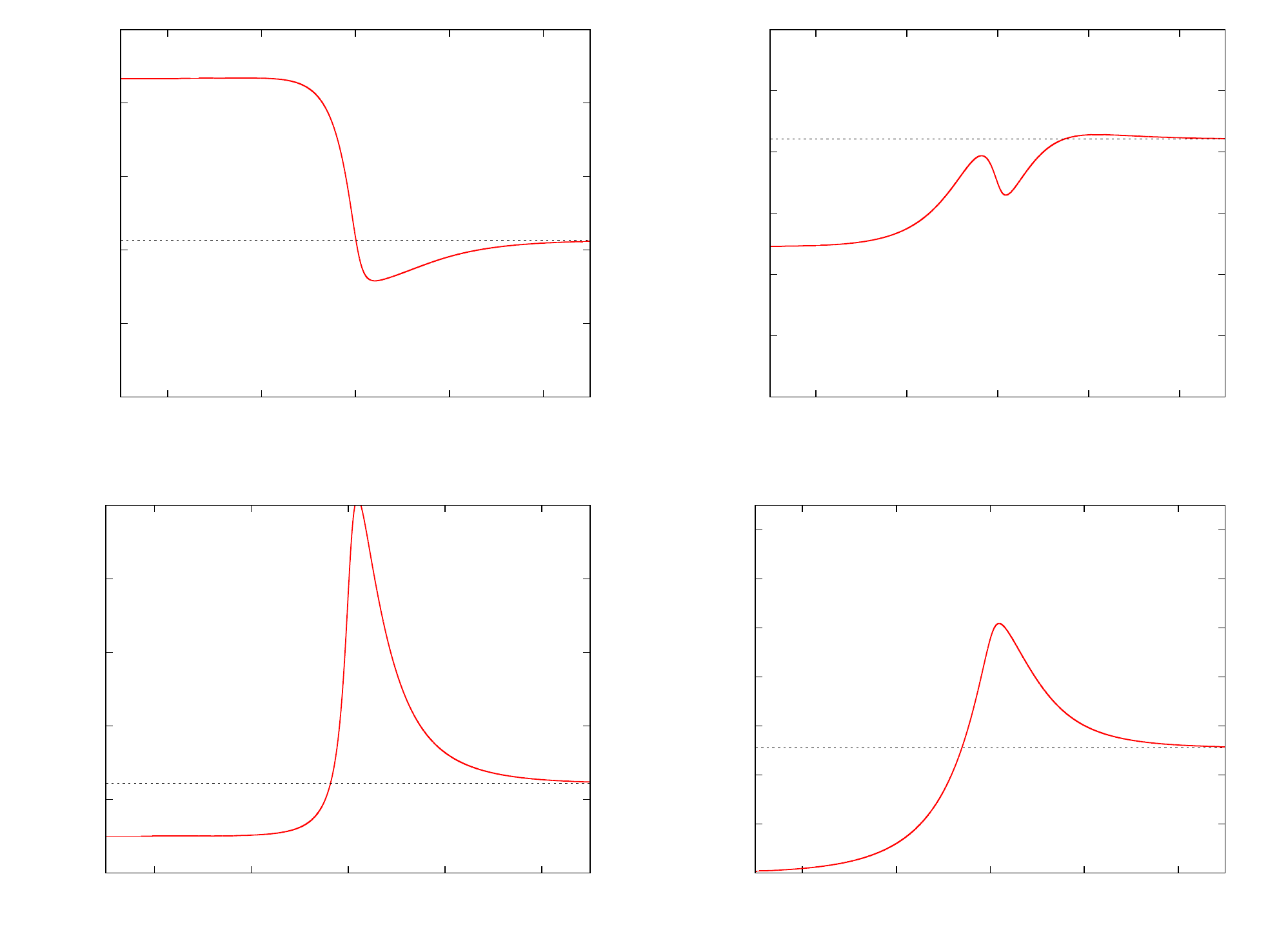}}%
    \gplfronttext
  \end{picture}%
\endgroup

%% file: Li_AdS_AdS_1.tex
\begingroup
  \makeatletter
  \providecommand\color[2][]{%
    \GenericError{(gnuplot) \space\space\space\@spaces}{%
      Package color not loaded in conjunction with
      terminal option `colourtext'%
    }{See the gnuplot documentation for explanation.%
    }{Either use 'blacktext' in gnuplot or load the package
      color.sty in LaTeX.}%
    \renewcommand\color[2][]{}%
  }%
  \providecommand\includegraphics[2][]{%
    \GenericError{(gnuplot) \space\space\space\@spaces}{%
      Package graphicx or graphics not loaded%
    }{See the gnuplot documentation for explanation.%
    }{The gnuplot epslatex terminal needs graphicx.sty or graphics.sty.}%
    \renewcommand\includegraphics[2][]{}%
  }%
  \providecommand\rotatebox[2]{#2}%
  \@ifundefined{ifGPcolor}{%
    \newif\ifGPcolor
    \GPcolortrue
  }{}%
  \@ifundefined{ifGPblacktext}{%
    \newif\ifGPblacktext
    \GPblacktexttrue
  }{}%
  \let\gplgaddtomacro\g@addto@macro
  \gdef\gplbacktext{}%
  \gdef\gplfronttext{}%
  \makeatother
  \ifGPblacktext
    \def\colorrgb#1{}%
    \def\colorgray#1{}%
  \else
    \ifGPcolor
      \def\colorrgb#1{\color[rgb]{#1}}%
      \def\colorgray#1{\color[gray]{#1}}%
      \expandafter\def\csname LTw\endcsname{\color{white}}%
      \expandafter\def\csname LTb\endcsname{\color{black}}%
      \expandafter\def\csname LTa\endcsname{\color{black}}%
      \expandafter\def\csname LT0\endcsname{\color[rgb]{1,0,0}}%
      \expandafter\def\csname LT1\endcsname{\color[rgb]{0,1,0}}%
      \expandafter\def\csname LT2\endcsname{\color[rgb]{0,0,1}}%
      \expandafter\def\csname LT3\endcsname{\color[rgb]{1,0,1}}%
      \expandafter\def\csname LT4\endcsname{\color[rgb]{0,1,1}}%
      \expandafter\def\csname LT5\endcsname{\color[rgb]{1,1,0}}%
      \expandafter\def\csname LT6\endcsname{\color[rgb]{0,0,0}}%
      \expandafter\def\csname LT7\endcsname{\color[rgb]{1,0.3,0}}%
      \expandafter\def\csname LT8\endcsname{\color[rgb]{0.5,0.5,0.5}}%
    \else
      \def\colorrgb#1{\color{black}}%
      \def\colorgray#1{\color[gray]{#1}}%
      \expandafter\def\csname LTw\endcsname{\color{white}}%
      \expandafter\def\csname LTb\endcsname{\color{black}}%
      \expandafter\def\csname LTa\endcsname{\color{black}}%
      \expandafter\def\csname LT0\endcsname{\color{black}}%
      \expandafter\def\csname LT1\endcsname{\color{black}}%
      \expandafter\def\csname LT2\endcsname{\color{black}}%
      \expandafter\def\csname LT3\endcsname{\color{black}}%
      \expandafter\def\csname LT4\endcsname{\color{black}}%
      \expandafter\def\csname LT5\endcsname{\color{black}}%
      \expandafter\def\csname LT6\endcsname{\color{black}}%
      \expandafter\def\csname LT7\endcsname{\color{black}}%
      \expandafter\def\csname LT8\endcsname{\color{black}}%
    \fi
  \fi
  \setlength{\unitlength}{0.0500bp}%
  \begin{picture}(11338.00,8502.00)%
    \gplgaddtomacro\gplbacktext{%
      \csname LTb\endcsname%
      \put(946,4955){\makebox(0,0)[r]{\strut{} 0}}%
      \put(946,5611){\makebox(0,0)[r]{\strut{} 0.2}}%
      \put(946,6268){\makebox(0,0)[r]{\strut{} 0.4}}%
      \put(946,6924){\makebox(0,0)[r]{\strut{} 0.6}}%
      \put(946,7581){\makebox(0,0)[r]{\strut{} 0.8}}%
      \put(946,8237){\makebox(0,0)[r]{\strut{} 1}}%
      \put(1917,4735){\makebox(0,0){\strut{} 5}}%
      \put(2965,4735){\makebox(0,0){\strut{} 10}}%
      \put(4014,4735){\makebox(0,0){\strut{} 15}}%
      \put(5062,4735){\makebox(0,0){\strut{} 20}}%
      \put(176,6596){\rotatebox{-270}{\makebox(0,0){\strut{}$\varphi$}}}%
      \put(3175,4405){\makebox(0,0){\strut{}$\rho$}}%
    }%
    \gplgaddtomacro\gplfronttext{%
    }%
    \gplgaddtomacro\gplbacktext{%
      \csname LTb\endcsname%
      \put(6747,4955){\makebox(0,0)[r]{\strut{} 0.2}}%
      \put(6747,5502){\makebox(0,0)[r]{\strut{} 0.25}}%
      \put(6747,6049){\makebox(0,0)[r]{\strut{} 0.3}}%
      \put(6747,6596){\makebox(0,0)[r]{\strut{} 0.35}}%
      \put(6747,7143){\makebox(0,0)[r]{\strut{} 0.4}}%
      \put(6747,7690){\makebox(0,0)[r]{\strut{} 0.45}}%
      \put(6747,8237){\makebox(0,0)[r]{\strut{} 0.5}}%
      \put(7691,4735){\makebox(0,0){\strut{} 5}}%
      \put(8707,4735){\makebox(0,0){\strut{} 10}}%
      \put(9722,4735){\makebox(0,0){\strut{} 15}}%
      \put(10738,4735){\makebox(0,0){\strut{} 20}}%
      \put(5845,6596){\rotatebox{-270}{\makebox(0,0){\strut{}$e^{-2H}$}}}%
      \put(8910,4405){\makebox(0,0){\strut{}$\rho$}}%
    }%
    \gplgaddtomacro\gplfronttext{%
    }%
    \gplgaddtomacro\gplbacktext{%
      \csname LTb\endcsname%
      \put(946,704){\makebox(0,0)[r]{\strut{} 0}}%
      \put(946,1251){\makebox(0,0)[r]{\strut{} 0.5}}%
      \put(946,1798){\makebox(0,0)[r]{\strut{} 1}}%
      \put(946,2346){\makebox(0,0)[r]{\strut{} 1.5}}%
      \put(946,2893){\makebox(0,0)[r]{\strut{} 2}}%
      \put(946,3440){\makebox(0,0)[r]{\strut{} 2.5}}%
      \put(946,3987){\makebox(0,0)[r]{\strut{} 3}}%
      \put(1917,484){\makebox(0,0){\strut{} 5}}%
      \put(2965,484){\makebox(0,0){\strut{} 10}}%
      \put(4014,484){\makebox(0,0){\strut{} 15}}%
      \put(5062,484){\makebox(0,0){\strut{} 20}}%
      \put(176,2345){\rotatebox{-270}{\makebox(0,0){\strut{}$\partial_{\rho} F$}}}%
      \put(3175,154){\makebox(0,0){\strut{}$\rho$}}%
    }%
    \gplgaddtomacro\gplfronttext{%
    }%
    \gplgaddtomacro\gplbacktext{%
      \csname LTb\endcsname%
      \put(6615,704){\makebox(0,0)[r]{\strut{} 0}}%
      \put(6615,1361){\makebox(0,0)[r]{\strut{} 0.2}}%
      \put(6615,2017){\makebox(0,0)[r]{\strut{} 0.4}}%
      \put(6615,2674){\makebox(0,0)[r]{\strut{} 0.6}}%
      \put(6615,3330){\makebox(0,0)[r]{\strut{} 0.8}}%
      \put(6615,3987){\makebox(0,0)[r]{\strut{} 1}}%
      \put(7586,484){\makebox(0,0){\strut{} 5}}%
      \put(8634,484){\makebox(0,0){\strut{} 10}}%
      \put(9683,484){\makebox(0,0){\strut{} 15}}%
      \put(10731,484){\makebox(0,0){\strut{} 20}}%
      \put(5845,2345){\rotatebox{-270}{\makebox(0,0){\strut{}$\beta$}}}%
      \put(8844,154){\makebox(0,0){\strut{}$\rho$}}%
    }%
    \gplgaddtomacro\gplfronttext{%
    }%
    \gplbacktext
    \put(0,0){\includegraphics{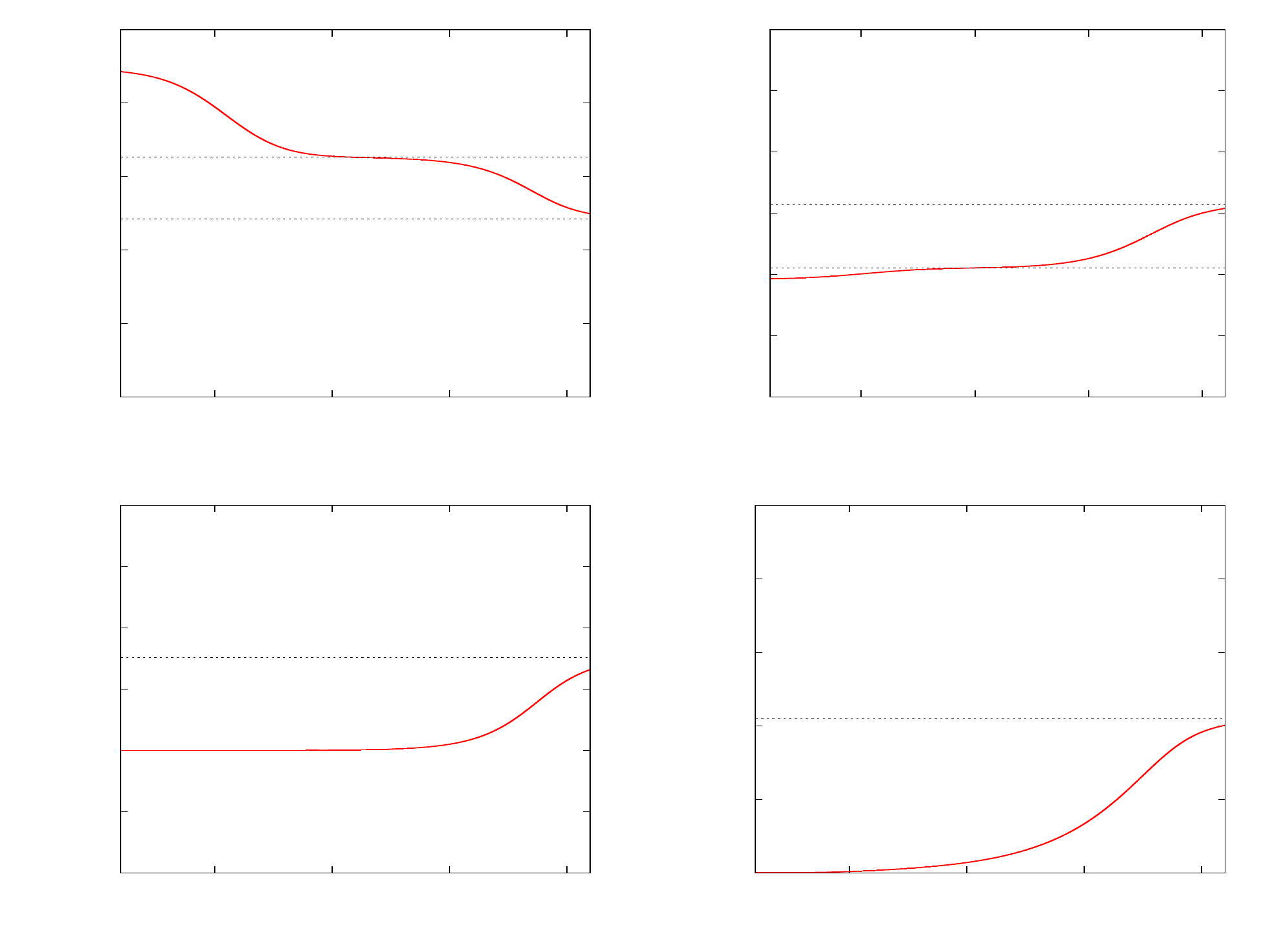}}%
    \gplfronttext
  \end{picture}%
\endgroup